\documentclass[galaxies,article,submit,moreauthors,pdftex]{mdpi}
\usepackage{graphicx,color}
\usepackage{amsmath}
\usepackage{natbib}
\usepackage{amssymb}
\usepackage{bm}
\usepackage{ulem,cancel}
\usepackage{mathtools}	%needed for dcase environment, loads amsmath package
\usepackage{enumitem}
\usepackage{upgreek}
\preto{\abstractkeywords}{\nolinenumbers}	%to disable line numbers to satisfy arXiv rules

\newcommand{\Exp}[1]{{\rm e}^{#1}}

\newcommand{\bfx}{\bm{x}}

\newcommand{\eps}{\epsilon}

\newcommand{\f}{_\mathrm{0}}					   	%forcing
\newcommand{\diss}{_\mathrm{d}}			   		%dissipation, subscript
\newcommand{\atilde}{\alpha}

\newcommand{\Atilde}{\widetilde{A}}

\newcommand{\ktilde}{\widetilde{k}}

\newcommand{\Strouhal}{\mathrm{St}}

\newcommand{\ergs}{\mathrm{erg}\,\mathrm{s}^{-1}}
\newcommand{\ergpersperkpccube}{\mathrm{erg}\,\mathrm{s}^{-1}\,\mathrm{kpc}^{-3}}
\newcommand{\SN}{_\mathrm{SN}}
\newcommand{\ESN}{E_0}
\newcommand{\SB}{_\mathrm{SB}}
\newcommand{\uSN}{^\mathrm{SN}}
\newcommand{\uSB}{^\mathrm{SB}}
\newcommand{\OB}{_\mathrm{SB}}
\newcommand{\sound}{_\mathrm{s}}

\newcommand{\blowout}{_\mathrm{b}}
\newcommand{\eddy}{^\mathrm{e}}
\newcommand{\renov}{^\mathrm{r}}
\newcommand{\E}{E}
\newcommand{\Etilde}{\widetilde{\E}}
\newcommand{\inj}{^\mathrm{i}}
\newcommand{\dis}{^\mathrm{d}}

\newcommand{\outflow}{^\mathrm{out}}
\makeatletter
\newsavebox{\@brx}

\newcommand{\llangle}[1][]{\savebox{\@brx}{\(\m@th{#1\langle}\)}%
  \mathopen{\copy\@brx\kern-0.5\wd\@brx\usebox{\@brx}}}
\newcommand{\rrangle}[1][]{\savebox{\@brx}{\(\m@th{#1\rangle}\)}%
  \mathclose{\copy\@brx\kern-0.5\wd\@brx\usebox{\@brx}}}

  \newcommand{\cmcmcm}{\,{\rm cm^{-3}}}
  \newcommand{\cmcube}{\,{\rm cm^{-3}}}
  
  \newcommand{\erg}{\,{\rm erg}}

  \newcommand{\kms}{\,{\rm km\,s^{-1}}}
  \newcommand{\kmskpc}{\,{\rm km\,s^{-1}\,kpc^{-1}}}

  \newcommand{\kpc}{\,{\rm kpc}}
  \newcommand{\pc}{\,{\rm pc}}
  
  \newcommand{\Myr}{\,{\rm Myr}}

  \newcommand{\p}{\,{\rm pc}}

\definecolor{webgreen}{rgb}{0,.5,0}
\definecolor{webbrown}{rgb}{.6,0,0}
\definecolor{purple}{rgb}{0.5,0,.5}
\Title{Parameters of the Supernova-Driven Interstellar Turbulence}
\Author{Luke~Chamandy$^{1}$*\orcidA{} \& 
Anvar~Shukurov$^{2}$\orcidB{} }
\AuthorNames{Luke Chamandy and Anvar Shukurov}
\address{$^{1}$ \quad Department of Physics and Astronomy, University of Rochester, Rochester NY, 14627, USA\\
$^{2}$ \quad School of Mathematics, Statistics \& Physics, Newcastle University, Newcastle upon Tyne, NE1 7RU, UK}
\corres{Correspondence: lchamandy@pas.rochester.edu, anvar.shukurov@newcastle.ac.uk}

\abstract{
Galactic dynamo models take as input certain parameters of the interstellar turbulence, most essentially the correlation time $\tau$, root-mean-square turbulent speed $u$, and correlation scale $l$.  However, these quantities are difficult, or, in the case of $\tau$, impossible, to directly observe, and theorists have mostly relied on order of magnitude estimates.  Here we present an analytic model to derive these quantities in terms of a small set of more accessible parameters.  In our model, turbulence is assumed to be driven concurrently by isolated supernovae (SNe) and superbubbles (SBs), but clustering of SNe to form SBs can be turned off if desired, which reduces the number of model parameters by about half.  In general, we find that isolated SNe and SBs can inject comparable amounts of turbulent energy into the interstellar medium, but SBs do so less efficiently.  This results in rather low overall conversion rates of SN energy into turbulent energy of $\sim1$--$3\%$.  The results obtained for $l$, $u$ and $\tau$ for model parameter values representative of the Solar neighbourhood are consistent with those determined from direct numerical simulations.  Our analytic model can be combined with existing dynamo models to predict more directly the magnetic field properties for nearby galaxies or for statistical populations of galaxies in cosmological models.
}

\keyword{turbulence; galaxies: ISM; ISM: kinematics and dynamics; dynamo; galaxies: spiral; galaxies: magnetic fields}

\begin{document}

\section{Introduction}
\label{sec:intro}
Turbulence affects a wide range of physical processes in the interstellar medium (ISM) of spiral galaxies,
including the turbulent dynamo.
Its spectrum extends over a wide range of scales \citep{Armstrong+95, Chepurnov+Lazarian10},
apparently maintained by diverse physical effects.
Turbulent flows are characterized by certain physical parameters,
which are defined through some sort of averaging.
These include the velocity correlation scale (or integral scale) $l$, 
which is similar to, but smaller than, the scale of the force driving the turbulence, 
the root-mean-square (RMS) turbulent speed $u$, and the turbulent correlation time $\tau$.

In this work, our aim is to obtain scaling relations for the turbulence parameters $l$, $u$ and $\tau$ 
in terms of other quantities like gas density, sound speed, and supernova rate density. 
The latter parameters are often readily computed from observations or models, 
so such relations can provide a missing link. 
We are motivated by one application in particular:
the galactic dynamo, which is responsible for amplifying a galaxy's magnetic field up to an energy 
comparable with that of the turbulence.
The properties of the magnetic field as it evolves and saturates are predicted to depend on the values of the parameters $u$, $\tau$ and $l$
\citep{Ruzmaikin+88,Beck+96,Brandenburg+Subramanian05a,Shukurov07,Beck+19}.

Stellar feedback probably contributes most of the energy injected into turbulence, 
at least in the Milky Way and in nearby star-forming galaxies \citep{Klessen+Glover14,Krumholz+18,Bacchini+20}.
This energy is mainly supplied by supernovae (SNe), as this source generally dominates over other stellar contributions
like winds \citep[e.g.][]{Leitherer+99,El-badry+19}
and the expansion of HII regions due to radiation pressure and ionized gas pressure \citep[e.g.][]{Kim+18}.
However, in some cases sources of interstellar turbulence other than SNe are likely to be important
\citep{Maclow+Klessen04,Elmegreen+Scalo04,Klessen+Glover14,Falceta-goncalves+14,Falceta-goncalves+15,Vazquez-semadeni15,Krumholz+18}.
Galaxies accrete gas---if gas were not replenished in this way then star formation would be quenched too early to explain observations.
Accretion onto the galactic disc could predominantly occur in the disc outer region, 
but inward flows can transport this gas to smaller radii \citep[e.g.][]{Schmidt+16}.
This potential source of turbulence can be difficult to estimate, but could dominate in some cases \citep{Klessen+Glover14,Krumholz+18}.
Spiral arms may also contribute to driving turbulence through various proposed mechanisms, 
but this contribution is likely subdominant \citep{Elmegreen+Scalo04,Klessen+Glover14}.
The magnetorotational instability could also drive interstellar turbulence, tapping energy from galactic differential rotation. 
This mechanism may be important in the outer disc where the star formation rate is small \citep{Sellwood+Balbus99}, 
but is otherwise likely to be subdominant \citep{Maclow+Klessen04,Klessen+Glover14,Bacchini+20}.
For simplicity, we will assume in this work that the turbulence is driven exclusively by SNe.

To 
derive the correlation scale $l$, 
we first estimate the turbulence driving (or injection) scale.
The driving scale is usually assumed to be of the order 
of the typical size of a supernova remnant (SNR).
However, this assumption ignores that SNe tend to be clustered together in OB associations, 
leading to the formation of much larger structures known as superbubbles (SBs),
which can drive turbulence on larger scales \citep[e.g.][]{Norman+Ferrara96}.

The ISM 
has a multi-phase structure, with a cold phase concentrated into dense clouds near the midplane,
a more diffuse warm phase, and a transient low density hot phase.
For simplicity, our model treats the ISM as a uniform medium, neglecting its various phases.
As such, the turbulence parameters we derive can be interpreted as volume- and time-averaged quantities.

There have been quite a few observational studies attempting to measure various length scales 
associated with turbulence, be it in the Milky Way or nearby galaxies,
using differing methodologies and assumptions 
\citep[e.g.][]{Elmegreen+01,Stanimirovic+Lazarian01,Haverkorn+04,Dib+Burkert05,Chepurnov+10,Dutta+13}.
However, interpreting such observations to extract particular physical scales used in theoretical models,
and the dependence of these scales on other parameters, 
is still very challenging, and estimates are prone to large uncertainties.

Modeling the dependence of the RMS turbulent speed $u$ on parameters like the SN rate density 
(which can be related to the star formation rate surface density) is desirable. 
Some progress along these lines has recently been made. 
For example, to estimate the magnetic field strength as a function of gas density, star formation rate density and gas scale height, 
\citet{Schober+16} assume proportionality between the turbulent magnetic and kinetic energy densities.
To estimate the turbulent kinetic energy density, 
they derive an expression for the turbulent velocity 
by assuming equal injection and dissipation rate densities of turbulence.
To compute the injection rate, they introduce the parameter $f\SN$, which is the fraction of SN energy converted into turbulence.
To compute the dissipation rate, a turbulent length scale is required, which they assume to be equal to the gas scale height.
Our approach is similar to that of \citet{Schober+16}, but we calculate the fraction of SN energy converted into turbulence, 
rather than treating it as a parameter, and we model the turbulent correlation scale using the similarity solutions of SNRs and SBs.

A detailed analytic model that includes the calculation of the turbulent velocity was presented by \citet{Krumholz+18}.
In addition to turbulence driving by SNe, 
their model includes driving by gravity, whereby potential energy of radially inflowing gas is converted into turbulent energy.
Net radial inflow is initiated by torques exerted by non-axisymmetric structures, 
that are in turn formed due to gravitational instability in the disc.
However, they find that this driver tends to be subdominant for local spirals and dwarfs, 
though it tends to dominate for high-redshift galaxies and local ultraluminous infrared galaxies (ULIRGs).
In their model, they compute the 1D turbulent velocity dispersion that can be sustained from star formation feedback alone,
by balancing energy injection rate and dissipation rate surface densities.
Like \citet{Schober+16}, they express the dissipation by setting the turbulent scale equal to the gas scale height
(see \citet{Bacchini+20} for another, very recent, model which takes the turbulent scale to be twice the scale height).
To express the injection rate surface density, \citet{Krumholz+18} assume a momentum injection per unit mass of stars formed,
which is based, in part, on results from simulations of single SNe,
and assume that the injection happens once motions driven by stellar feedback slow to be comparable to the overall ambient velocity dispersion.
Again, the main difference in our approach is that we compute the turbulent scale and SN energy injection using the SNR and SB similarity solutions.%
\footnote{\citet{Elstner+Gressel12} used MHD simulations of the local ISM 
to obtain empirical scaling relations for certain turbulent transport coefficients,
with independent parameters being the star formation rate surface density, midplane gas density,
and angular velocity of gas about the galactic centre.
These could potentially be combined with standard analytic expressions for the turbulent transport coefficients 
to obtain scaling relations for $\tau$ and $u$.}

The quantity $u$, or, more precisely, its 1D counterpart, can in principle be directly observed as a velocity dispersion
\citep[e.g][]{Kulkarni+Fich85,Chemin+09,Tamburro+09}.
However, disentangling the contribution to line broadening due to turbulence from other contributions, such as thermal broadening 
and cloud-cloud dispersions, introduces uncertainty \citep{Mogotsi+16}.
Assuming the turbulence to be isotropic, values for the 1D velocity dispersion can be multiplied by $\sqrt{3}$,
which typically yields estimates in the range $10$--$30\kms$ for the warm gas.

Also elusive is a reliable estimate of the turbulent correlation time $\tau$,
which cannot be observed directly since it is of order $10\Myr$.
It is usually assumed to be equal to the eddy turnover time $\tau\eddy\simeq l/u$.
But this assumption is not justified when the time scale for the flow to renovate $\tau\renov$
is smaller than $\tau\eddy$, in which case one would expect $\tau\simeq\tau\renov$ \citep[e.g.~Ch.~VI of Reference][]{Ruzmaikin+88}.
This scenario occurs if the SN rate density is sufficiently high that, 
on average, successive SN or SB shocks pass through a random point in the ISM 
before a typical energy-carrying eddy has had a chance to undergo significant distortion.

A possible remedy is to estimate these parameters from simulations
\citep{Avillez+Breitschwerdt07,Gressel+08a,Gent+13a,Hollins+17}.
De~Avillez and Breitschwerdt~\citep{Avillez+Breitschwerdt07} present hydrodynamic and magnetohydrodynamic (MHD) simulations 
of the multi-phase ISM in the Solar neighbourhood. 
In their models, turbulence is driven by SN explosions, 50--60\% of them clustered, 
and they observe SBs of up to $500\pc$ in size. 
The correlation scale of the turbulent motions in the ISM is about $75\pc$ in their models.

More recently \citet{Hollins+17} conducted MHD ISM simulations in a shearing periodic box,
with parameters suitable for the Solar neighbourhood, but without SN clustering.
For the warm gas at the galactic midplane, 
they obtain a velocity correlation scale of $l\approx60\pc$ and 
an RMS turbulent speed of $u=8\kms$,  which corresponds to an eddy turnover time of $\tau\eddy\simeq l/u\approx7\Myr$.
If no phase separation is applied they find $l\approx74\pc$, $u\approx13\kms$ and $\tau\eddy\approx6\Myr$.
Away from the midplane, at height $|z|=0.4\kpc$, they find $l\approx87\pc$ and $u\approx3\kms$ for the warm phase,
and $l\approx117\pc$ and $u\approx4\kms$ for all the gas, unseparated by phase.
These lead to much larger values of $\tau\eddy\approx30\Myr$ at $|z|=0.4\kpc$ than at the midplane,
but they compute the correlation time as $\tau\approx5\Myr$, finding it to be almost independent of $|z|$.

Although such studies are valuable, their underlying setups can differ in important but subtle ways, 
and they are too expensive to be able to probe a large swath of parameter space,
and, ideally, flesh out relations between turbulence parameters and other galaxy parameters.
Thus, there is a need for transparent calculations that improve upon order-of-magnitude estimates 
\citep{Ruzmaikin+88,Cox90,Shukurov07,Breitschwerdt+05},
and this is the gap we strive to fill, to some extent, in this work.

The paper is organized in the following way.
In Section~\ref{sec:turb} we present our main calculation of the turbulence parameters, 
along with the necessary theoretical background.
We then explore the parameter space in Section~\ref{sec:pspace}.
In Section \ref{sec:SNSB}, we discuss the results of models that assume that all SNe are isolated or all SNe reside in SBs.
We go on to briefly consider a variation of the model that places more emphasis on the thin cloud layer in Section~\ref{sec:xi}.
In Section~\ref{sec:discussion}, we discuss the limitations of our model and opportunities for extending it.
Finally, we summarize and conclude in Section~\ref{sec:conclusions}.

%_________________________________________________
\section{Estimation of Interstellar Turbulence Parameters}
\label{sec:turb}
Our approach for calculating the turbulence parameters $l$, $u$, and $\tau$ is summarized as follows.
Using a standard scaling relation for the outer radius and speed of an SNR,
and assuming that the SNR injects its energy into the ISM once it slows to the ambient sound speed,
we derive a scaling relation for the injection scale $l\SN$ of turbulence driven by isolated SNe.
We then repeat this procedure for SBs, which have their own standard scaling relation, to obtain an injection scale $l\SB$;
however, we in addition include the possibility that the SB blows out of the disc before it can slow to the ambient sound speed.
Combining driving by SNe and SBs additively allows us to write down the energy injection rate density.
We then assume a simple form for the spectrum which allows us to compute analytically the integral scale $l$.
Given $l$, we can write the spectral energy transfer rate density, equal to the turbulent energy dissipation rate density, 
in terms of the turbulent velocity $u$.
We next balance energy injection and dissipation rates, and solve for $u$ to obtain a scaling relation for the latter.
The next step is to estimate the correlation time $\tau$, 
which we take to be equal to the smaller of the eddy turnover time $\tau\eddy$
and the average time for the flow at a given position to renovate due to the passage of an SN or SB blast wave, $\tau\renov$.
Table~\ref{tab:params} summarizes the independent parameters used in our scaling relations.
These are free to take on any values, but we have included a rough range.
If desired, our model can be further simplified by including only one of the driving channels, either isolated SNe or SBs.

%---------------------------------------------------------------------------------------------------
\begin{table*}
  \begin{center}
    \caption{The underlying ISM parameters with adopted values. 
             `Usage' designates whether the parameter is needed to model isolated SN driving or SB driving.
             (If either SBs or isolated SNe are neglected, then $f\SB=0$ or $f\SB=1$, respectively.)
             The range suggests plausible values for Milky Way-like galaxies, but the model is not restricted to this range. 
             The fiducial values refer to typical estimates for the Solar neighbourhood of the Milky Way,
             averaged across the Galactic disc \citep{Ferriere01}.
            }
    \label{tab:params}
    \begin{tabular}{@{}lccccc@{}}
      \hline
   							   	&Usage		&Symbol		&Unit			&Range			&Fiducial \\
      \hline                                            	 
      Ambient sound speed		 			&SN \& SB	&$c\sound$	&$\!\kms$   		&$10$--$20$		&$10$	  \\
      Ambient gas number density				&SN \& SB	&$n$		&$\!\cmcube$		&$0.1$--$1$	&$0.1$	  \\
      SN rate per unit volume                   		&SN \& SB	&$\nu$        	&$\!\kpc^{-3}\Myr^{-1}$ &$25$--$100$		&$50$	  \\
      Initial SN energy		     		   		&SN \& SB	&$\ESN$ 	&$\erg$ 		&$10^{50}$--$10^{51}$   &$10^{51}$\\
      Factor used in estimate of integral scale			&SN \& SB 	&$C$		&--		  	&$3/8$--$1$		&$3/4$	  \\
      Fraction of SNe clustered into OB associations		&SN \& SB	&$f\OB$		&--			&$0.5$--$0.75$		&$0.75$	  \\
      Disk scale height						&SB		&$H$		&$\!\kpc$		&$0.2$--$1$		&$0.4$    \\
      Number of SNe residing in an SB				&SB		&$N\OB$		&--			&$10^2$--$10^3$ 	&$10^2$	  \\
      Fraction of the SB energy that is mechanical 		&SB		&$\eta$		&--			&$0.05$--$0.1$	        &$0.1$    \\
      SB horiz. radius at blowout, as fraction of $H$       	&SB		&$\xi$		&--			&$\tfrac{1}{3}$--$1$	&$1$	  \\
      \hline
    \end{tabular}
  \end{center}
\end{table*}

%=======
\subsection{Similarity Solutions}
\label{sec:similarity}
%.......................................................................................................
\subsubsection{SNRs}
\label{sec:similarity_SNRs}
Toward the end of its life, 
an SNR experiences a momentum-conserving snowplough (MCS) phase,
such that its shell expands according to \citep{Woltjer72,Cox72}
\begin{equation}
  \label{approx}
  R\SN\simeq A t^a \quad \mathrm{and} \quad \dot{R}\SN\simeq a A t^{a-1},
\end{equation}
where $A=R\f^{1-a}(\dot{R}\f/a)^a$, $a=1/4$, $R\f\equiv R(t\f)$, $\dot{R}\f\equiv\dot{R}(t\f)$, 
the subscript `0' denotes the onset of the MCS, 
when the interior has cooled, 
and $t\gg t\f$ has been assumed.
Using the Sedov--Taylor similarity solution ($R\SN\propto t^{2/5}$)
and a prescription for radiative cooling,
\citet{Woltjer72} (see also Reference~\citep{Cox72}) estimates
$t\f = 0.036\Myr\,E_{51}^{4/17} n_1^{-9/17}$,
$R\f= 21\pc\,E_{51}^{5/17}n_1^{-7/17}$,
and $\dot{R}\f= 2.3\times10^2\kms\,E_{51}^{1/17}n_1^{2/17}$,
where $E_{51}= \ESN/(10^{51}\erg)$ is the initial energy of the SN explosion
in units of $10^{51}\erg$ and $n_1=n/(1\cmcube)$ is the ambient gas number density in $\cmcube$
\citep[see also Ch.~7 of Reference][]{Dyson+Williams97}.
Then, with $t_1= t/(1\Myr)$, we have 
\begin{equation}
  \label{R_SN}
  R\SN= 54\pc\,E_{51}^{4/17}n_1^{-19/68}t_1^{1/4} \quad \mathrm{and} \quad \dot{R}\SN= 14\kms\,E_{51}^{4/17}n_1^{-19/68}t_1^{-3/4}\,.
\end{equation}

\citet{Cioffi+88} argue that while the MCS phase is approached asymptotically at late times, 
the SNR will typically merge with the ISM before entering a full-fledged MCS phase.
They derive a slightly different solution, 
valid for $t>t\f=0.013\Myr\,E_{51}^{1/14}n_1^{-4/7}$,
corresponding to $R\f= 14\pc\,E_{51}^{2/7}n_1^{-3/7}$
and $\dot{R}\f= 4.1\times10^2\kms\, E_{51}^{1/14}n_1^{1/7}$.
For $t\gg t\f$ this leads to the relations
\begin{equation}
  \label{R_SN_CMB}
  R\SN= 56\pc\, E_{51}^{31/140} n_1^{-9/35} t_1^{3/10} \quad \mathrm{and} \quad \dot{R}\SN= 16\kms\, E_{51}^{31/140} n_1^{-9/35} t_1^{-7/10},
\end{equation}
which are very similar to equations~\eqref{R_SN}.
Below we choose to adopt equations~\eqref{R_SN},
but it is trivial to replace these with equations~\eqref{R_SN_CMB}, with $a=3/10$.
This leads to only minor differences in the results.

The ISM contains magnetic fields and is multi-phase, hence inhomogeneous; these details are neglected in our model.
\citet{Kim+Ostriker15a} simulated the expansion of a radiative SNR 
into a uniform medium or a two-phase ambient medium containing cold clouds embedded in a warm neutral medium.
For the former, they also considered the case where the environment was filled with 
an initially uniform magnetic field.
They found that magnetic fields do not affect the evolution of the SNR unless the field is very strong (plasma $\beta\sim0.1$).
Further, they found that the radial momentum injection by the SNR into the environment is only $5\%$ smaller
for a two-phase medium than for a uniform medium with the same mean density.
Meanwhile, \citet{Martizzi+15} found that the asymptotic radial momentum of an SNR
is typically smaller by about $30\%$ in an inhomogeneous medium compared
to a homogeneous medium of the same mean density.
This decrement is caused by extra cooling due to the inter-mixing of cold clouds and hot shocked gas.
As these differences are relatively small, 
we model the ambient medium as being of uniform density.
These studies do suggest, however, that it is important to include the different ISM phases in estimates of $n$,
and in our model, $n$ could be thought of as a volume-averaged value in the galactic disc.

%.............................................................................
\subsubsection{SBs}
\label{sec:similarity_SBs}
Following \citet{Maclow+Mccray88}, 
we use the similarity solution of \citet{Weaver+77} and \citet{Weaver+78} 
to model the evolution of an SB in a homogeneous medium:
\begin{equation}
\label{SB_R}
  R\SB= \Atilde t^{\atilde}
      = 0.27\kpc\;L_{38}^{1/5}n_1^{-1/5}t_{10}^{3/5} \quad \mathrm{and} \quad \dot{R}\SB = 16\kms L_{38}^{1/5}n_1^{-1/5}t_{10}^{-2/5}\,,
\end{equation}
where $\Atilde=(125L\SB/154\uppi\rho)^{1/5}$, $\atilde=3/5$, 
and $\rho=(14/11)m_\mathrm{H}n$ has been assumed to allow for helium.
Further, 
$t_{10}= t/(10\Myr)$ and $L_{38}= L\SB/(10^{38}\ergs)$,
where $L\SB$ is the equivalent mechanical luminosity of the SNe in the OB association, 
\begin{equation}  \label{L_SB1}
  L\SB= \eta\frac{N\OB \ESN}{t\OB}\,,
\end{equation}
and where $N\OB$ is the number of SNe contributing to the SB over its lifespan 
$t\OB$ and $\eta$ is the fraction of the injected energy converted into bulk kinetic energy of the SB.

A detailed discussion of the features and applicability of the above similarity solution 
is given by \citet{Maclow+Mccray88} and \citet{Breitschwerdt+05}. 
In particular, radiative cooling of the SB interior can be neglected for the Milky Way parameter values
but may be important in denser media, thicker discs, or smaller OB associations.
We have included an efficiency factor $\eta\le1$ to account for the fact 
that not all of the energy from SNe ends up as mechanical energy of the SB.
The value of $\eta$ is currently not well-constrained.
\citet{Yadav+17} compare 3D and 1D simulations of varying resolution.
They conclude that $\eta$ (which in their definition is the ratio 
of the total combined thermal and bulk kinetic energy to energy injected by SNe) 
increases with resolution and is not converged at the highest resolutions.
They attribute this to the exclusion in the simulations of explicit diffusion processes
needed to obtain radiative layers thick enough to be resolved.
Nevertheless, their work suggests that $\eta$ 
is a decreasing function of the ambient density $n$ 
and time since the onset of SB expansion.
These results are generally consistent with those of \citet{El-badry+19}, 
who use 1D simulations and analytical solutions to model the evolution of an SB.
To model mixing of hot and cold gas due to 3D hydrodynamic instabilities,
which would affect cooling in the shell-bubble interface, 
they include an explicit diffusion term with adjustable diffusivity.%
\footnote{\citet{Fierlinger+16} performed 1D simulations of SN explosions of massive single stars
in a dense ambient medium $\sim100\cmcmcm$, and found that the preceding wind-blown bubble phase leads to a strong overall reduction in radiative cooling losses.
Consistent with the works mentioned above, they find that the importance of cooling depends on resolution, 
and argue that numerical diffusive mixing in their simulations approximates mixing by turbulent diffusion in nature.}
We adopt $\eta=0.1$ as a fiducial value for an ambient density of $n=0.1\cmcube$, 
and $\eta=0.05$ for $n=1\cmcube$.
These values are generally consistent with simulations, 
and are also comparable to mechanical efficiencies estimated for isolated SNe
(e.g., Ch.~7 of \citealt{Dyson+Williams97}, and Section~\ref{sec:SNSB} below).

The solutions presented above have been obtained for a homogeneous ambient medium, and, 
strictly speaking, they do not apply when $R\SB$ becomes comparable to the pressure scale height.
We address this limitation below by considering the case where the SB blows out of the disc.
In 3D MHD numerical simulations with a realistic stratification of the ambient medium and an imposed magnetic field,
\citet{Stil+09} find that the above self-similarity solution 
approximates the SB evolution rather accurately.
On the other hand, \citet{Ntormousi+17} simulate two SBs colliding within a turbulent ISM. 
While the expected $R\SB\propto t^{3/5}$ expansion is reproduced without magnetic fields or for a mean magnetic field that is parallel to the collision axis,
they find that the presence of a mean magnetic field component perpendicular to the collision axis can result in a flatter SB expansion with time.
Furthermore, \citet{Breitschwerdt+05} argue, using MHD simulations of the local ISM, 
that analytical solutions are rather inadequate for modeling SB evolution in a multi-phase ISM.
In this work we choose, for simplicity and tractability, to treat the ISM as a uniform medium and to ignore the effects of magnetic fields.

%__________________________________________________________________

\subsection{Energy Injection Scales}
\label{sec:lSN_lSB}
\subsubsection{SNRs}
An SNR merges with the ISM at $t\approx t\sound\uSN$,
where $t\sound\uSN$ is the time at which the expansion velocity of the SNR shell equals the ambient sound speed $c\sound$.%
\footnote{Alternatively, we could have assumed that SNRs and SBs break up when their expansion velocity reduces to $u$.
This would complicate the model by introducing an additional feedback, 
and is not expected to lead to important differences, since $u/c\sound$ is of order unity (Section~\ref{sec:pspace}).
As such, we leave this idea for future work.}
Then we obtain 
\begin{equation}
  \label{ts}
  t\sound\uSN= \left(\frac{a A}{c\sound}\right)^{1/(1-a)}
             =  3.5\Myr\; E_{51}^{16/51}n_{0.1}^{-19/51}c_{10}^{-4/3},
\end{equation}
where $n_{0.1}= n/{0.1\cmcube}$ and $c_{10}= c\sound/(10\kms)$.
%Nevertheless, 
The turbulent driving scale can be taken to be of order the SNR radius at that time,
\begin{equation}
  \label{l_SN}
    l\SN\approx R\SN(t\sound\uSN)= \left(\frac{a A^{1/a}}{c\sound}\right)^{a/(1-a)}
        = 141\pc\; E_{51}^{16/51}n_{0.1}^{-19/51}c_{10}^{-1/3}.
\end{equation}

\subsubsection{SBs}
Similarly, the time $t\sound\uSB$ when the expansion speed of an SB reduces to the ambient sound speed, 
and hence it comes into pressure balance with the surrounding medium, 
follows from the second of equations~\eqref{SB_R}, assuming $\dot{R}\SB=c\sound$.
Unlike a typical SNR, an SB may be able to blow out of the disc at some time $t\blowout\uSB<t\sound\uSB$.
Then the SB time scale appropriate to the turbulent flow in the disc is given by $t\blowout\uSB$ rather than $t\sound\uSB$.
For the mechanical luminosity $L\SB$, this leads to
\begin{equation}
  \label{L_SB2}
  L\SB= \eta\frac{N\SB \ESN}{\min(t\sound\uSB,t\blowout\uSB)}.
\end{equation}
The self-similar solution presented above yields, from $\dot{R}\SB( t\sound\uSB)=c\sound$,
\begin{equation}
  \label{SB_ts}
  t\sound\uSB = 31\Myr\; \eta_{0.1}^{1/3}N_{100}^{1/3}E_{51}^{1/3}n_{0.1}^{-1/3}c_{10}^{-5/3},
\end{equation} 
where $\eta_{0.1}=\eta/0.1$ and $N_{100}= N\SB/100$. 
This corresponds to
\begin{equation}
  \label{SB_Rs}
    R\SB(t\sound\uSB) = 0.53\kpc\, \eta_{0.1}^{1/3}N_{100}^{1/3}E_{51}^{1/3}n_{0.1}^{-1/3}c_{10}^{-2/3}.
\end{equation}

The evolution of an SB in a stratified ISM \citep{Lockman+86} is discussed by \citet{Maclow+Mccray88}
(see also Refs.~\citep{Maclow+89} and \citep{Maclow+Ferrara99}). 
These authors identify the time when an SB blows out of the disc 
with the moment when its shell begins to accelerate away from the midplane.
They suggest that this occurs when the SB radius at the midplane is about equal to the gas density scale height $H$, 
and the SB extends to a height $z\approx3H$ above the galactic midplane. 
\citet{Maclow+Mccray88} adopt $H=500\p$, 
their estimate of the scale height of the Lockman layer of neutral hydrogen. 
They argue convincingly that it is this diffuse layer, 
rather than the thinner cloud layer with scale height $\sim100\pc$,
that has the more important influence on the dynamics of the SB.
More recent estimates of the local scale height of the Lockman layer 
suggest $300$--$400\pc$ \citep{Ferriere01}, and we adopt $H=400\pc$ as our fiducial value.

As with turbulence driven by isolated SNe, 
we identify the injection scale of turbulence with the final radius reached by an SB near the midplane,
\begin{equation}
  \label{l_SB}
  l\SB=    \min\left[R\SB(t\sound\uSB), \xi H\right],
\end{equation}
where the second case corresponds to blowout,
and $\xi$ is a parameter of order unity.
The scale $\xi H$ is the turbulent driving scale in the case that the SB blows out of the disc.
Note that the vertical size of the SB when it blows out is not required,
though it is expected to be equal to a $\mathrm{few}\,\times H$ \citep{Maclow+Mccray88}.
Below we set $\xi=1$ for simplicity and because this is consistent with the 
horizontal extents of SBs at blowout in the simulations of \citet{Maclow+Mccray88},
but we return to consider the case $\xi=1/3$ in Section~\ref{sec:xi}.

Hence, $l\SB=R\SB(t\sound\uSB)$ is replaced by $l\SB=\xi H$ if the time to attain blowout $t\blowout\uSB$
is less than the time to slow to the ambient sound speed $t\sound\uSB$. 
For a flared disc, we might have $H\sim0.2\kpc$ near the disc centre, 
and $H\sim1\kpc$ near the outskirts of the star-forming disc.
Since $R\SB(t\blowout\uSB)=\xi H$, we then have 
\begin{equation}
  \label{tb}
  t\blowout\uSB= \left(\frac{\xi H}{\Atilde\blowout}\right)^{1/\atilde}
               = 15\Myr\; \eta_{0.1}^{-1/2}N_{100}^{-1/2}E_{51}^{-1/2}n_{0.1}^{1/2}\xi^{5/2}H_{400}^{5/2},
\end{equation}
where $H_{400}= H/(0.4\kpc)$.
For the equivalent mechanical luminosity $L\SB$, we then obtain
\begin{equation}
  L_{38}= 
    \begin{dcases}
        0.10 \eta_{0.1}^{2/3} N_{100}^{2/3}E_{51}^{2/3}n_{0.1}^{ 1/3}c_{10}^{5/3},   & \mbox{if } t\sound\uSB\le t\blowout\uSB;\\
        0.21 \eta_{0.1}^{3/2} N_{100}^{3/2}E_{51}^{3/2}n_{0.1}^{-1/2}\xi^{-5/2}H_{400}^{-5/2}, & \mbox{if } t\sound\uSB >  t\blowout\uSB,
    \end{dcases}                                                          
\end{equation}
which is close to the values typically assumed.

We assume that most of the energy is injected into the ISM once the SNR or SB reaches its maximum radius, 
respectively $l\SN$ or $l\SB$, and fragments.
The mass of the outer shell is equal to that swept up from the ambient medium.
A fraction $f\SB$ of SNe are assumed to contribute to SBs,
so that 
\begin{equation}
  \label{nu_SN}
  \nu\SN=(1-f\SB)\nu, 
\end{equation}
where $\nu$ is the rate per unit volume for all SNe in the galaxy.
The rate per unit volume of SBs is then
\begin{equation}
  \label{nu_SB}
  \nu\SB=\frac{f\SB\nu}{N\SB}.
\end{equation}
\citet{Higdon+Lingenfelter05} estimate that the fraction of SNe occurring in OB associations is $\sim 3/4$ for the Milky Way,
and we adopt this as our fiducial value for $f\SB$.

If the SB blows out, 
we assume that it immediately loses pressure support and slows to the ambient sound speed in the disc.
The energy injected is assumed to be equal to the bulk kinetic energy of swept up ambient matter $\sim\tfrac{2}{3}\uppi\rho \xi^3H^3 c\sound^2$.
Likewise, the SB would have injected $\sim\tfrac{2}{3}\uppi\rho R\SB^3(t\sound\uSB)c\sound^2$
had it been able to decelerate to the sound speed and break up without blowing out.
Therefore, the efficiency of conversion of SN energy to turbulent energy in the disc
by SBs is reduced by the factor $\xi^3H^3/R\SB^3(t\sound\uSB)$.
This implies that the value of the scale height can be critical in determining the efficiency of energy conversion.

\subsection{Energy Conversion Efficiency}
\label{sec:efficiency}
The efficiency of turbulent energy injection is given by
\begin{equation}
  \label{epsilon}
  \eps=\frac{\dot{\E}\inj}{\nu \ESN}, 
\end{equation}
where $\dot{\E}\inj$ is the rate of turbulent energy injection per unit volume 
and $\nu \ESN$ is the rate of total energy per unit volume released by SNe.
The injection rate $\dot{\E}\inj$ is given by the sum of the contributions from isolated SNe and SBs,
\begin{equation}
  \label{energy_injection}
  \dot{\E}\inj= \dot{\E}\inj\SN+\dot{\E}\inj\SB
                   = \frac{2\uppi}{3}\rho c\sound^2\nu
                     \left[ (1-f\SB)l\SN^3 
                           +\frac{f\SB}{N\SB}l\SB^3
                     \right].
\end{equation}
where the swept up material is assumed to have the same average density $\rho$ for both isolated SNe and SBs.
While SBs expand to larger heights, where the ambient density is smaller,
it is the lateral expansion within the disc that is most relevant for our calculation,
so we regard this assumption as reasonable.
For fiducial parameter values this gives $\dot{\E}\inj\sim2.5\times10^{37}\ergpersperkpccube$,
which leads to an overall efficiency $\eps=0.016$, or $3.4\times10^{37}\ergpersperkpccube$ and $0.022$ if we instead adopt $n_{0.1}=10$.
This value of $\eps$ is still a few times lower than the value $\approx0.1$ obtained by \citet{Thornton+98}.
The latter could be an overestimate because their simulations end long before the shell velocity has reduced to the ambient sound speed \citep{Fierlinger+16}.
However, the values obtained in our model agree closely with those obtained by \citet{Bacchini+20}, 
who quote a median SN efficiency, for the galaxies modeled, of $\sim0.015$ in the warm atomic gas, and a range of $\sim0.01$--$0.03$.

\subsection{Energy input into an outflow}
\label{sec:outflows}
When an SB blows out, the part of its mechanical energy that is not deposited into the ISM is, 
in principle, available for driving a galactic outflow.
This may take the form of a galactic wind if the gas escapes, or fountain flow if it returns to the disc.
Thus, we estimate the power per unit volume of the ISM made available to drive outflows as
\begin{equation}
  \dot{\E}\outflow =
      \begin{dcases}
                     0,
                     \qquad \mbox{if } t\sound\uSB\le t\blowout\uSB;\\
                     \frac{2\uppi}{3}\rho \left[\left(\dot{R}\SB(t\blowout\uSB)\right)^2-c\sound^2\right]\nu\frac{f\SB}{N\SB}(\xi H)^3  
                  =\left[\left(\frac{\dot{R}\SB(t\blowout\uSB)}{c\sound}\right)^2-1\right]\dot{\E}\inj\SB,
                     \qquad \mbox{if } t\blowout\uSB< t\sound\uSB.
      \end{dcases}
\end{equation}
where $\dot{R}\SB(t\blowout\uSB)$ (Equation~\ref{SB_R}, right) is the SB expansion speed at the time at which blowout occurs, $t\blowout\uSB$ (Equation~\ref{tb}), 
and where we have made use of Equation~\eqref{l_SB} for $l\SB$.
For our fiducial values, we obtain $\dot{R}\SB(t\blowout\uSB)= 15\kms$ and $\dot{\E}\outflow=1.3\times10^{37}\ergpersperkpccube$,
which is equal to about half the value deposited into the ISM by isolated SNe and SBs, 
and about $1.25$ times the value deposited by SBs alone.
The estimate is sensitive to the degree of  SN clustering, disc thickness (and, hence, galactocentric distance) among other parameters.
We leave further exploration of the outflow properties for future work.

\subsection{Relative Contribution from Isolated SNe and SBs}
\label{sec:SN_SB}
The ratio of the rates of energy per unit volume injected by the two mechanisms is
\begin{equation}
  \label{Eratio}
    \frac{\dot{\E}\inj\SN}{\dot{\E}\inj\SB}= \frac{l^3\SN\nu\SN}{l^3\SB\nu\SB}
                     =
      \begin{dcases}
                     0.63 \left(\tfrac{3(1-f\SB)}{f\SB}\right)\eta_{0.1}^{-1}
                     E_{51}^{-1/17}n_{0.1}^{-2/17}c_{10},
                     \qquad \mbox{if } t\sound\uSB\le t\blowout\uSB;\\
                     1.47 \left(\tfrac{3(1-f\SB)}{f\SB}\right)
                     N_{100}E_{51}^{16/17}n_{0.1}^{-19/17} c_{10}^{-1}\xi^{-3}H_{400}^{-3},
                     \qquad \mbox{if } t\blowout\uSB< t\sound\uSB.
      \end{dcases}
\end{equation}
Thus, for typical parameter values isolated SNe and SBs inject comparable amounts of energy,
but the ratio can vary rather strongly between and within galaxies.
Likewise, \citet{Norman+Ferrara96} find that SBs and isolated SNe contribute about equal energies
(see their Figure~1 and discussion following their Equation~(3.13)).
Their model, however, is for highly supersonic turbulence,
obtained for the case where heating of the ISM by turbulent dissipation is negligible.
Turbulence in the warm phase of the ISM is known to be transonic \citep[e.g.][]{Hill+08,Iacobelli+14,Kim+Ostriker15b}.

\subsection{Turbulent Correlation Scale}
We are now in a position to estimate the correlation or integral scale $l$.
We adopt a simplified spectral model with constant spectral index $-\gamma$.
The energy per unit volume injected at scale $l\f$ with wavenumber $k\f=2\uppi/l\f$ is given by
\begin{equation}
  \label{E_1}
  \E =\displaystyle\int^\infty_0 \Etilde(k)dk=\Etilde\f\displaystyle\int^\infty_{k\f} \left(\frac{k}{k\f}\right)^{-\gamma}dk,
\end{equation}
where the spectral energy distribution $\Etilde(k)=\Etilde\f(k/k\f)^{-\gamma}$ for $k\ge k\f$ and $\Etilde(k)=0$ for $k<k\f$,
and $\Etilde\f$ is a constant.
We have assumed that $\Etilde(k)\rightarrow0$ abruptly for $k<k\f$ and we also assume that $\gamma>0$,
so the integral converges.
Defining $\ktilde\equiv k/k\f$ we then have
\begin{equation}
  \label{E_2}
  \E=k\f\Etilde\f\displaystyle\int^\infty_1 \ktilde^{-\gamma}d\ktilde.
\end{equation}
Applying these expressions individually to the spectrum of turbulence driven by SBs or isolated SNe,
and assuming each spectrum to have the same spectral index $-\gamma$,
gives
\begin{equation}
  \label{E_SB}
  \E\SB=\Etilde\SB\displaystyle\int^\infty_{k\SB}\left(\frac{k}{k\SB}\right)^{-\gamma}dk
       =k\SB\Etilde\SB\displaystyle\int^\infty_1 \ktilde^{-\gamma}d\ktilde 
\end{equation}
and
\begin{equation}
  \label{E_SN}
  \E\SN=\Etilde\SN\displaystyle\int^\infty_{k\SN}\left(\frac{k}{k\SN}\right)^{-\gamma}dk
             =k\SN\Etilde\SN\displaystyle\int^\infty_1 \ktilde^{-\gamma}d\ktilde.
\end{equation}
Here, $\Etilde\SB$ and $\Etilde\SN$ are constants, $k\SB=2\uppi/l\SB$, $k\SN=2\uppi/l\SN$, 
and the injection scales $l\SB$ and $l\SN$, with $l\SN<l\SB$,
can be estimated from equations~\eqref{l_SB} and \eqref{l_SN}, respectively.
Since the total energy density is simply equal to the sum of the two contributions,
it follows from Equation~\eqref{E_1} and the first equalities of equations~\eqref{E_SB} and \eqref{E_SN} that
\begin{equation}
  \label{spectrum}
  \Etilde(k)= 
    \begin{dcases}
      0                                                                                                , & \mbox{if } k<k\SB;\\
      \Etilde\SB \left(\frac{k}{k\SB}\right)^{-\gamma}                                                 , & \mbox{if } k\SB\le k<k\SN;\\
      \Etilde\SB \left(\frac{k}{k\SB}\right)^{-\gamma} +\Etilde\SN\left(\frac{k}{k\SN}\right)^{-\gamma}, & \mbox{if } k\ge k\SN.
    \end{dcases}
\end{equation}

The correlation scale for the overall spectrum can be approximated as
\begin{equation}
  \label{l_1}
  l=2\uppi C\frac{\displaystyle\int^\infty_{k\SB}k^{-1}\Etilde(k)dk}{\displaystyle\int^\infty_{k\SB}\Etilde(k)dk},
\end{equation}
with $C$ a dimensionless parameter of order unity.
For solenoidal (zero divergence) turbulence, \citet{Monin+Yaglom75} (Chapter~12) 
obtain $C=3/8$ and $3/4$ for longitudinal and transverse correlations, respectively.  
For potential (zero curl) turbulence, they obtain $C=0$ and $3/8$ for longitudinal and transverse correlations, respectively.
Interstellar turbulence is in reality expected to contain potential (or compressible) modes as well as solenoidal modes,
with the latter modes containing about twice as much energy as the former \citep{Brandenburg+Nordlund11},
and they are coupled to one another \citep[see, e.g. Reference][for a discussion]{Elmegreen+Scalo04}.
We choose $C=3/4$ as our fiducial value but also provide some examples which use $C=3/8$.

To make progress, we need an expression that relates $\Etilde\SN$ and $\Etilde\SB$.
It follows from comparison of equations~\eqref{E_SB} and \eqref{E_SN} that 
\begin{equation}
  \label{Etilde_ratio1}
  \frac{\Etilde\SN}{\Etilde\SB}=\frac{l\SN}{l\SB}\frac{\E\SN}{\E\SB}. 
\end{equation}
In a steady state, the energy injection and dissipation rates balance, 
and can be assumed to be equal to the spectral energy transfer rate.
If we assume that this holds separately for energy injected by isolated SNe and SBs,
we can write $\dot{\E}\inj\SN\sim\E\SN/\tau\eddy$ and $\dot{\E}\inj\SB\sim\E\SB/\tau\eddy$.
Thus, we assume that the ratio of the turbulent energy densities supplied by isolated SNe and SBs
is equal to the ratio of the energy injection rate densities, i.e. $\E\SN/\E\SB=\dot{\E}\inj\SN/\dot{\E}\inj\SB$.
Then from Equation~\eqref{Etilde_ratio1} we obtain
\begin{equation}
  \label{Etilde_ratio2}
  \frac{\Etilde\SN}{\Etilde\SB}=\frac{l\SN}{l\SB}\frac{\dot{\E}\inj\SN}{\dot{\E}\inj\SB},
\end{equation}
where the ratio $\dot{\E}\inj\SN/\dot{\E}\inj\SB$ is obtained from Equation~\eqref{Eratio}.

Below we assume $\gamma>1$.
Evaluating the integrals in Equation~\eqref{l_1} assuming the spectral energy distribution~\eqref{spectrum} we obtain
\begin{equation}
  \displaystyle\int^\infty_{k\SB}k^{-1}\Etilde(k)dk
  = \frac{\Etilde\SB}{\gamma}\left(1+\frac{l\SN}{l\SB}\frac{\dot{\E}\inj\SN}{\dot{\E}\inj\SB}\right)
  \quad \mathrm{and} \quad 
  \displaystyle\int^\infty_{k\SB}\Etilde(k)dk
  = \frac{k\SB\Etilde\SB}{\gamma-1}\left(1+\frac{\dot{\E}\inj\SN}{\dot{\E}\inj\SB}\right),
\end{equation}
where we have made use of Equation~\eqref{Etilde_ratio2}.
Substituting the evaluated integrals into Equation~\eqref{l_1}, 
we obtain
\begin{equation}
  \label{l}
  l= \left(\frac{\gamma-1}{\gamma}\right)Cl\SB\left(\frac{1+(l\SN/l\SB)\dot{\E}\inj\SN/\dot{\E}\inj\SB}{1+\dot{\E}\inj\SN/\dot{\E}\inj\SB}\right).
\end{equation}
Henceforth we choose $\gamma=5/3$, appropriate for Kolmogorov turbulence.
\footnote{The arguments can be adapted to other spectra.
One possibility is to model the spectrum as a broken power law with a different spectral index, say $\gamma=2$,
for scales larger than the sonic scale $l\sound$, and $\gamma=5/3$ for scales smaller than $l\sound$ \citep{Schober+15,Fouxon+Mond20}.
However, the effect on the estimates provided would be rather negligible, 
particularly considering the myriad of other uncertainties and assumptions on which our model relies.}
With the choice $\gamma=5/3$, we have $(\gamma-1)/\gamma=2/5$,
$l\rightarrow \tfrac{2}{5}Cl\SB$ as $\dot{\E}\inj\SN\rightarrow0$ or $l\SN\rightarrow l\SB$,
and $l\rightarrow \tfrac{2}{5}Cl\SN$ as $\dot{\E}\inj\SN\rightarrow\infty$.
For our fiducial values, Equation~\eqref{l} with $\gamma=5/3$ and $C=3/4$ gives $l=74\pc$.

\begin{figure*}                     
  \centering
  %figures produced by ~/Turbulence_estimates/spectrum.pro
  \includegraphics[width=0.9\columnwidth,clip=true,trim=   0 35  0  0]{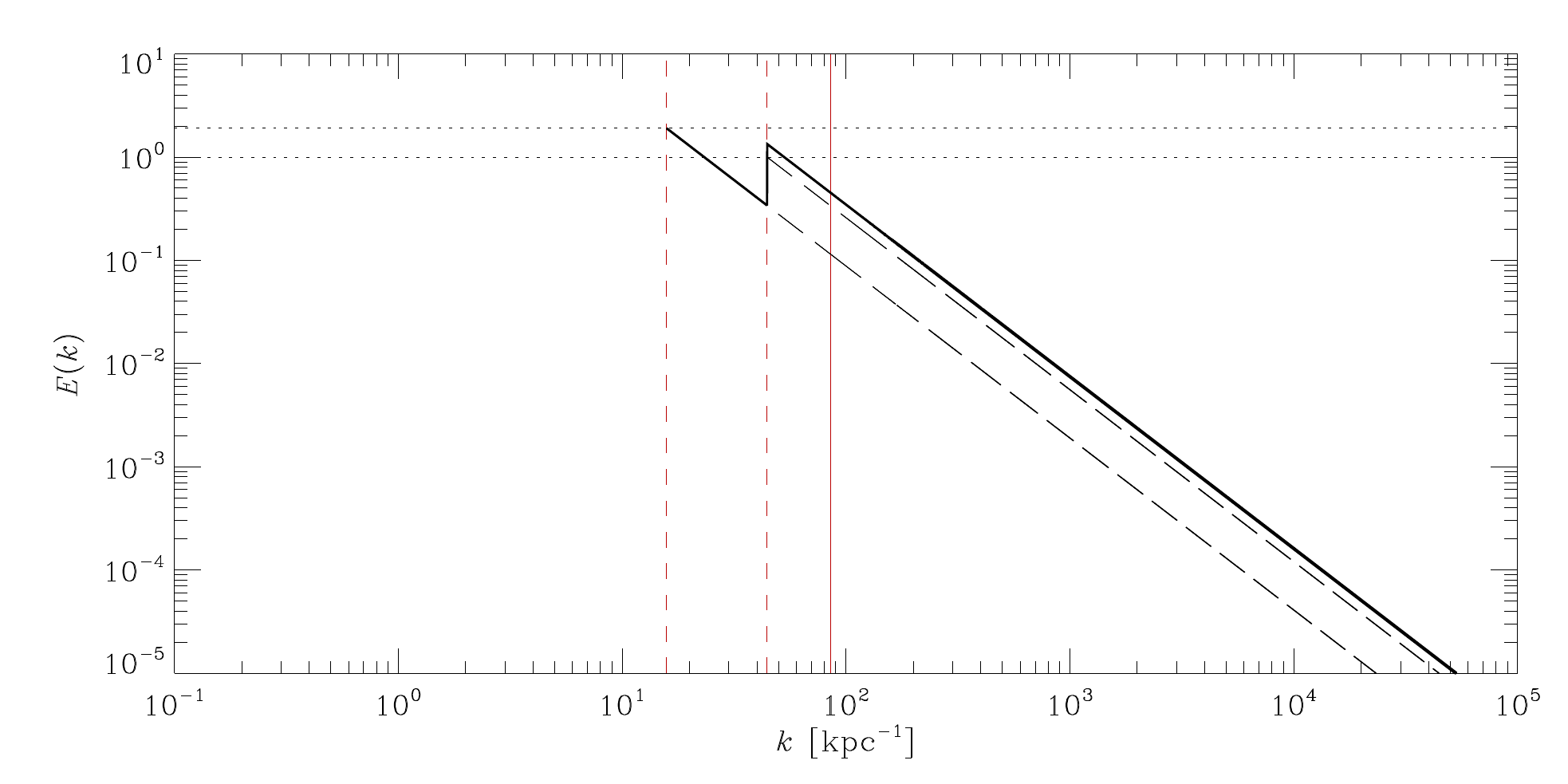}\\
  \includegraphics[width=0.9\columnwidth,clip=true,trim=   0  0  0 10]{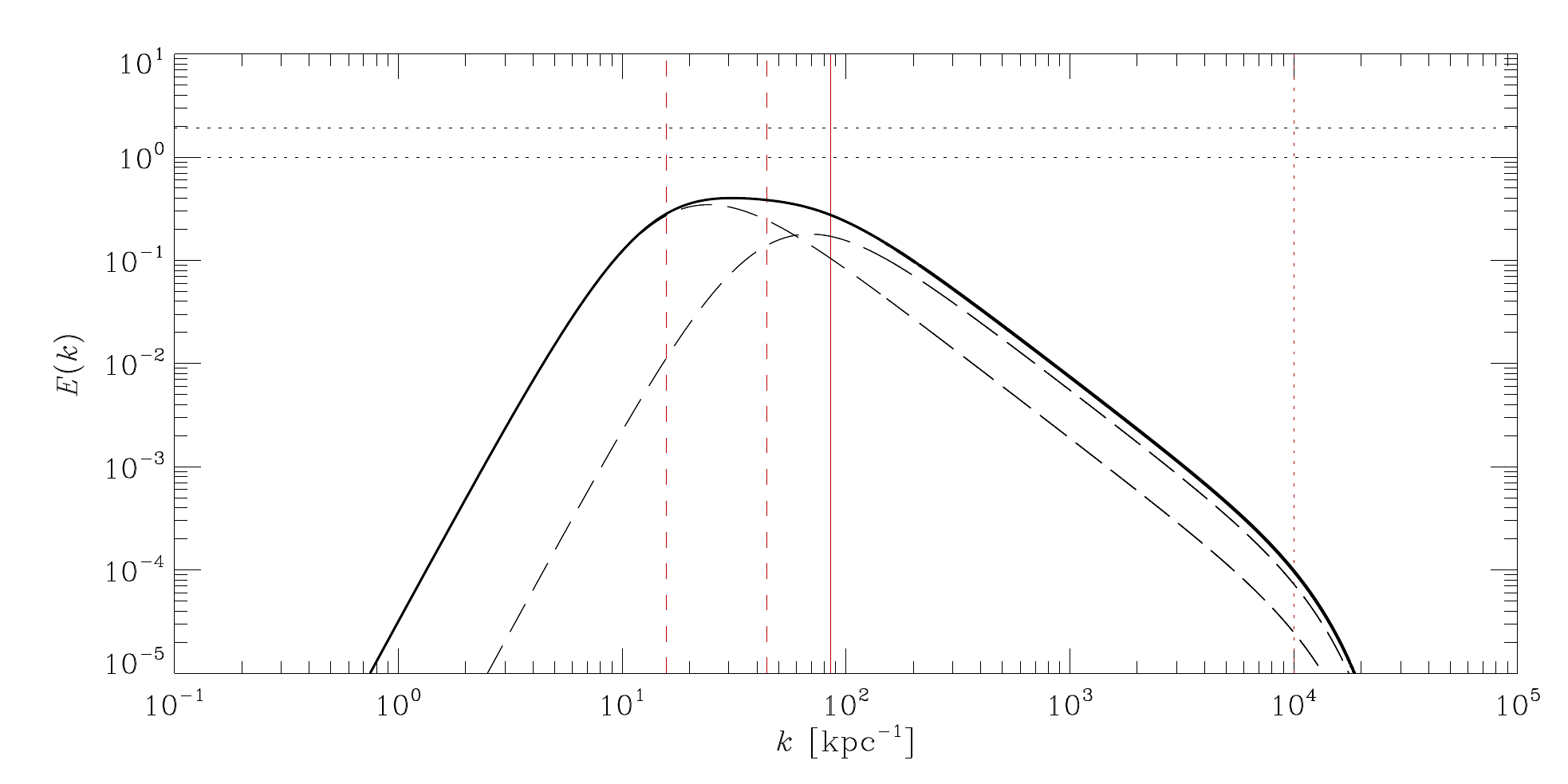}
  \caption{Top: The simple idealized spectrum of Equation~\eqref{spectrum}
           normalized to the peak value for isolated SNe,
           $E(k)\equiv \Etilde(k)/\Etilde\SN$.
           Bottom: The modified von Karman spectrum of Equation~\eqref{vonKarman},
           with dissipation scale chosen as $k\diss=10^4\kpc^{-1}$, 
           and represented by a vertical dotted line.
           Long-dashed curves show individual spectra of isolated SNe and SBs,
           while the solid curve shows the total spectrum.
           Fiducial parameter values of Table~\ref{tab:params} have been used, along with $\gamma=5/3$.
           Dotted horizontal lines reference the peaks of the isolated SN and SB spectra, 
           and vertical dashed lines their respective injection scales $l\SN$ and $l\SB$.
           The wavelength corresponding to the correlation scale of turbulence, $2\pi/l$, 
           with $l$ given by Equation~\eqref{l} with $C=3/4$, is marked by a vertical solid line.
           \label{fig:spectrum}
          }
\end{figure*}

In Figure~\ref{fig:spectrum},
we plot an example of the energy spectrum using the fiducial parameter values of Table~\ref{tab:params} and $\gamma=5/3$.
SB and isolated SNe spectra are shown by dashed curves,
with the combined spectrum shown by a solid curve.
The top panel shows the simple energy spectrum \eqref{spectrum}.
By construction, the ratio of the areas under the isolated SNe and SB spectra
is equal to $\dot{\E}\inj\SN/\dot{\E}\inj\SB$.
On the bottom we illustrate the more realistic 
modified von Karman spectrum employed by \citet{Wilkin+07},
\begin{equation}
  \label{vonKarman}
  \Etilde(k)= \Etilde\f\left(\frac{k}{k\f}\right)^4\left[1 +\left(\frac{k}{k\f}\right)^2\right]^{(-\gamma-4)/2}\Exp{-k^2/2k\diss^2},
\end{equation}
where the dissipative wave number $k\diss$ is chosen, arbitrarily, as $10^4\kpc^{-1}$, and $\gamma=5/3$.
The form of the spectrum at $k\to0$, $\Etilde\propto k^4$, 
is characteristic of both solenoidal and potential velocity fields \citep[p.~52 in Reference~][]{Monin+Yaglom75}; 
this part of the spectrum has negligible effect on the results.
(For the von~Karman spectrum, the ratio of the areas differs only slightly from $\dot{\E}\inj\SN/\dot{\E}\inj\SB$ in the fiducial case.)
In each panel of Figure~\ref{fig:spectrum}, the integral scale of turbulence $l$ is represented by a vertical solid line.

%--------------------------------------------------------------------------------------------------------------------------------------
\subsection{RMS Turbulent Velocity}
\label{sec:u}
We now estimate the RMS turbulent velocity.
The energy dissipation rate per unit volume is equal to the spectral energy transfer rate,\footnote{%
An extra factor greater than $1$ and $\lesssim1.5$ could be included in the numerator 
to account for the expectation that some of the injected energy is converted to magnetic energy via dynamo action.
Given that we have so far neglected magnetic fields, we neglect this factor here; 
it would have the mild effect of reducing $u$ by at most about 10\%.}
\begin{equation}
  \label{energy_dissipation}
  \dot{\E}\dis=\frac{\tfrac{1}{2}\rho u^2}{\tau\eddy}= \frac{\rho u^3}{2l},
\end{equation}
where the eddy turnover time (comparable to the lifetime of the largest eddies) 
is estimated as the ratio of the integral scale and RMS turbulent speed,
\begin{equation}
  \label{tau_eddy}
  \tau\eddy=\frac{l}{u}.
\end{equation}
Assuming a statistical steady state, equating the injection rate per unit volume from Equation~\eqref{energy_injection} 
and dissipation rate per unit volume from Equation~\eqref{energy_dissipation},
and solving for $u$, we obtain
\begin{equation}
  \label{u}
  u= \left(\frac{2l\dot{\E}\inj}{\rho}\right)^{1/3}=\left[ \frac{4\uppi}{3}l c\sound^2\nu\left(
           (1-f\SB)l\SN^3+\frac{f\SB}{N\SB}l\SB^3
           \right)
     \right]^{1/3},
\end{equation}
where the right-hand side is independent of the density.
This result is consistent with that obtained by, for example, \citet{Schober+16},
who also obtain the scaling $u\propto\nu^{1/3}$.
\citet{Krumholz+18} compute the 1D turbulent velocity dispersion that can be sustained by star formation feedback (SNe) alone. 
Multiplying the approximate range they obtain by $\sqrt{3}$, assuming isotropy, gives $u\sim10$--$17\kms$.
Their no-transport models assume that turbulence is driven by star formation feedback alone, 
and that this is dominated by SN feedback, as assumed in our model.
Assuming the star formation rate surface density to be proportional to $\nu$,\footnote{The proportionality constant
would depend on the stellar mass function and contain the scale height of the SN distribution.}
their fixed star formation efficiency per free-fall time no-transport model predicts $u$ to be independent of $\nu$,
while their fixed Toomre $Q$ no-transport model predicts $u\propto\nu^{1/2}$.
The power law index of our model ($1/3$) thus lies in between that of those models ($0$ and $1/2$).
We caution that in our model $u\propto\nu^{1/3}$ only assuming that certain other parameters, like $c\sound$, are independent of $\nu$.

For the fiducial parameter values Equation~\eqref{u} gives $u=12\kms$. 
Note that if we instead assumed that all SN explosions lead to isolated SNRs, 
then we would have $f\SB=0$
and we would obtain $u=[(8\uppi C/15)c\sound^2\nu l\SN^4]^{1/3}=13\kms$ for the fiducial parameter values, 
whereas if all SNe resided in SBs ($f\SB=1$), we would obtain $u=[(8\uppi C/15)c\sound^2\nu l\SB^4/N\SB]^{1/3}=12\kms$.

\subsection{Correlation Time}
We estimate the correlation time $\tau$ as the 
minimum of the turnover time $\tau\eddy$ of energy-carrying eddies,
and the time $\tau\renov$ for the flow to renovate due to the passage of an SN or SB blast wave,
\begin{equation}
  \label{tau}
  \tau=\min(\tau\renov,\tau\eddy).
\end{equation}
This prescription guarantees that the Strouhal number, which we define here as $\Strouhal=\tau/\tau\eddy$,
will be less than or equal to $1$ in our model.

The renovation rate is equal to the sum of the rates from isolated SNe and SBs,
\begin{equation}
  \label{tau_renov}
  \tau\renov= \left(\frac{1}{\tau\SN\renov} +\frac{1}{\tau\SB\renov}\right)^{-1}.
\end{equation}
We take $\tau\SN\renov$ to be equal to the average time between successive 
SNR shells (blast waves) passing through a given point $\bfx$.
In a statistical steady state, the rate at which blast waves cross $\bfx$ is equal to the rate of isolated SN explosions 
within a sphere of radius $l\SN=A(t\sound\uSN)^a$, since more distant SNRs will break up before reaching $\bfx$.
We then obtain
\[
  \frac{4}{3}\uppi l\SN^3\nu\SN\tau\SN\renov= 1,
\]
or, solving for $\tau\SN\renov$,
\begin{equation}
  \label{tau_SN_renov}
    \tau\SN\renov= \left(\frac{4}{3}\uppi l^3\SN\nu\SN\right)^{-1}
                   = 6.8\Myr\; \left(\frac{1}{4(1-f\SB)}\right)
                      \nu_{50}^{-1}E_{51}^{-16/17}n_{0.1}^{19/17}c_{10},
\end{equation}
where $\nu_{50}=\nu/(50\kpc^{-3}\Myr^{-1})$ and we have made use of equations~\eqref{l_SN} and \eqref{nu_SN} in the last equality.
The same result can be obtained by computing how long it takes for SNRs to fill the volume, neglecting overlapping.
Using the same approach to estimate the renovation time due to SBs, we find
\begin{equation}
  \label{tau_SB_renov}
    \tau\SB\renov= \left(\frac{4}{3}\uppi l^3\SB\nu\SB\right)^{-1}
                   =
      \begin{dcases}
        4.3 \Myr\;\left(\tfrac{f\SB}{3/4}\right)^{-1}\nu_{50}^{-1}
                 \eta_{0.1}^{-1}
                 E_{51}^{-1} 
                 n_{0.1}
                 c_{10}^2,
                 \qquad &\mbox{if } t\sound\uSB\le t\blowout\uSB;\\
        9.9 \Myr\;\left(\tfrac{f\SB}{3/4}\right)^{-1}\nu_{50}^{-1}
                 N_{100}
                 \xi^{-3} 
                 H_{400}^{-3}, 
                 &\mbox{if } t\sound\uSB> t\blowout\uSB,
      \end{dcases}
\end{equation}
where we have made use of equations~\eqref{l_SB} and \eqref{nu_SB}.
For the fiducial parameter values, $t\blowout\uSB<t\sound\uSB$, so $\tau\SB\renov=9.9\Myr$,
and $\tau\renov=[1/(6.8\Myr) +1/(9.9\Myr)]^{-1}=4.0\Myr$.
On the other hand, $\tau\eddy=l/u=5.9\Myr$ for fiducial parameter values.
In this case $\tau=\tau\renov$ since $\tau\renov<\tau\eddy$.

\begin{figure*}                     
  %figures produced by ~/Turbulence_estimates/Revision1/turbp.pro which uses turbcalc.pro
  \includegraphics[width=50mm,clip=true,trim=  0 0 0 0]{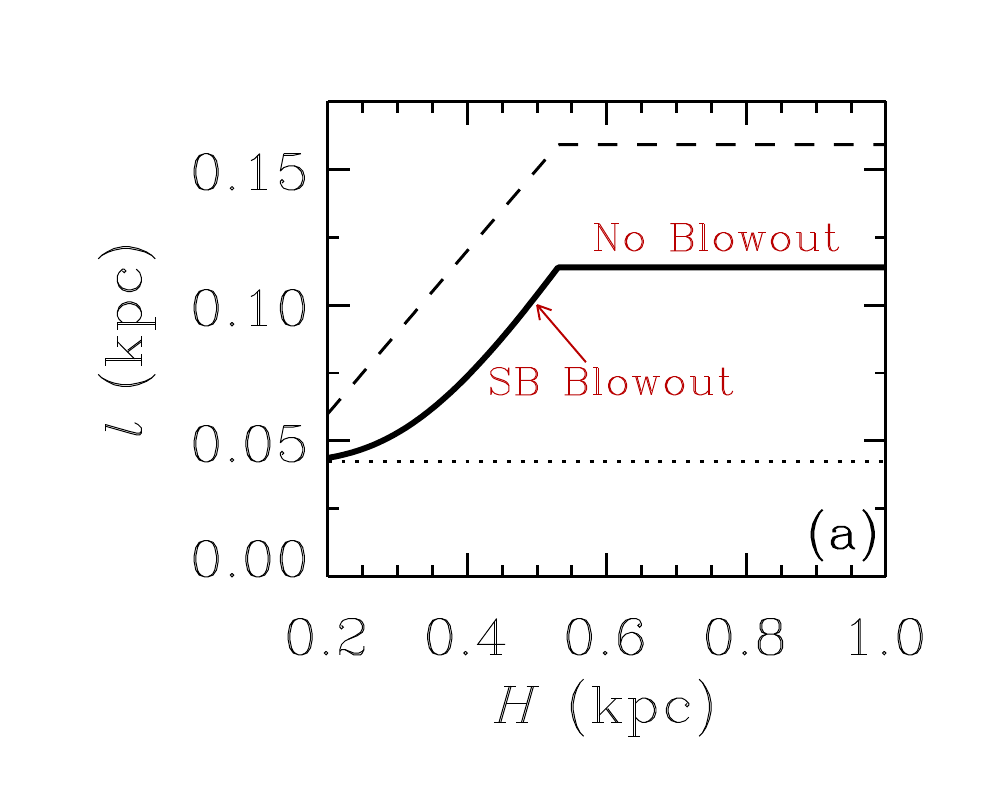}
  \includegraphics[width=50mm,clip=true,trim=  0 0 0 0]{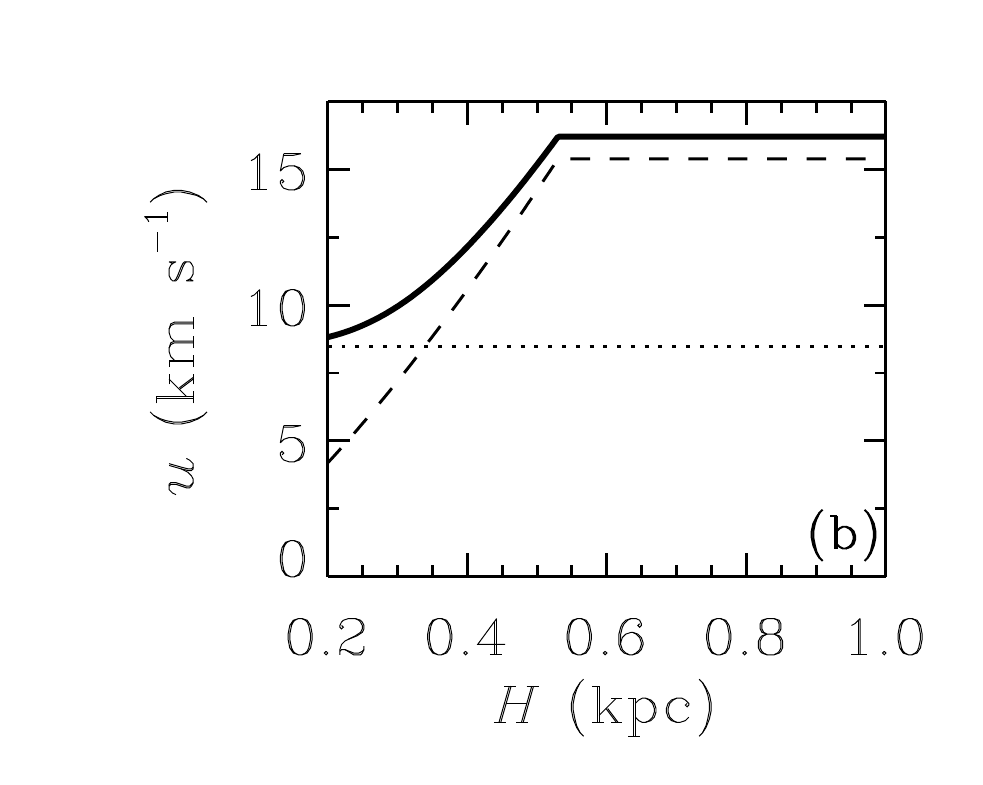}
  \includegraphics[width=50mm,clip=true,trim=  0 0 0 0]{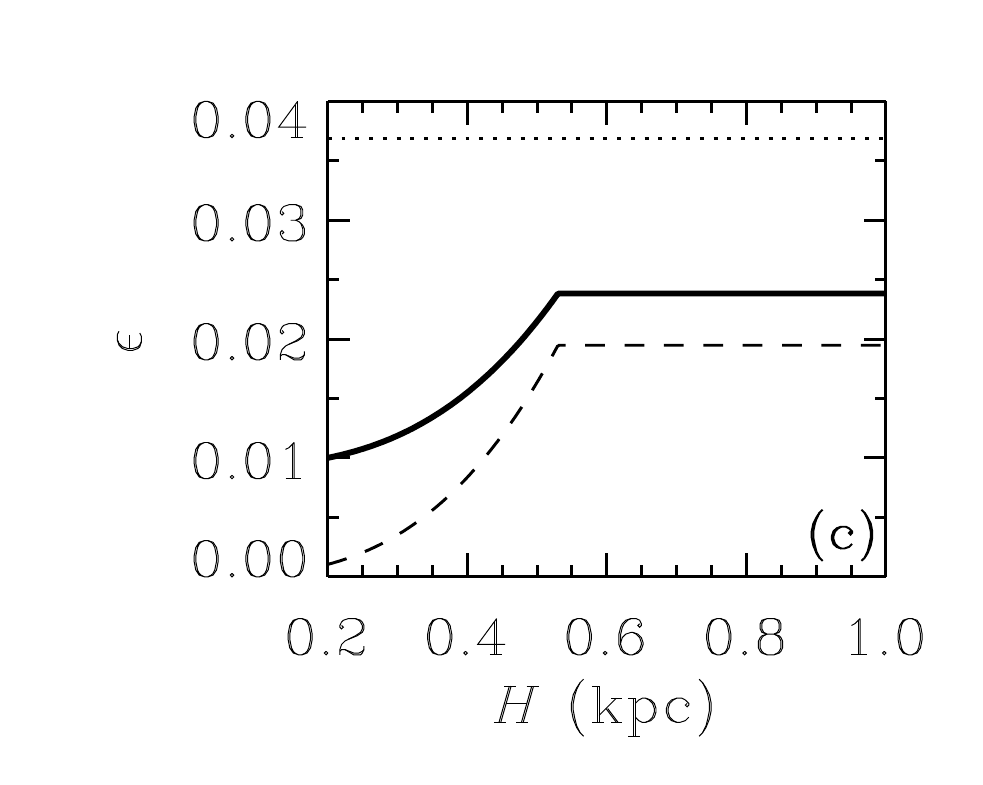}\\
  \includegraphics[width=50mm,clip=true,trim=  0 0 0 0]{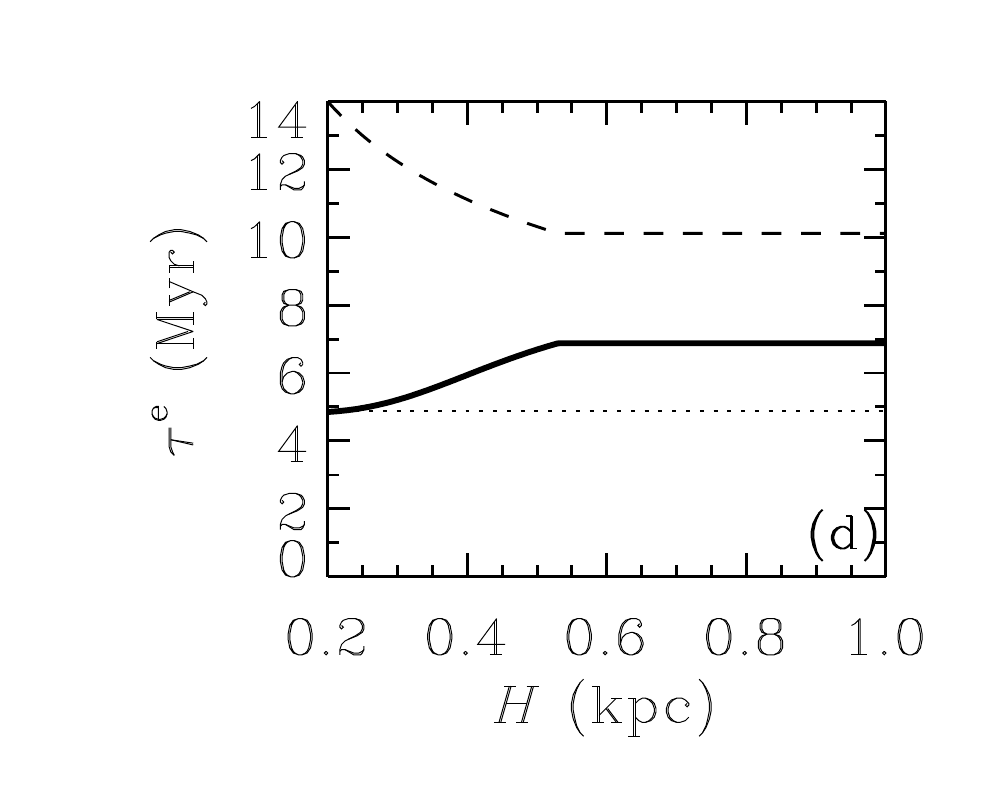}
  \includegraphics[width=50mm,clip=true,trim=  0 0 0 0]{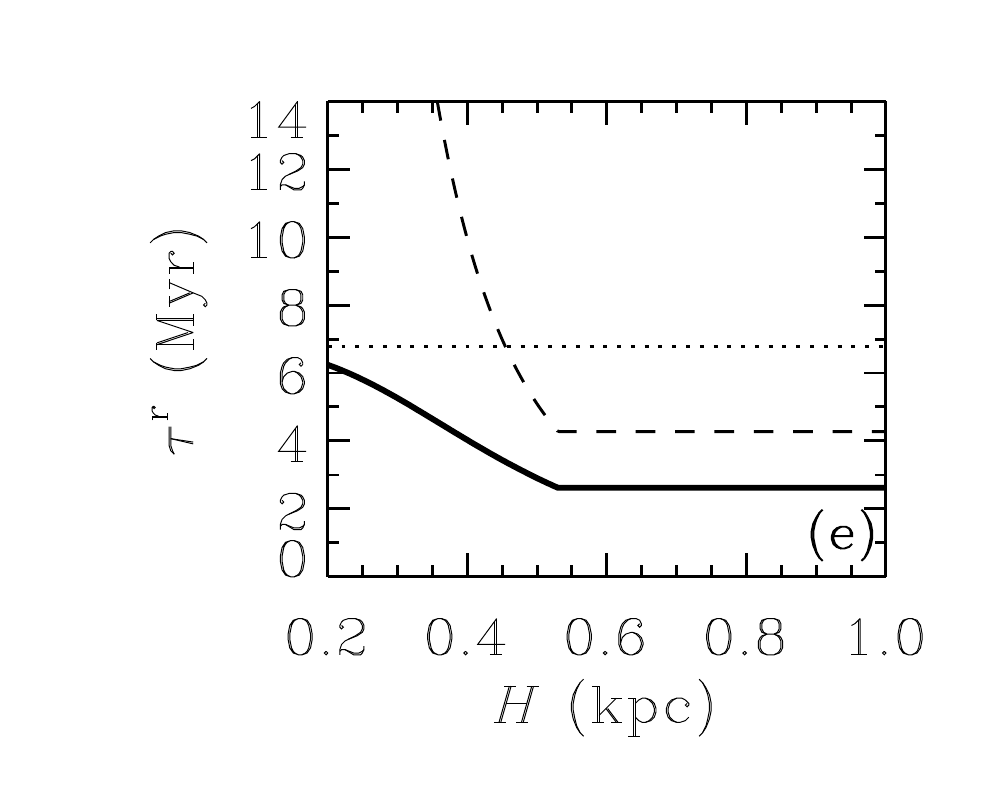}
  \includegraphics[width=50mm,clip=true,trim=  0 0 0 0]{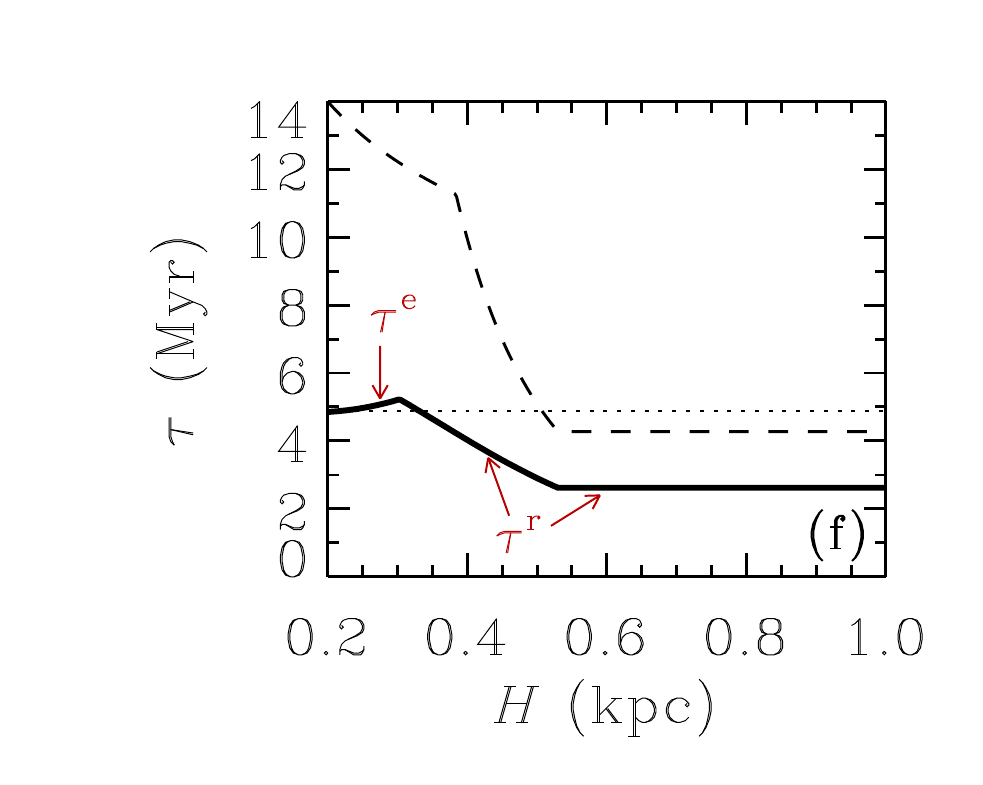}
  \caption{Results for the fiducial parameter values of Table~\ref{tab:params} 
           ($n=0.1\cmcube$, $c\sound=10\kms$, $\nu=50\kpc^{-3}\Myr^{-1}$, $N\SB=100$, $\eta=0.1$ and $C=3/4$)
           plotted against gas scale height $H$
           (fiducial values quoted in the text assume $H=0.4\kpc$).
           Dashed curves show the solution obtained if only the SB component is included, 
           dotted lines show the solution obtained if only the isolated SN component is included
           (whose evolution is independent of $H$ in the model), 
           and solid curves show the solution obtained by including both SB and isolated SN components.
           \label{fig:fiducial}
          }
\end{figure*}

\subsection{Graphical Example}
\label{sec:fiducial}
We illustrate the fiducial case (Table~\ref{tab:params}) in Figure~\ref{fig:fiducial}, where we plot solutions as functions of $H$.
Solid curves show the full solutions, while dashed and dotted curves show the contributions from SBs and isolated SNe, respectively.
The breaks in the curves at $H=0.53\kpc$ are caused by the transition from SBs blowing out of the disc when the disc is thin,
to being unable to blow out when the disc is thick.
In Figure~\ref{fig:fiducial}f, there is a second break at $H=0.30\kpc$ where $\tau=\min(\tau\renov,\tau\eddy)$ transitions from $\tau=\tau\eddy$ for a thinner disc
to $\tau=\tau\renov$ for a thicker disc.

%------------------------------------------------------------------------------------------------------------------------
\section{Exploration of the Parameter Space}
\label{sec:pspace}

\begin{figure*}                     
  %figures produced by ~/Turbulence_estimates/turbp.pro which uses turbcalc.pro
  \includegraphics[width=39.2mm,clip=true,trim=  25 54 17 28]{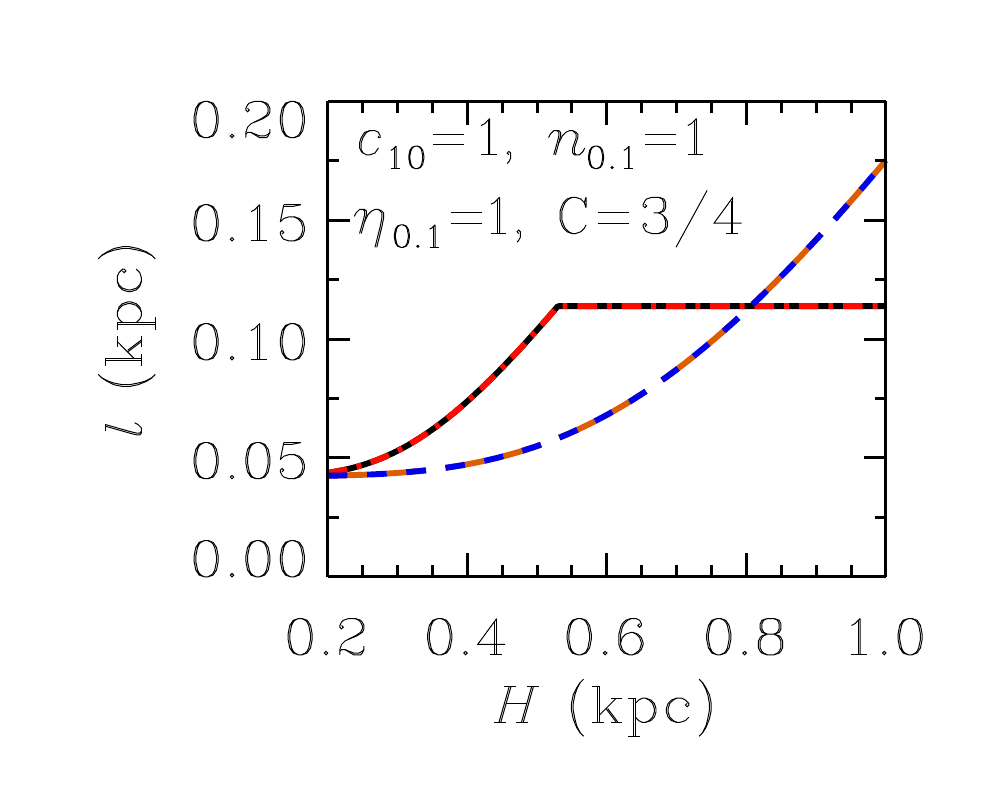}
  \includegraphics[width=28mm  ,clip=true,trim=  93 54 17 28]{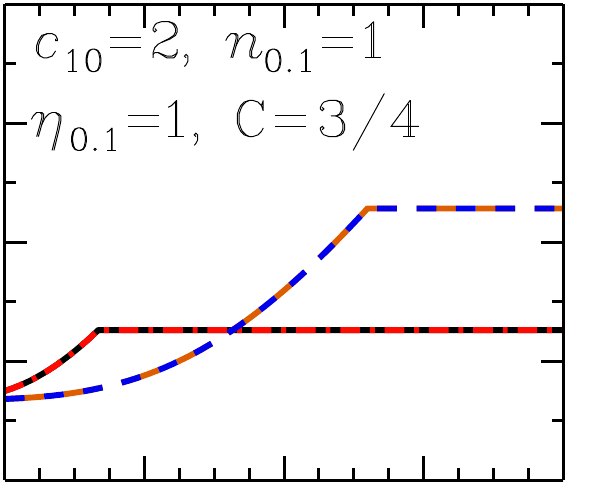}
  \includegraphics[width=28mm  ,clip=true,trim=  93 54 17 28]{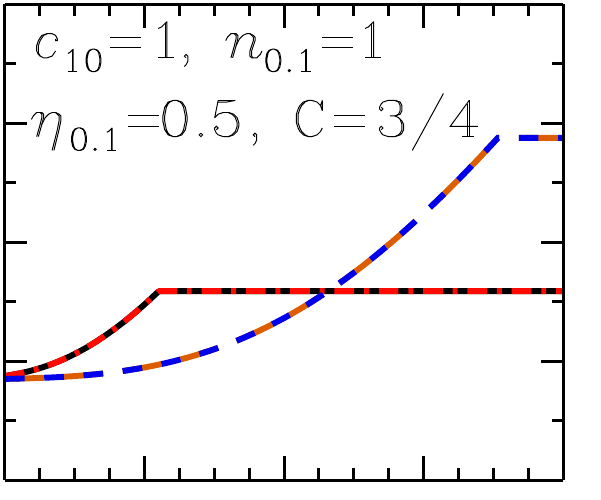}
  \includegraphics[width=28mm  ,clip=true,trim=  93 54 17 28]{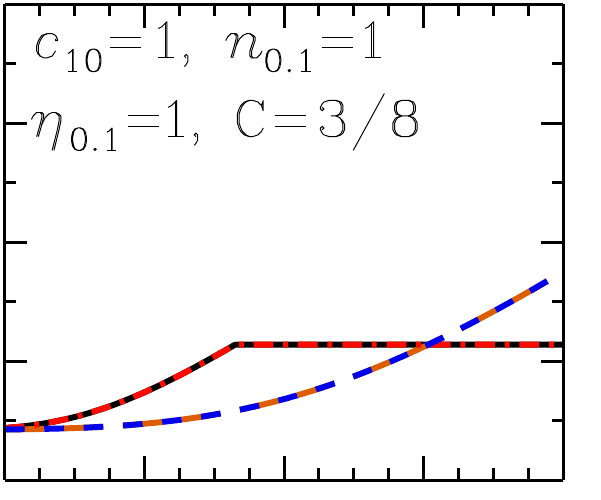}
  \includegraphics[width=28mm  ,clip=true,trim=  93 54 17 28]{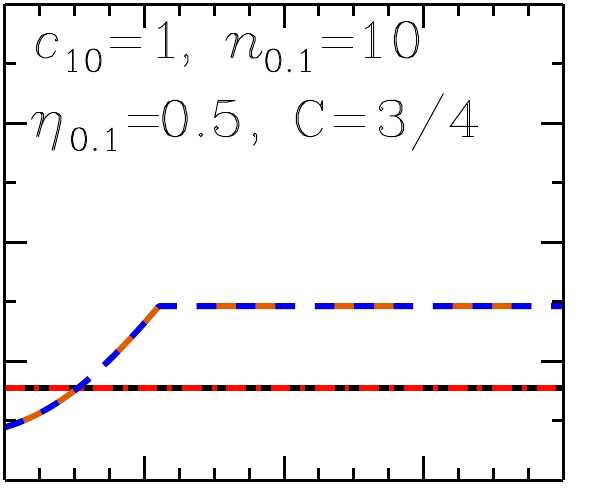}\\
  \includegraphics[width=39.2mm,clip=true,trim=  25 54 17 28]{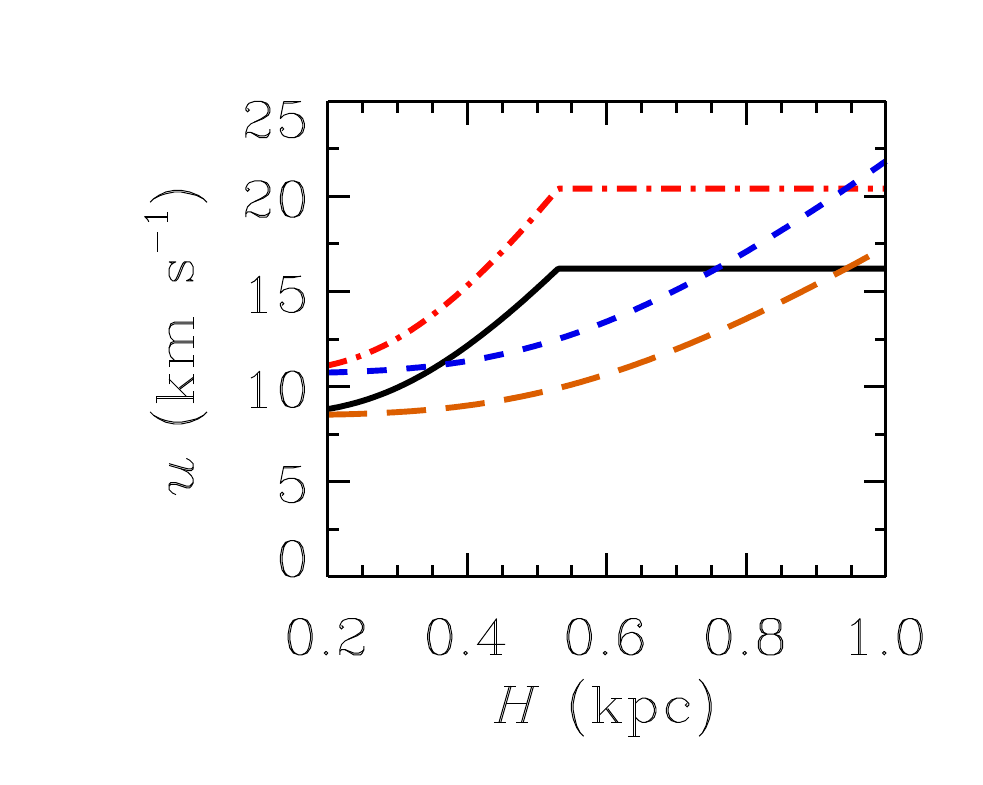}
  \includegraphics[width=28mm  ,clip=true,trim=  93 54 17 28]{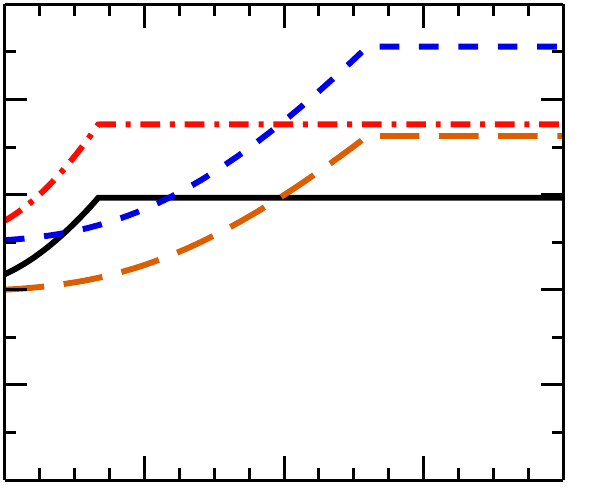}
  \includegraphics[width=28mm  ,clip=true,trim=  93 54 17 28]{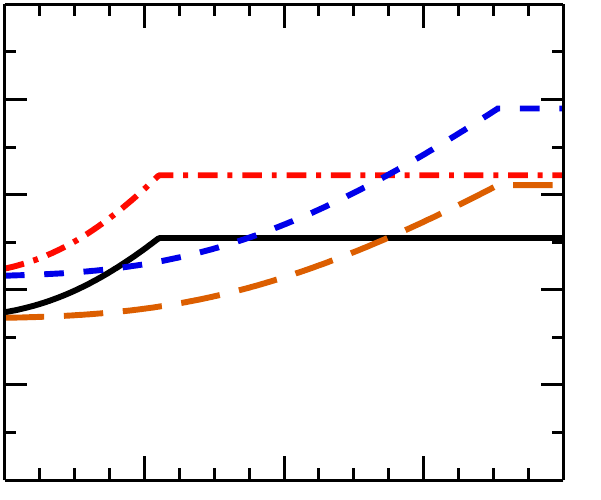}
  \includegraphics[width=28mm  ,clip=true,trim=  93 54 17 28]{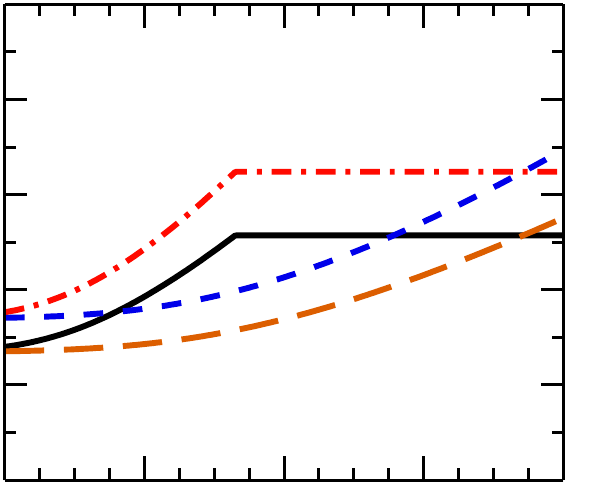}
  \includegraphics[width=28mm  ,clip=true,trim=  93 54 17 28]{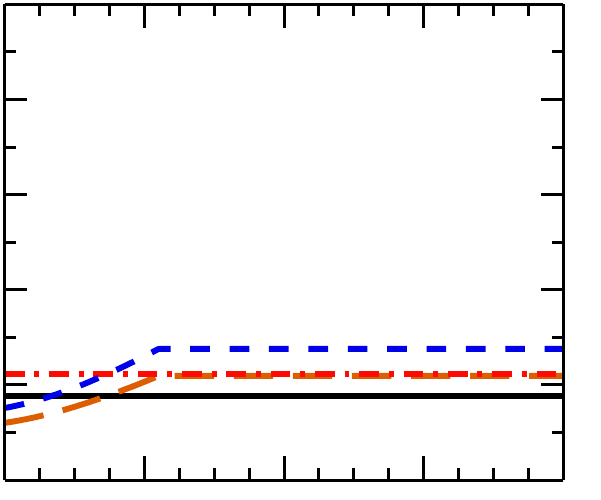}\\
  \includegraphics[width=39.2mm,clip=true,trim=  25 54 17 28]{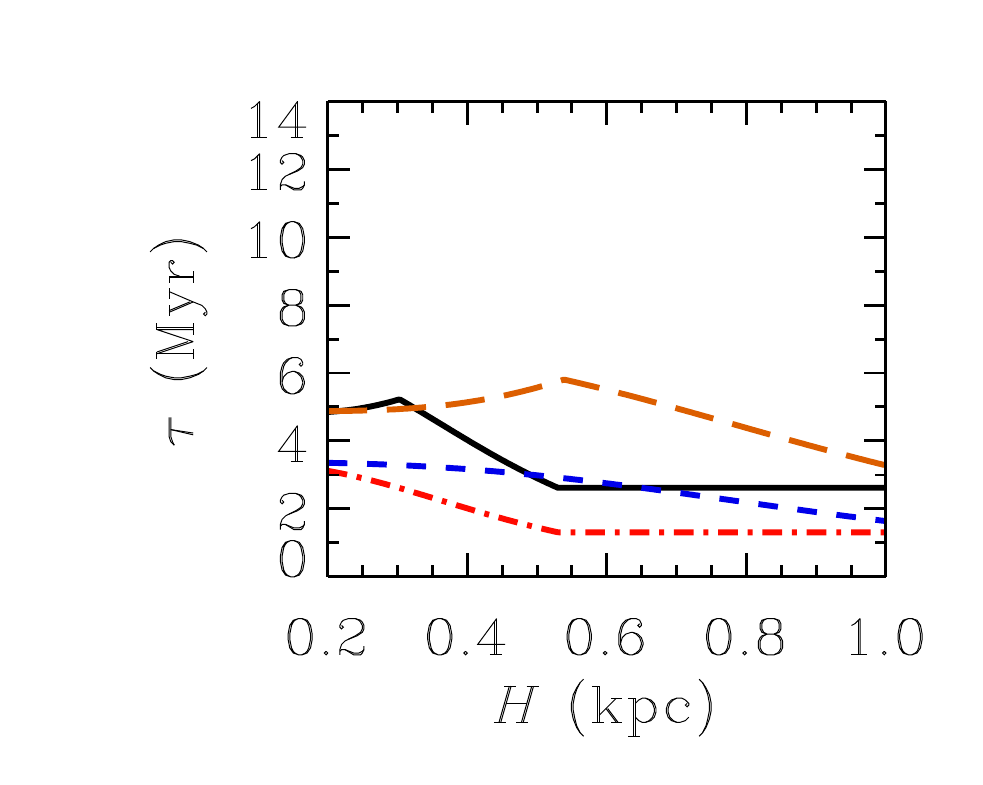}
  \includegraphics[width=28mm  ,clip=true,trim=  93 54 17 28]{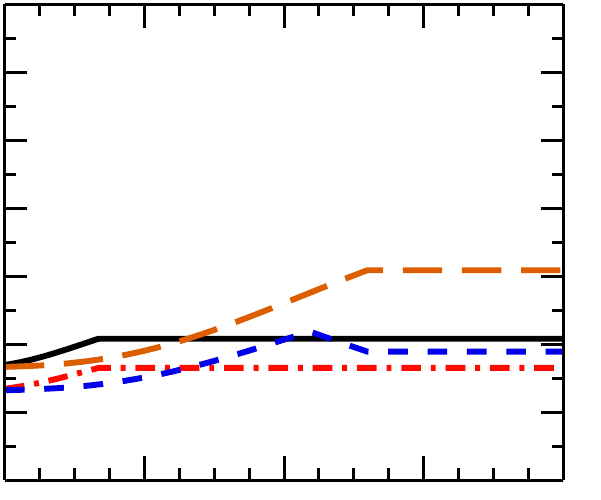}
  \includegraphics[width=28mm  ,clip=true,trim=  93 54 17 28]{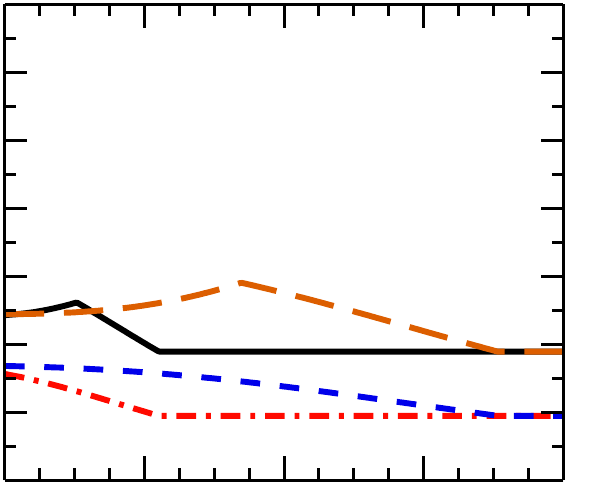}
  \includegraphics[width=28mm  ,clip=true,trim=  93 54 17 28]{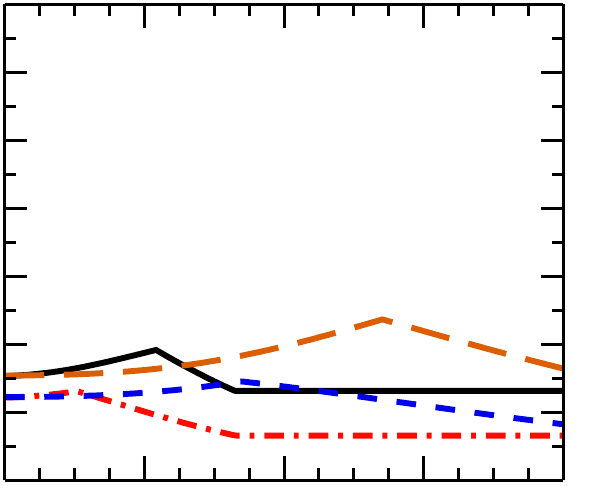}
  \includegraphics[width=28mm  ,clip=true,trim=  93 54 17 28]{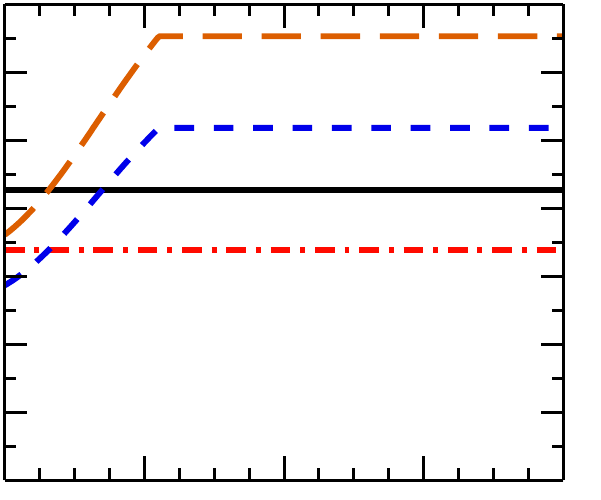}\\
  \includegraphics[width=39.2mm,clip=true,trim=  25 54 17 28]{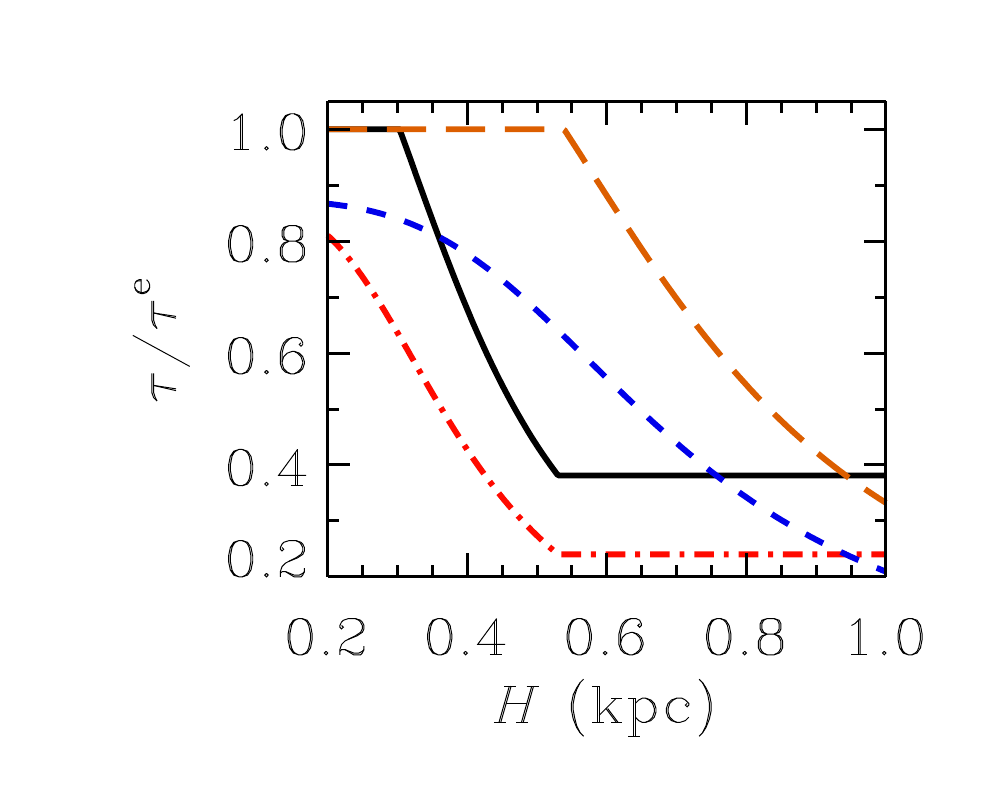}
  \includegraphics[width=28mm  ,clip=true,trim=  93 54 17 28]{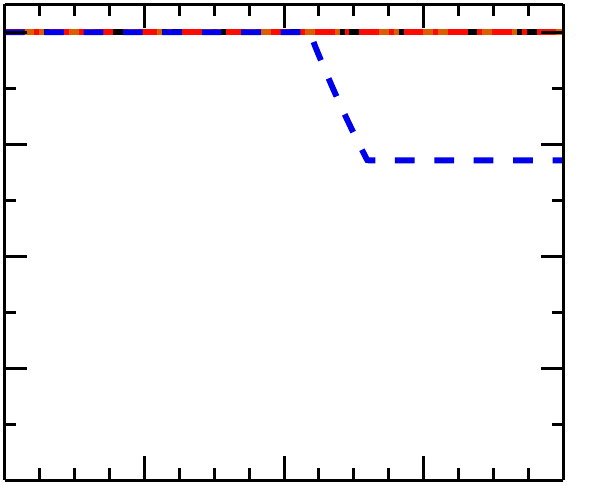}
  \includegraphics[width=28mm  ,clip=true,trim=  93 54 17 28]{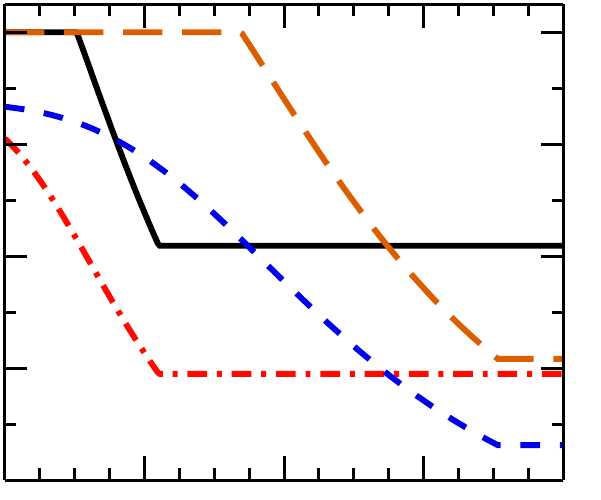}
  \includegraphics[width=28mm  ,clip=true,trim=  93 54 17 28]{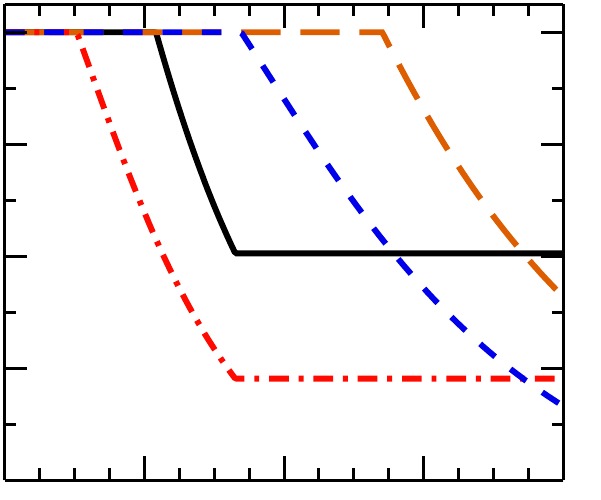}
  \includegraphics[width=28mm  ,clip=true,trim=  93 54 17 28]{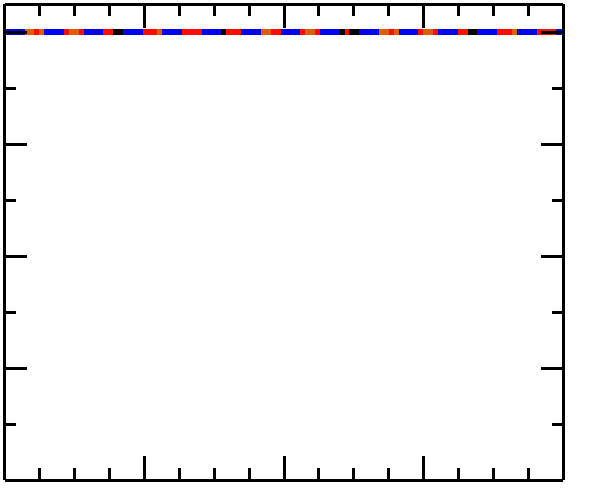}\\
  \includegraphics[width=39.2mm,clip=true,trim=  25 54 17 28]{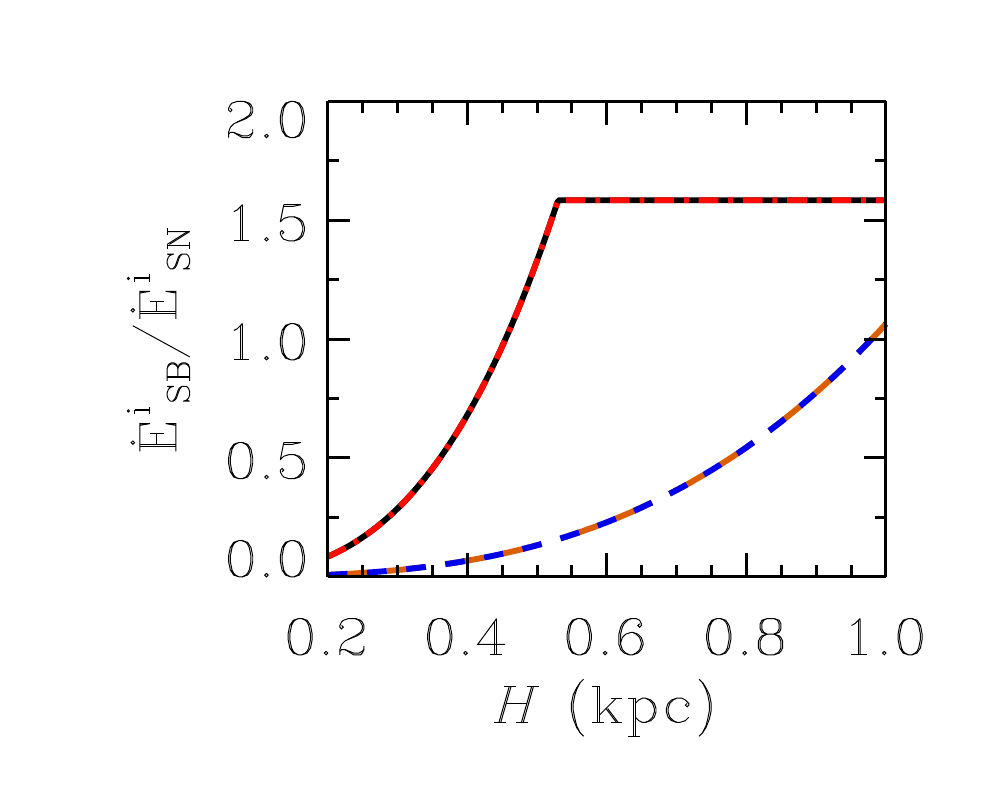}
  \includegraphics[width=28mm  ,clip=true,trim=  93 54 17 28]{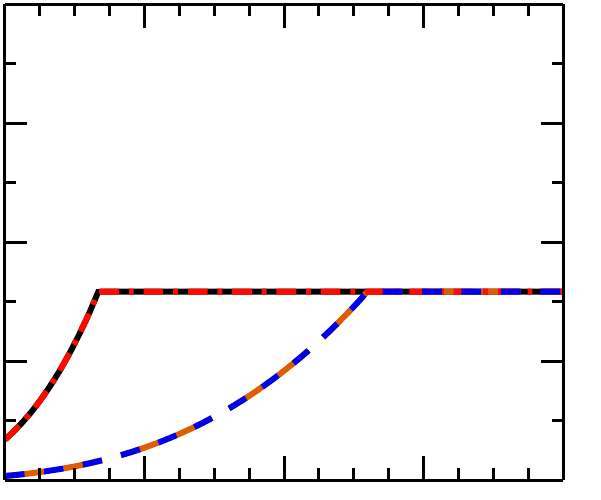}
  \includegraphics[width=28mm  ,clip=true,trim=  93 54 17 28]{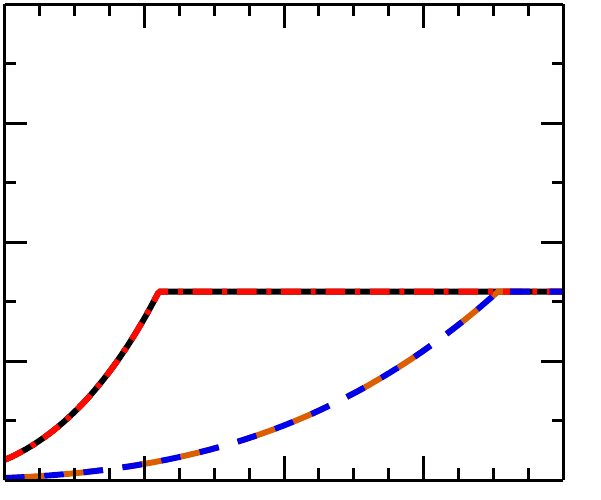}
  \includegraphics[width=28mm  ,clip=true,trim=  93 54 17 28]{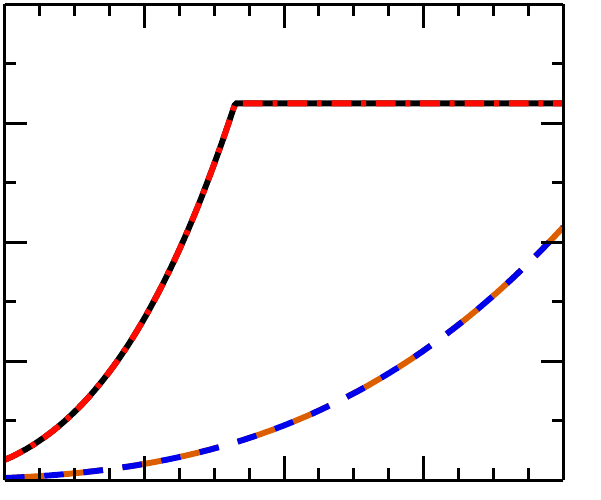}
  \includegraphics[width=28mm  ,clip=true,trim=  93 54 17 28]{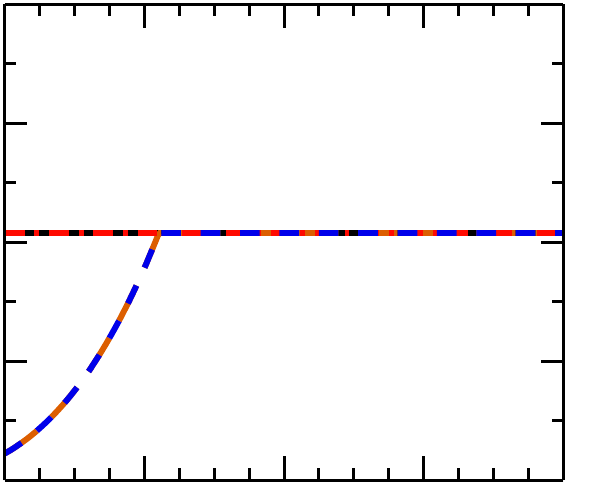}\\
  \includegraphics[width=39.2mm,clip=true,trim=  25 14 17 28]{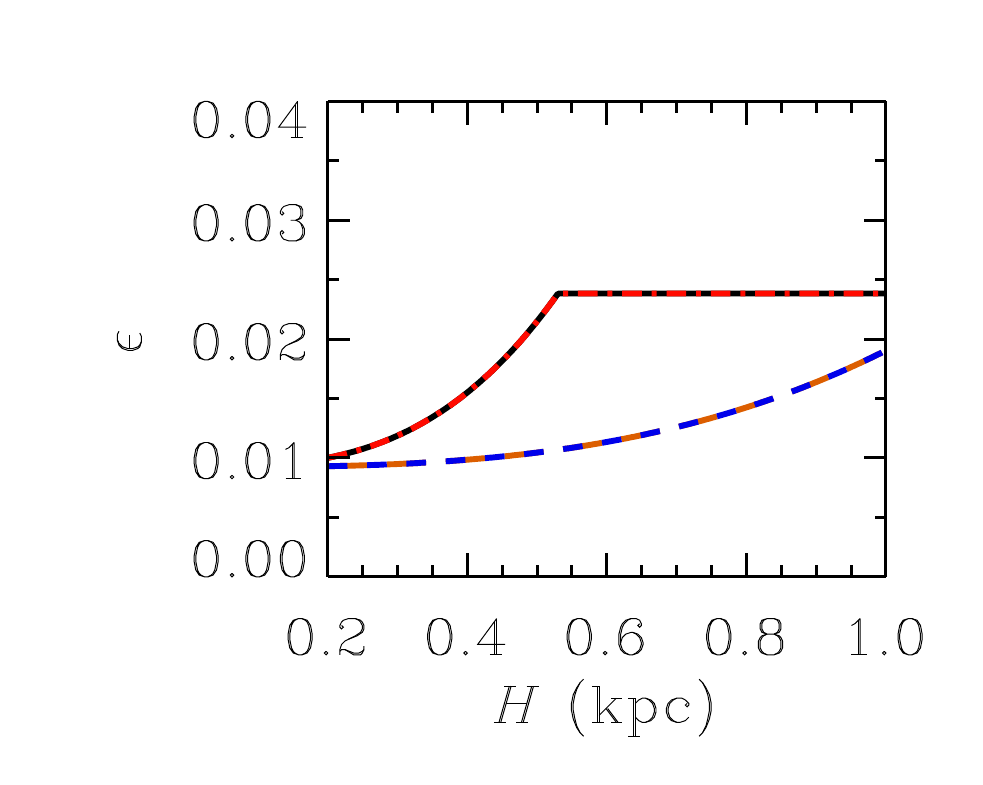}
  \includegraphics[width=28mm  ,clip=true,trim=  93 14 17 28]{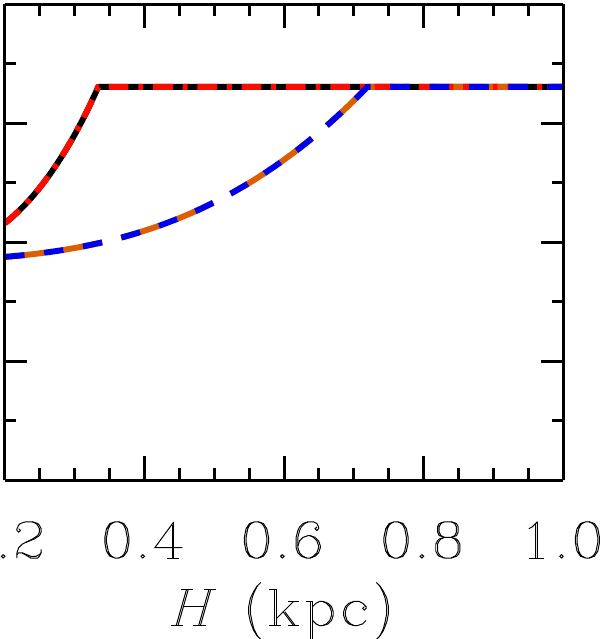}
  \includegraphics[width=28mm  ,clip=true,trim=  93 14 17 28]{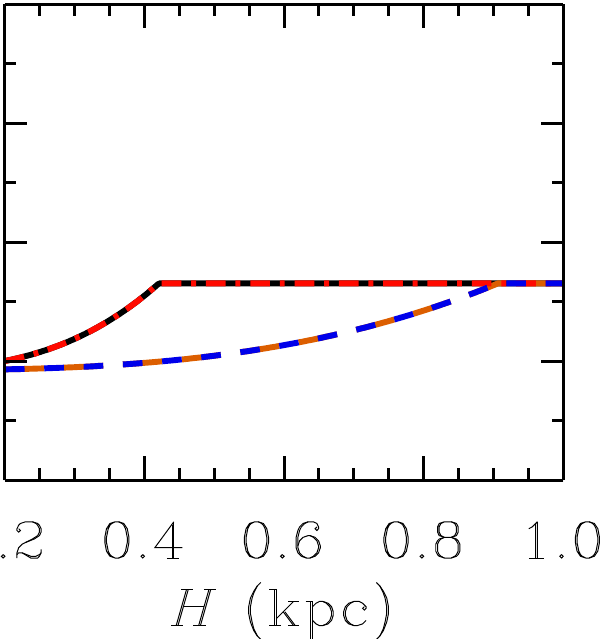}
  \includegraphics[width=28mm  ,clip=true,trim=  93 14 17 28]{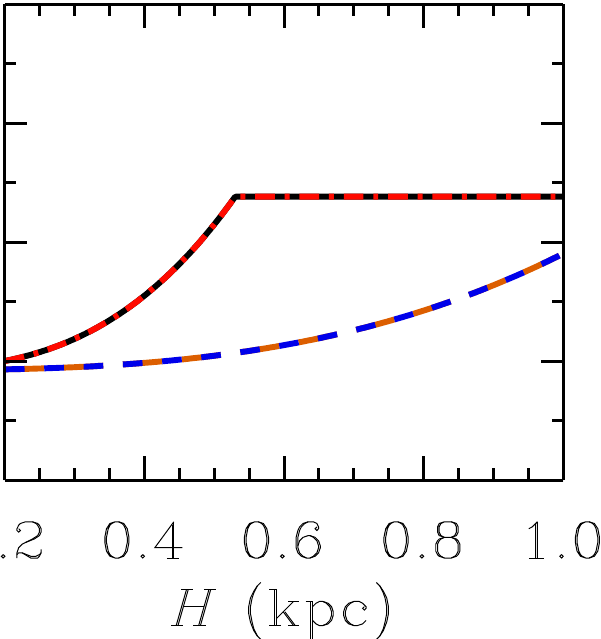}
  \includegraphics[width=28mm  ,clip=true,trim=  93 14 17 28]{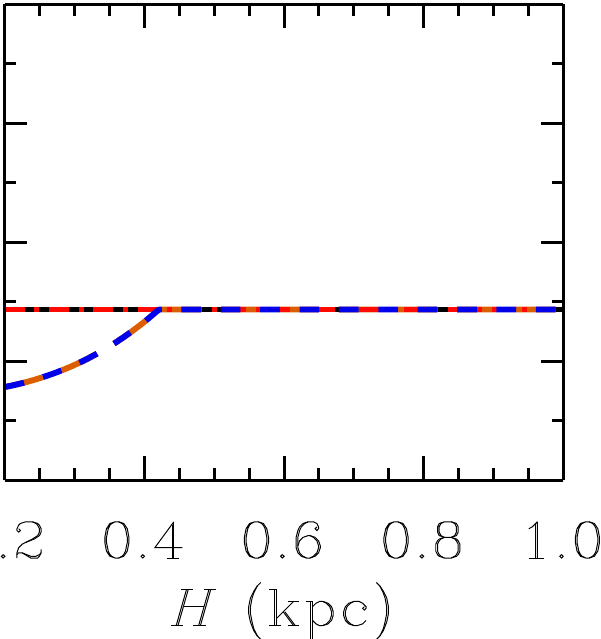}\\
  \caption{Turbulence parameters estimated for various choices of the model parameters in Table~\ref{tab:params}.
           The horizontal axis is the density scale height.
           From the top to bottom row, 
           the vertical axis shows 
           (i) the correlation scale $l$, 
           (ii) the RMS turbulent velocity $u$, 
           (iii) the correlation time scale $\tau=\min(\tau\eddy,\tau\renov)$,
           (iv) the Strouhal number $\Strouhal=\tau/\tau\eddy= \tau u/l$,
           (v) the ratio of energy density per unit time injected by SBs and isolated SNe $\dot{\E}\inj\SB/\dot{\E}\inj\SN$,
           and (vi) the efficiency of converting SN energy to turbulent energy $\eps=\dot{\E}\inj/(\nu \ESN)$.
           Each column corresponds to a different set of values 
           for the parameters $c_{10}$, $n_{0.1}$, $\eta_{0.1}$ and $C$,
           with values indicated on the top row plots.
           In each panel, four cases are shown:
               $\nu_{50}=1$ and $N\SB=100$ (solid black), 
               $\nu_{50}=1$ and $N\SB=1000$ (long-dashed orange), 
               $\nu_{50}=2$ and $N\SB=100$ (dashed-dotted red), 
           and $\nu_{50}=2$ and $N\SB=1000$ (short-dashed blue).
           The solid black curve in the leftmost column with $H=0.4\kpc$ corresponds to our fiducial case. 
           \label{fig:pspace}
          }
\end{figure*}

We plot several example solutions in Figure~\ref{fig:pspace}, and refer to rows and columns by number starting from top left.
Panels show a given turbulence parameter as a function of disc scale height $H$
(see the figure caption for a detailed description).
The fiducial case is illustrated by the solid black curves of the left-hand column, at $H=0.4\kpc$.

The correlation scale $l$ (top row) lies in the range $20\pc<l<175\pc$.
It increases with scale height if $H$ is small enough that SBs blow out, and is otherwise independent of $H$.
The dependence is stronger than linear because of the term $\dot{\E}\inj\SN/\dot{\E}\inj\SB\propto H^{-3}$ 
in the denominator of Equation~\eqref{l} (the same ratio is suppressed by the factor $l\SN/l\SB$ in the numerator, so is less consequential there).
Since $l$ is independent of the SN rate density $\nu$, dashed-dotted red and solid black curves overlap, as do short-dashed blue and long-dashed orange curves.
Since $l\propto C$, its value is somewhat sensitive to the choice of $C$ (compare first and fourth columns).

The turbulent velocity $u$ (second row) is found to occupy the range $3\kms\le u\le 23\kms$,
but $u\ge7\kms$ for $n=0.1\cmcmcm$. 
These values are in good agreement with observations and simulation results \citep[e.g. Reference][]{Gent+13a}.
If other sources of turbulence, in addition to SNe, were important,
then there would be additional contributions to $\dot{\E}\inj$, so $\dot{\E}\inj$ would increase.
If, further, such contributions injected energy on similar or larger scales compared to SNe, 
so that $l$ remained roughly the same or larger,
then, from the first equality in Equation~\eqref{u}, we would expect $u$ to increase.
Therefore, if additional sources of turbulence were important, our model would predict a larger value of $u$.
Gravitationally-driven inward radial transport is one such source that is likely to be important, 
at least at high cosmological redshift \citep{Krumholz+18}.
The lower end of the range of $u$ corresponds to dense gas (compare third and fifth columns), 
mainly owing to smaller $l\SN$ and $l\SB$.
The upper end of the range corresponds to large $H$, which prevents or delays blowout, 
allowing more power to be deposited into the ISM.
Also, $u$ increases with $\nu$ since more power is deposited ($u\propto\nu^{1/3}$).
Note that doubling the sound speed from $10\kms$ to $20\kms$ can actually
lead to a slight reduction in $u$ because expanding SBs slow to the ambient sound speed earlier (compare first and second columns). 
If all SNe were to reside in SBs, and the SBs did not blow out, 
then $l\propto l\SB$ and we would have $u\propto c\sound^{-2/9}$.
If all SNe were isolated, then we would have $l\propto l\SN$ and $u\propto c\sound^{2/9}$.
On the other hand, at small $H$, SBs blow out and the injection scale from SBs
is independent of $c\sound$, while the injected energy from SBs
scales as $c\sound^2$, so $u$ always increases with $c\sound$ in this case.

For the correlation time (third row), we find $1\Myr\le\tau\le13\Myr$,
and $\tau\le6\Myr$ for $n=0.1\cmcmcm$,
which agrees with order-of-magnitude estimates \citep{Shukurov07} and simulations \citep{Hollins+17}.
The smallest values of $\tau$ are obtained for large values of $\nu$,
and the largest values of $\tau$ are obtained for large values of $n$. 
In the fourth row, we plot $\Strouhal=\tau/\tau\eddy$, which is $\le1$ by design in our model.
This quantity is found to lie in the range $0.2$--$1$, 
which implies that $\tau\renov$ and $\tau\eddy$ are generally comparable to one another.
At small scale height $\tau\eddy<\tau\renov$ since the expansion of SBs, and hence the renovation rate, is reduced by blowout.
For large values of $\nu$, $\tau\renov<\tau\eddy$ is more likely.
As $n$ increases, $\tau\renov/\tau\eddy$ also increases, leading to $\tau=\tau\eddy$ for the examples with $n=1\cmcmcm$.
Our results suggest that the common assumption 
that the correlation time is equal to the eddy turnover time
is generally reasonable to within a factor of $2$--$4$.

It is worth pointing out that separate relations would exist between the underlying parameters of our model.
For example, $\nu$ and $H$ are likely to be inversely related.
Firstly, for a flared disc $H$ is larger in the disc outer region, where the SN rate surface density is low.
Secondly, the SN rate volume density decreases with $H$ for a given SN rate surface density. 
Likewise $H$ and $n$ are expected to be inversely related, and $\nu$ would be expected to increase with $n$.

We next turn to the ratio of energies injected by SBs and isolated SNe, $\dot{\E}\inj\SB/\dot{\E}\inj\SN$,
plotted in the fifth row.
As this ratio can be less than or greater than unity, both types of driving can be important.
The ratio is independent of the overall supernova rate $\nu$,
but depends on the fraction $f\SB$ of SNe contributing to SBs,
from Equation~\eqref{Eratio}. 
In the case where $H$ is small enough that SBs blow out,
$\dot{\E}\inj\SN/\dot{\E}\inj\SB$ varies linearly with $N\SB$ and approximately linearly with $\ESN$.
This is because the more energetic the SB, the fraction of its energy that is ``wasted'' 
when the supernova blows out of the disc increases.
In particular, for $N\SB=1000$, $n=0.1\cmcmcm$, and $H<0.5\kpc$,
isolated SNe strongly dominate the energy injection.
This becomes clearer by computing the efficiency $\eps$ of conversion 
of SN energy into turbulent energy, shown in the bottom row. 
For the examples plotted, the efficiency ranges between about $1\%$ and $4\%$,
and is smaller when SBs experience blowout.

%-----------------------------------------------------------------------------------------------------
\section{Relative Importance of Isolated SNe and SBs and Effect of SN Clustering}
\label{sec:SNSB}
\begin{figure*}                     
  %figures produced by ~/Turbulence_estimates/turbp.pro which uses turbcalc.pro
  \includegraphics[width=39.2mm,clip=true,trim=  25 54 17 28]{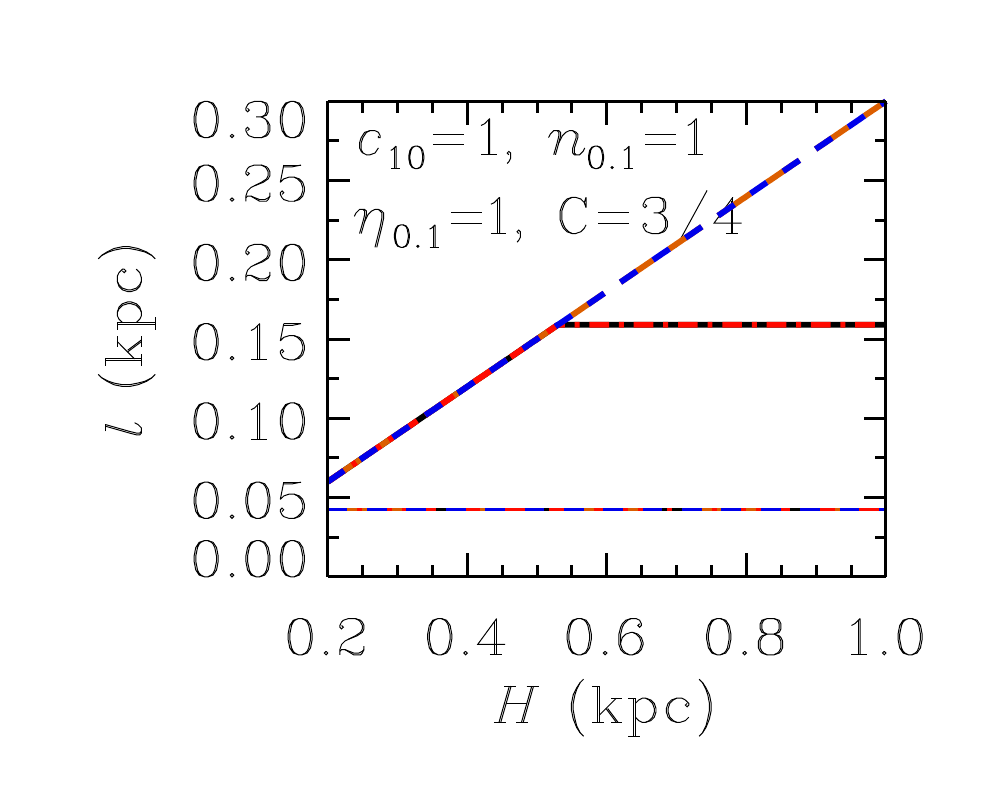}
  \includegraphics[width=28mm  ,clip=true,trim=  93 54 17 28]{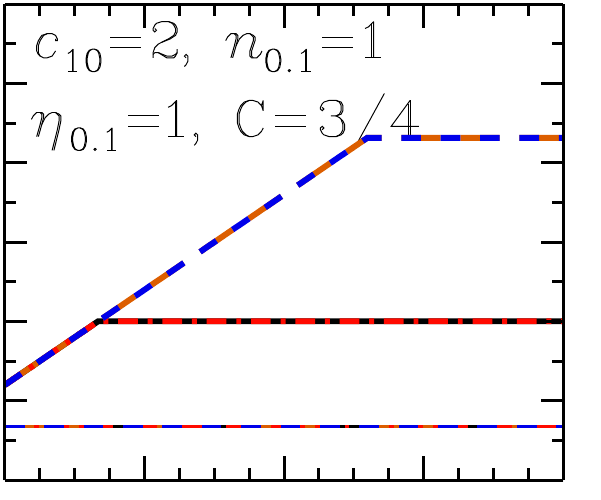}
  \includegraphics[width=28mm  ,clip=true,trim=  93 54 17 28]{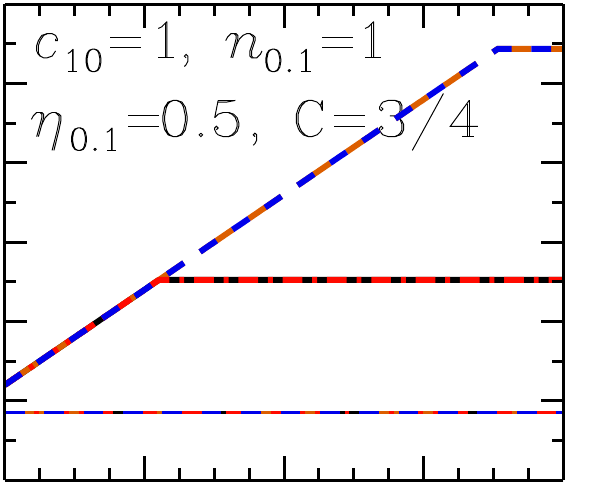}
  \includegraphics[width=28mm  ,clip=true,trim=  93 54 17 28]{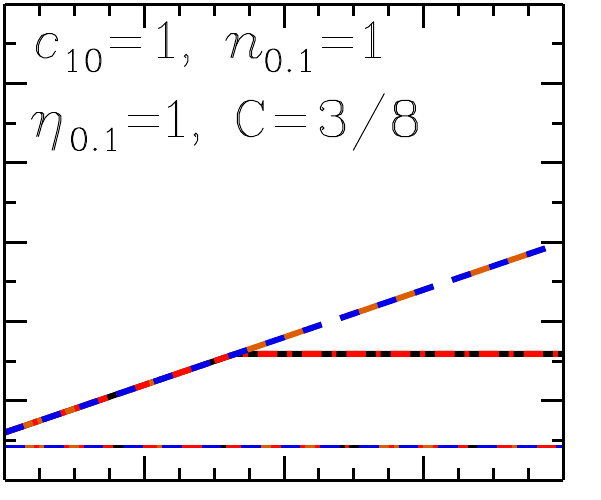}
  \includegraphics[width=28mm  ,clip=true,trim=  93 54 17 28]{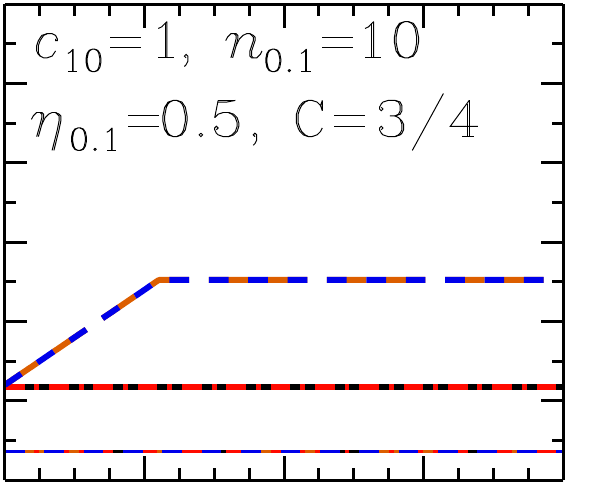}\\
  \includegraphics[width=39.2mm,clip=true,trim=  25 54 17 28]{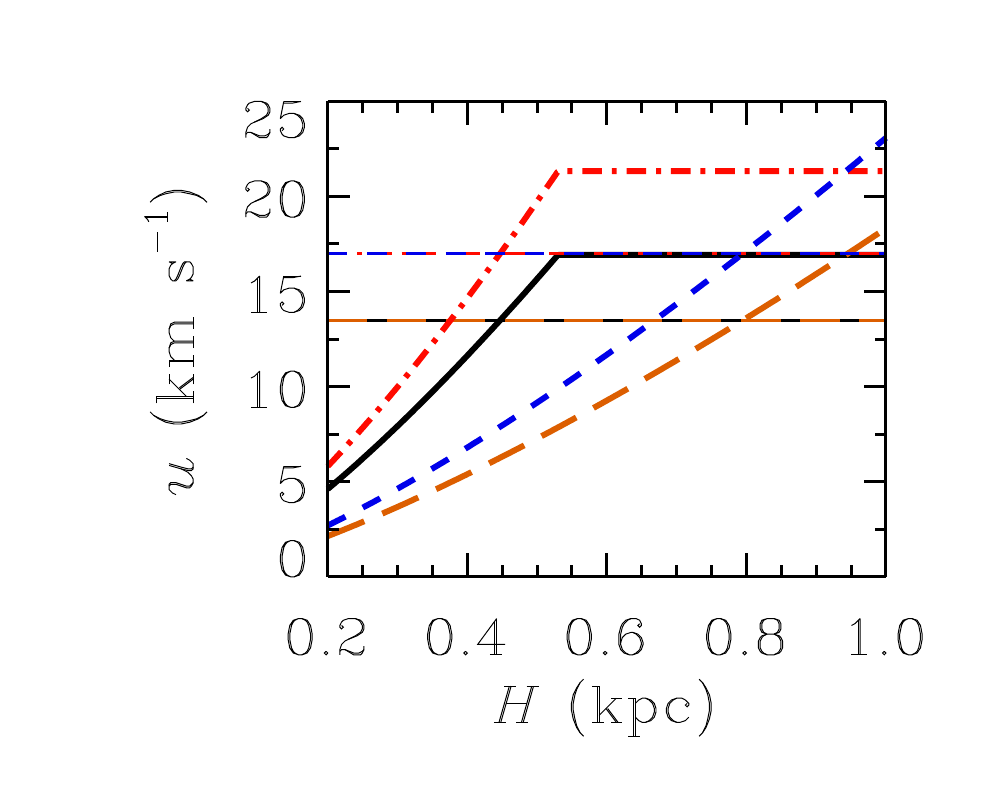}
  \includegraphics[width=28mm  ,clip=true,trim=  93 54 17 28]{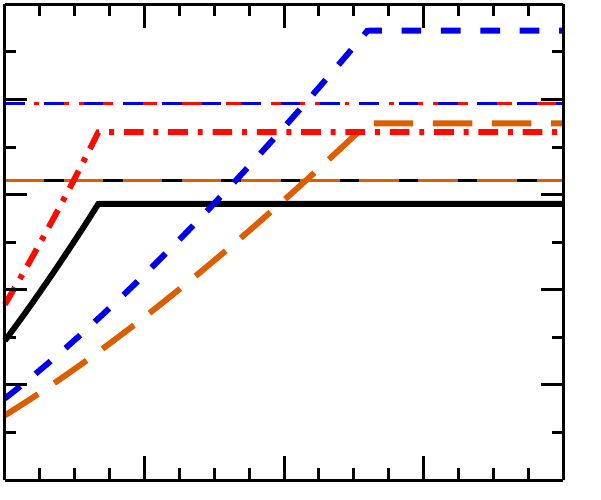}
  \includegraphics[width=28mm  ,clip=true,trim=  93 54 17 28]{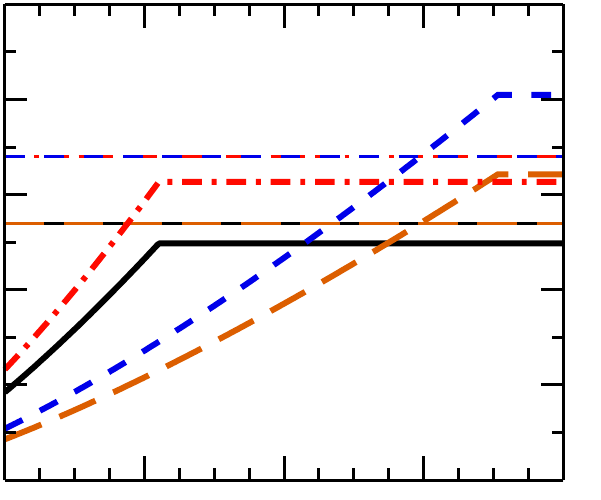}
  \includegraphics[width=28mm  ,clip=true,trim=  93 54 17 28]{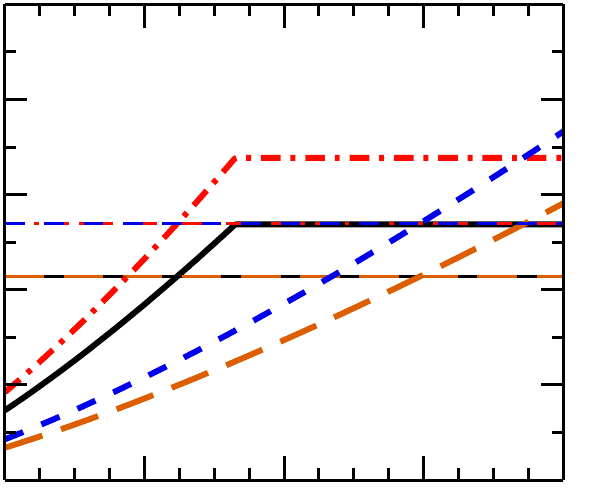}
  \includegraphics[width=28mm  ,clip=true,trim=  93 54 17 28]{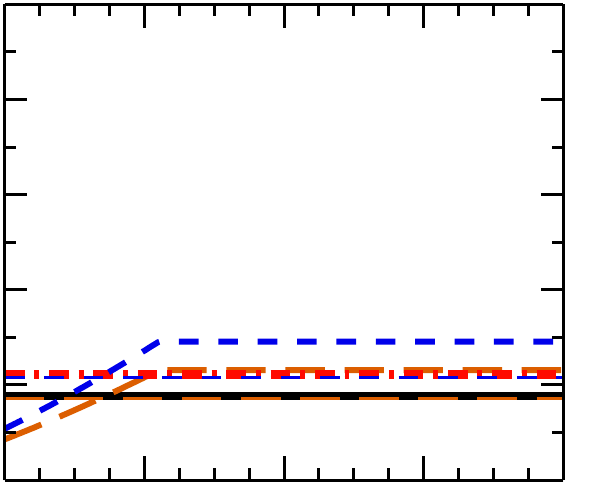}\\
  \includegraphics[width=39.2mm,clip=true,trim=  25 54 17 28]{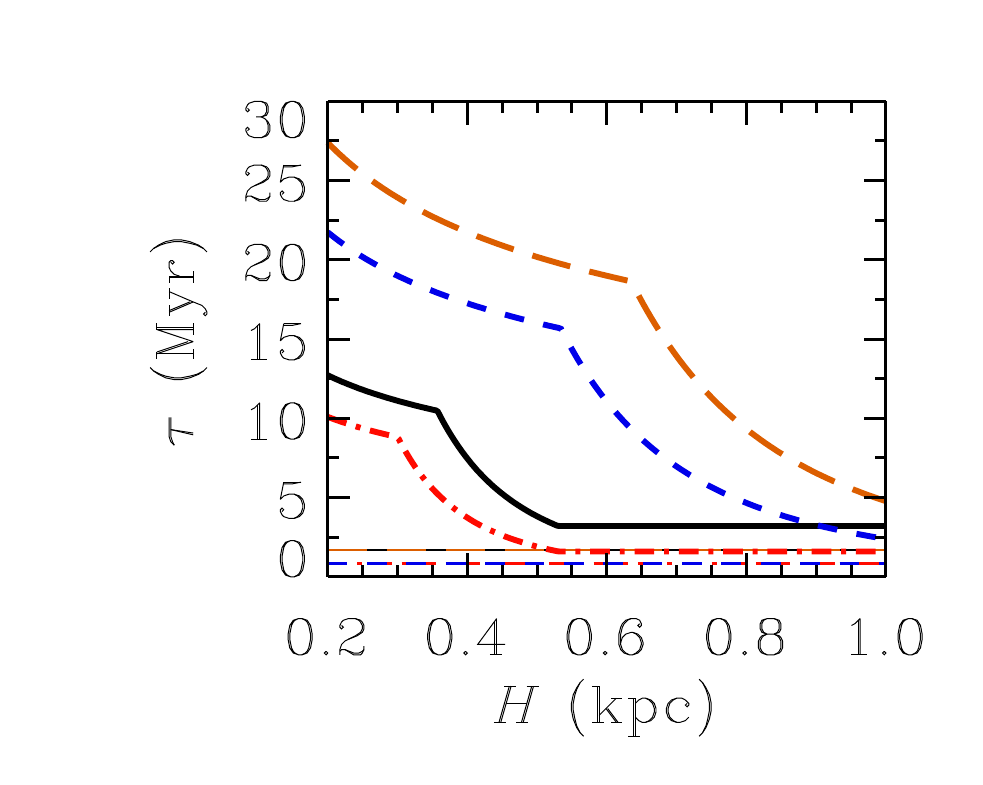}
  \includegraphics[width=28mm  ,clip=true,trim=  93 54 17 28]{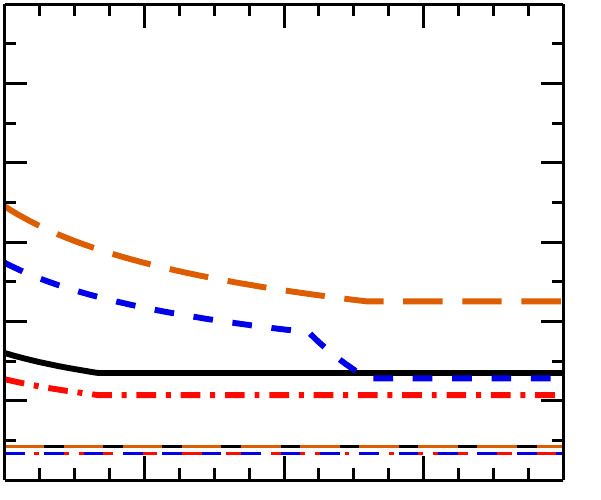}
  \includegraphics[width=28mm  ,clip=true,trim=  93 54 17 28]{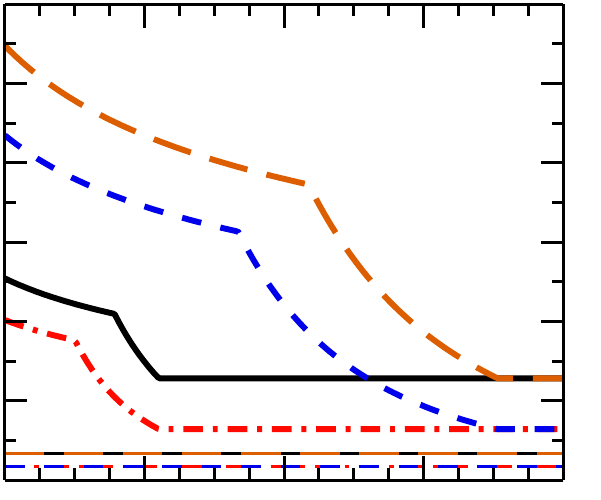}
  \includegraphics[width=28mm  ,clip=true,trim=  93 54 17 28]{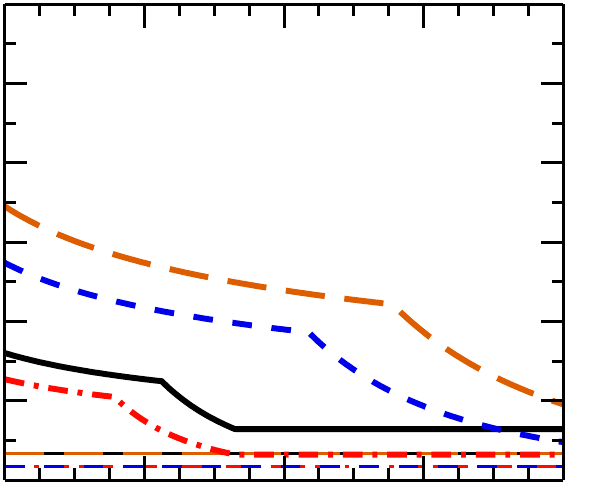}
  \includegraphics[width=28mm  ,clip=true,trim=  93 54 17 28]{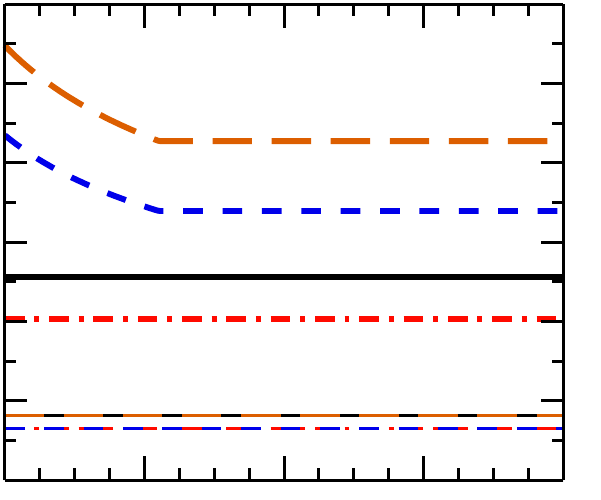}\\
  \includegraphics[width=39.2mm,clip=true,trim=  25 54 17 28]{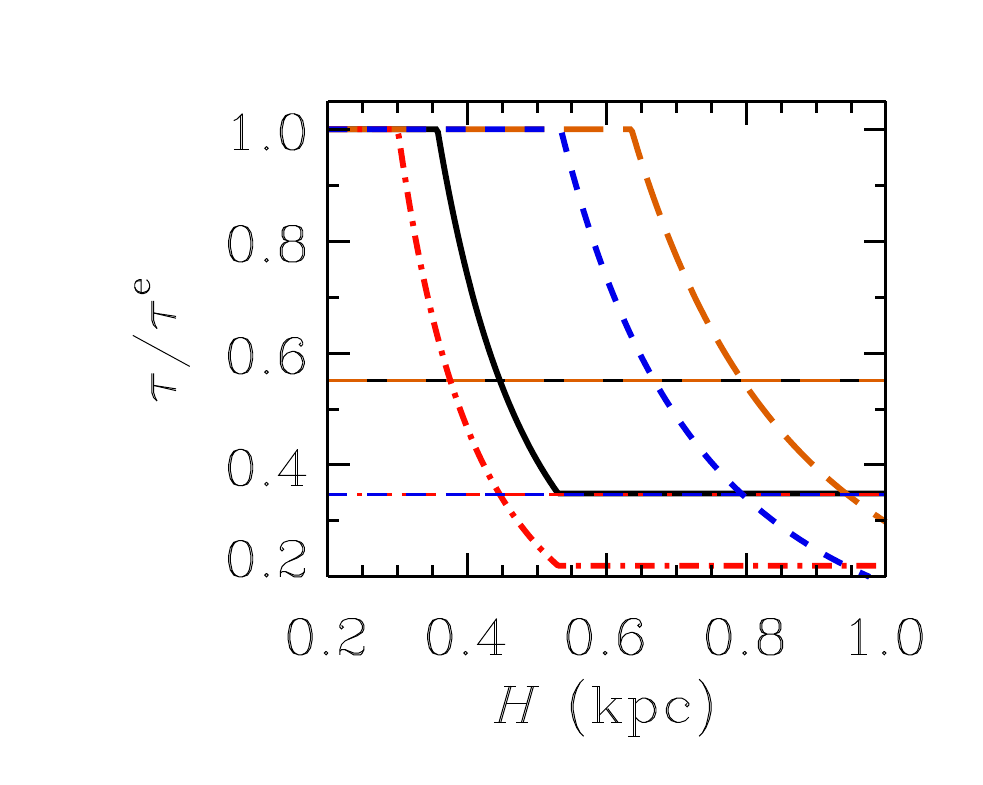}
  \includegraphics[width=28mm  ,clip=true,trim=  93 54 17 28]{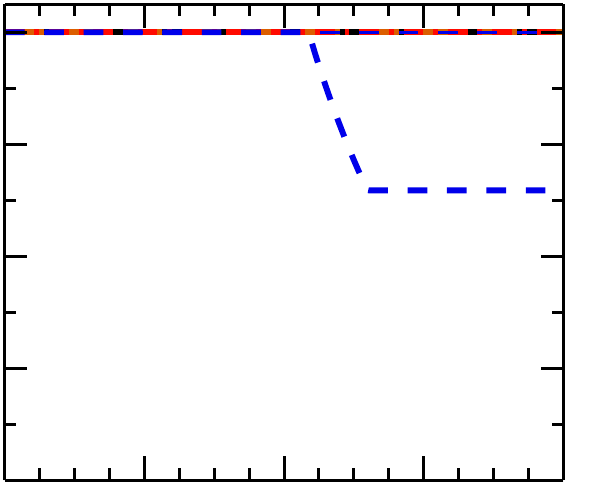}
  \includegraphics[width=28mm  ,clip=true,trim=  93 54 17 28]{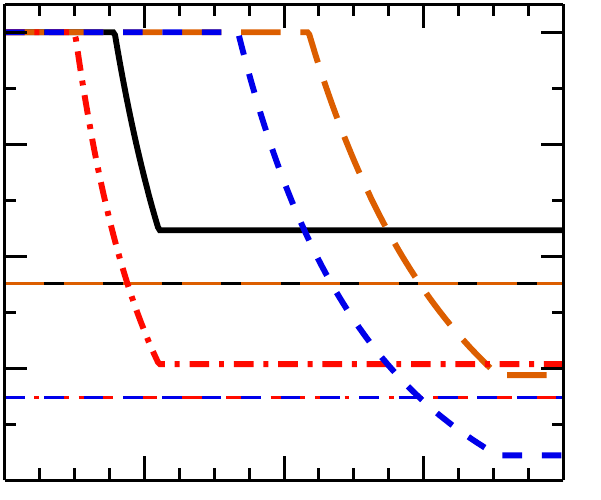}
  \includegraphics[width=28mm  ,clip=true,trim=  93 54 17 28]{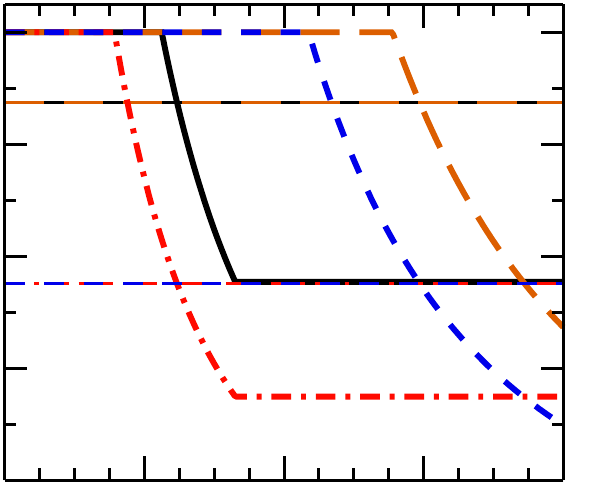}
  \includegraphics[width=28mm  ,clip=true,trim=  93 54 17 28]{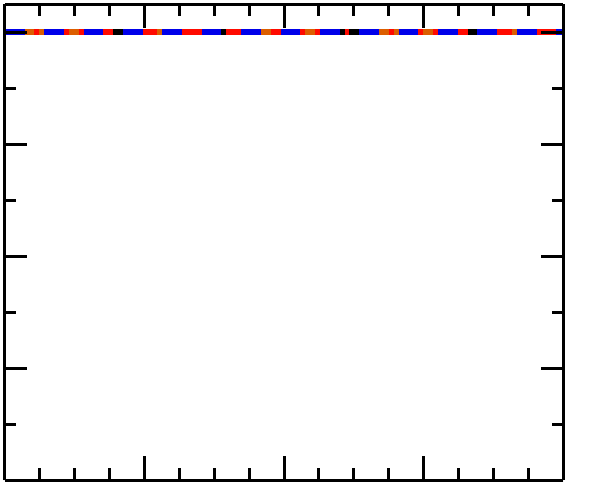}\\
  \includegraphics[width=39.2mm,clip=true,trim=  25 54 17 28]{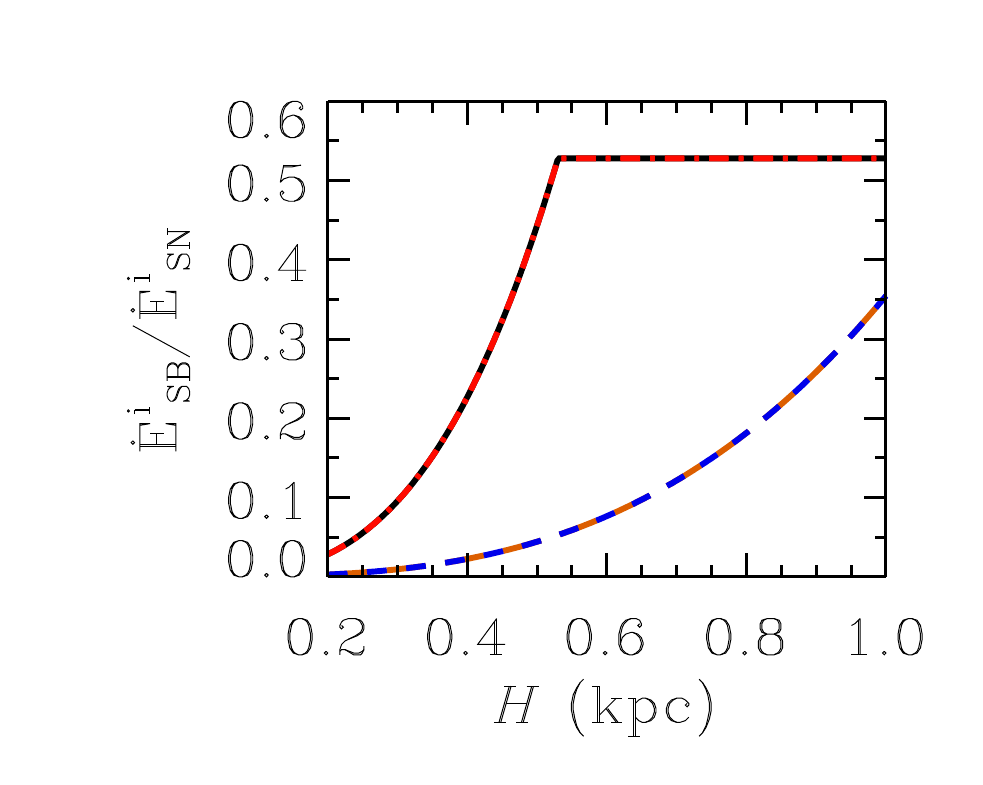}
  \includegraphics[width=28mm  ,clip=true,trim=  93 54 17 28]{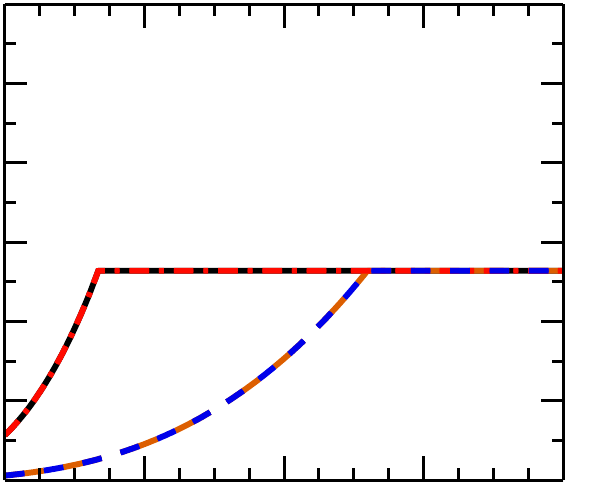}
  \includegraphics[width=28mm  ,clip=true,trim=  93 54 17 28]{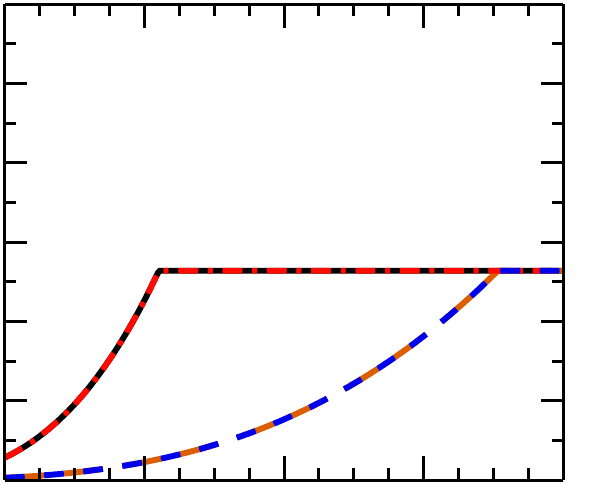}
  \includegraphics[width=28mm  ,clip=true,trim=  93 54 17 28]{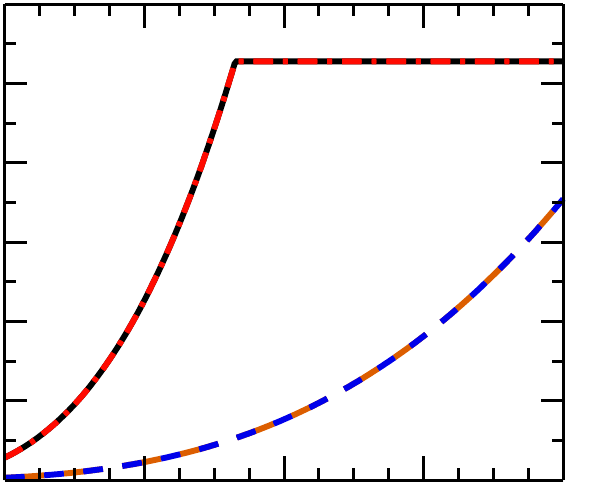}
  \includegraphics[width=28mm  ,clip=true,trim=  93 54 17 28]{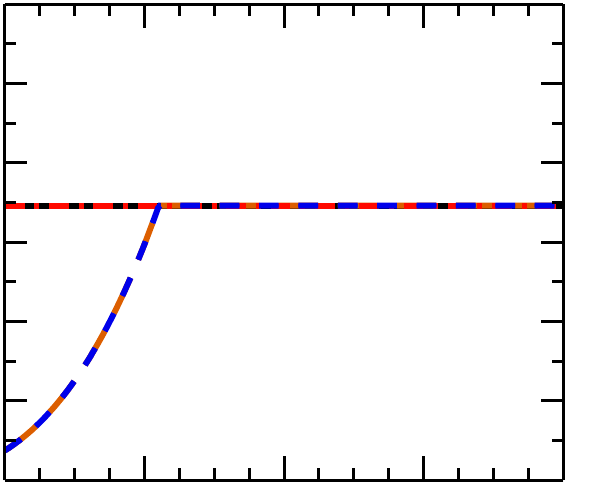}\\
  \includegraphics[width=39.2mm,clip=true,trim=  25 14 17 28]{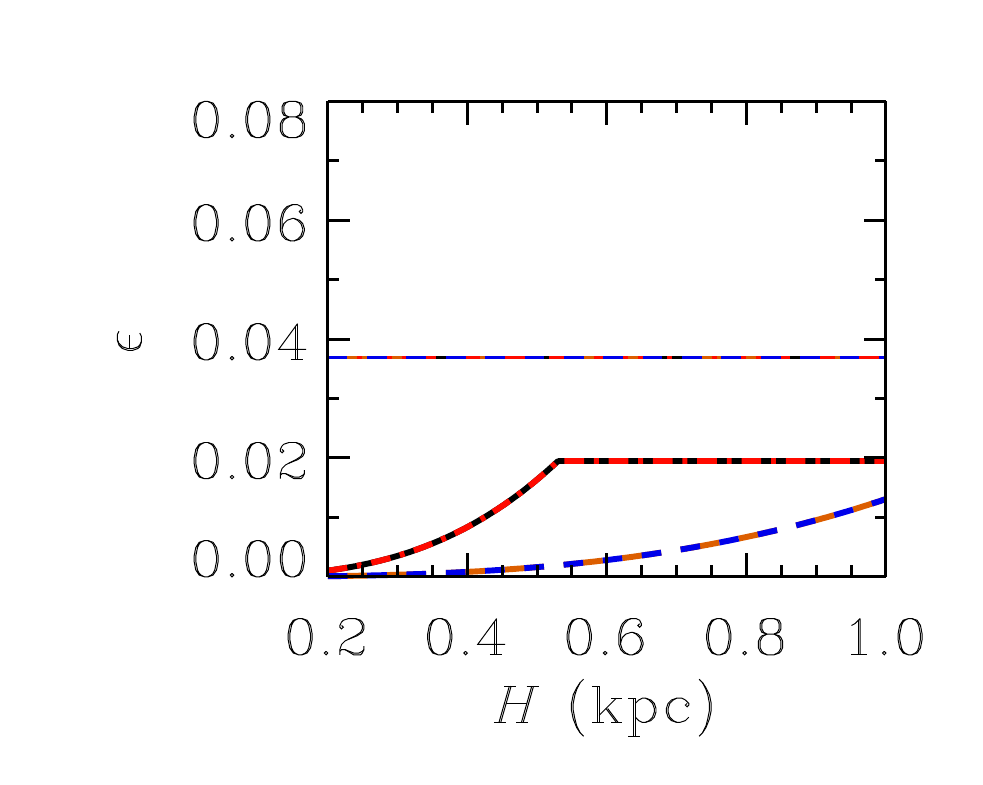}
  \includegraphics[width=28mm  ,clip=true,trim=  93 14 17 28]{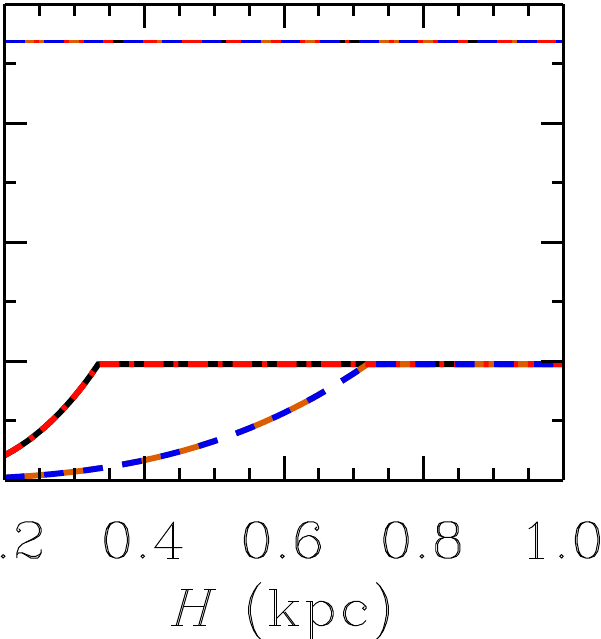}
  \includegraphics[width=28mm  ,clip=true,trim=  93 14 17 28]{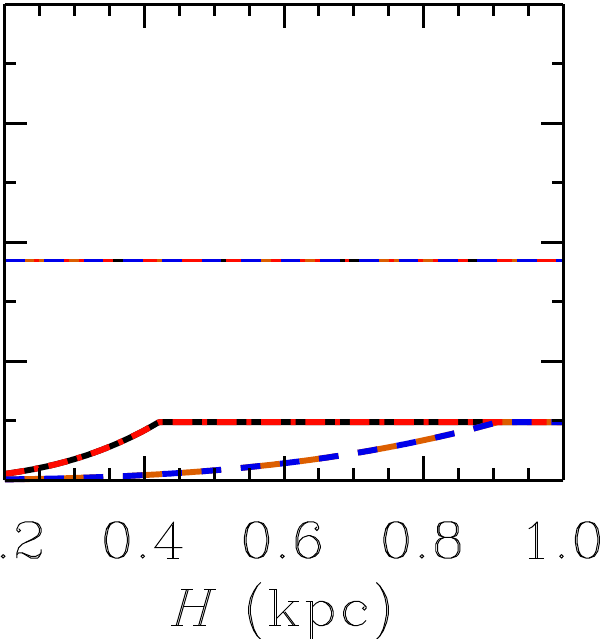}
  \includegraphics[width=28mm  ,clip=true,trim=  93 14 17 28]{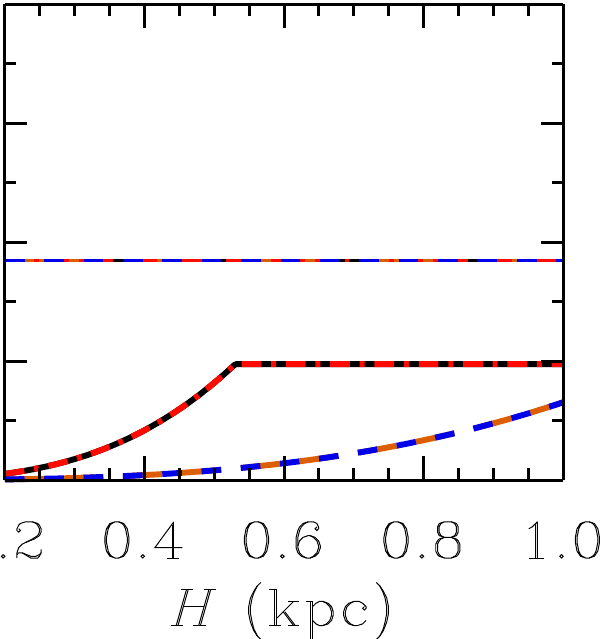}
  \includegraphics[width=28mm  ,clip=true,trim=  93 14 17 28]{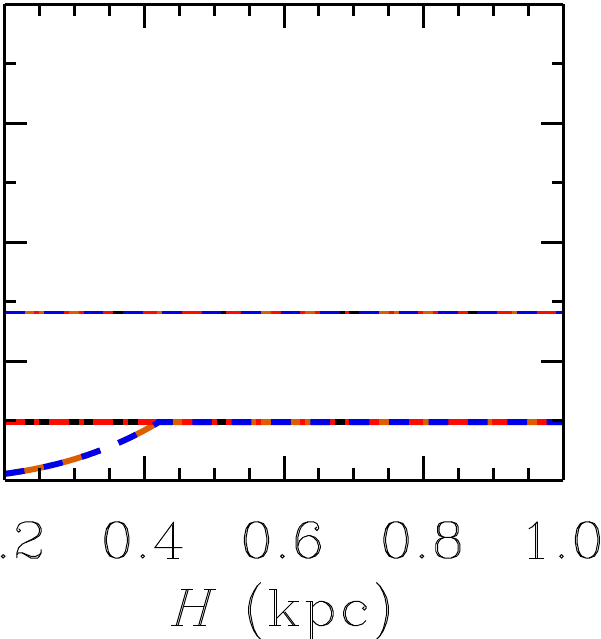}\\
  \caption{Similar to Figure~\ref{fig:pspace} but showing the cases
           where all SNe are isolated SNe (thin lines) 
           or all SNe reside in SBs (thick lines).
           \label{fig:pspace_pureSNSB}
          }
\end{figure*}

It is interesting to consider how the results change if all SNe 
are assumed to be isolated, or conversely, if all SNe are assumed to reside in SBs.
Thus, in Figure~\ref{fig:pspace_pureSNSB} we adopt the same underlying parameter values as for Figure~\ref{fig:pspace},
but now we plot the cases $f\SB=0$ (thin lines) and $f\SB=1$ (thick lines).

The correlation scale $l$ is considerably smaller for the pure isolated SN case (top row),
generally in the range $20$--$50\pc$
(in this case the thin lines coincide in each panel as $l$ does not depend on $N\SB$ or $\nu$).
For the pure SB case, $l$ increases linearly with $H$ (Equation~\ref{l_SB}) at small enough $H$ such that blowout occurs,
and can reach values as large as $300\pc$.

The value of $u$ (second row) is similar for the pure SB and pure SNR cases,
$u\sim10$--$20\kms$ for $n=0.1\cmcmcm$ and $u\sim5$--$7\kms$ for $n=1\cmcmcm$,
if blowout is avoided in the SB case. 
However if blowout does take place, values of $u$ can be much smaller in the pure SB case (as low as $2\kms$ for $H=200\pc$ and $N\SB=1000)$,
due to inefficient energy conversion from SNe to turbulence.
The pure SB low $H$ case is not very realistic though, because if even a small fraction of SNe were isolated,
that channel could dominate the energy injection, raising $u$ to values closer to the thin lines, 
as can be seen in Figure~\ref{fig:pspace} where $1/4$ of SNe are isolated.

The correlation time, like the correlation scale, is much smaller in the pure isolated SN case
as compared to the pure SB case, as seen in the third row.
In the fourth row, we see that $\tau$ is sometimes equal to $\tau\eddy$, and sometimes to $\tau\renov$, 
but for large $c\sound$ or large $n$, usually $\tau=\tau\eddy$.
The correlation time increases with stronger clustering of SNe (compare solid black to long-dashed orange and dash-dotted red to short-dashed blue),
and with reduced SN rate $\nu$ (compare red with black and blue with orange).%
\footnote{Note that we ignore possible correlations between the underlying parameters of the model.}

The ratio of injected energies for the pure isolated SN and pure SB cases,
shown in the fifth row, are similar to those shown in Figure~\ref{fig:pspace},
except lower by a factor of three.
This is because in the model plotted in Figure~\ref{fig:pspace}, 
$f\SB/(1-f\SB)=3$ times more SNe reside in SBs compared to those that are isolated.

In the bottom row we see that the conversion efficiency $\eps$ of SN energy
to turbulent energy is very low for the case $f\SB=1$ if $H$ is small.
The efficiency for $f\SB=1$ becomes larger with increasing $H$, 
but even neglecting blowout, SBs are less efficient than isolated SNe in transferring SN energy to turbulence.
Hence, clustering of SNe leads to less efficient turbulence driving in our model.
The efficiencies obtained for the pure isolated SN case ($f\SB=0$)
are generally consistent with the estimate of $4\%$ made by \citet[][Ch.~7]{Dyson+Williams97}.

If we ignore clustering of SNe ($f\SB=0$) so that all SNe are isolated,
we obtain, for fiducial parameter values, $l=42\pc$, $u=13\kms$, $\tau=\tau\renov=2\Myr$ and $\eps=0.037$,
whereas if all SNe are assumed to reside in SBs ($f\SB=1$) we obtain $l=120\pc$, $u=12\kms$, $\tau=\tau\renov=7\Myr$ and $\eps=0.008$.
If instead we retain $f\SB=3/4$ but only the isolated SNe are assumed to drive turbulence and those residing in SBs are neglected,
$l$ and $\eps$ are unchanged from the $f\SB=0$ case but $u=8\kms$ and $\tau=\tau\eddy=5\Myr$,
whereas if isolated SNe are neglected and only the SBs are assumed to drive turbulence,
$l$ and $\eps$ are unchanged from the $f\SB=1$ case but $u=11\kms$ and $\tau=\tau\renov=10\Myr$.

\section{Varying the Parameter $\xi$}
\label{sec:xi}
In our estimates above we parameterized the driving scale
of an SB that blows out as $\xi H$, 
with $H$ the gas scale height.
We then made the choice $\xi=1$,
since an SB which blows out has expanded to a horizontal scale $\sim H$,
according to simulations of \citet{Maclow+Mccray88} (c.f.~their figure 8). 
Those authors made use of a two-component density model from \citet{Lockman+86} 
which consists of a diffuse exponential component of scale height $H$ (the Lockman layer) 
and a thin Gaussian cloud layer with scale height a few times smaller than $H$.
They found in their simulations that while the vertical extent of an SB at blowout 
is a few times larger than $H$, 
it experiences a pinch due to the greater density at the midplane
so that its horizontal expanse is roughly equal to $H$.
In this section we consider adopting a smaller value of $\xi$, 
corresponding to a more severe pinch near the midplane.
In Equation~\eqref{l_SB} it is implied that an SB cannot expand to a horizontal scale $>\xi H$ 
because once it reaches that scale it blows out.
Before blowout, SB expansion is independent of $\xi$,
so a smaller $\xi$ causes SBs to blow out earlier.

\subsection{Motivation for Reducing $\xi$}
It appears to violate self-consistency that turbulence can be driven in our model 
at a scale a few times larger than the cloud layer,
since it would seem that the cloud layer would then be disrupted by such turbulence.
However, the current understanding of interstellar clouds is still fragmentary, 
and it is possible that they are transient and formed by compressions 
in the SN-shocked transonic gas \citep[e.g][]{Maclow+Klessen04,Vazquez-semadeni15}.
In any case,
the entrainment of clouds by SBs can lead to enhanced cooling in the SB interior.
Cloud evaporation in the SB interior can occur through thermal conduction,
while mixing between clouds and hot gas can lead to radiative cooling.
These processes cause a reduction in the interior pressure of the SB \citep{Mckee+Cowie77,Maclow+Mccray88,Kim+17}.
Since SBs interact energetically with the thin cloud layer, which has a scale height $h\sim130\pc$
in the Solar neighbourhood \citep{Dickey+Lockman90,Ferriere01},
the driving scale could plausibly be capped at $h\sim\tfrac{1}{3}H$ by setting $\xi$ to be $\sim1/3$ instead of $1$.
In this scenario an SB would experience a more severe pinch than predicted by \citet{Maclow+Mccray88},
presumably leading to narrower chimneys, as seen in simulations by \citet{Avillez+Breitschwerdt04},
where the horizontal size of the chimneys of hot gas produced by clusters of SNe in the local ISM is about $150$--$200\pc$.

\begin{figure*}                     
  %figures produced by ~/Turbulence_estimates/turbp.pro which uses turbcalc.pro
  \includegraphics[width=39.2mm,clip=true,trim=  25 54 17 28]{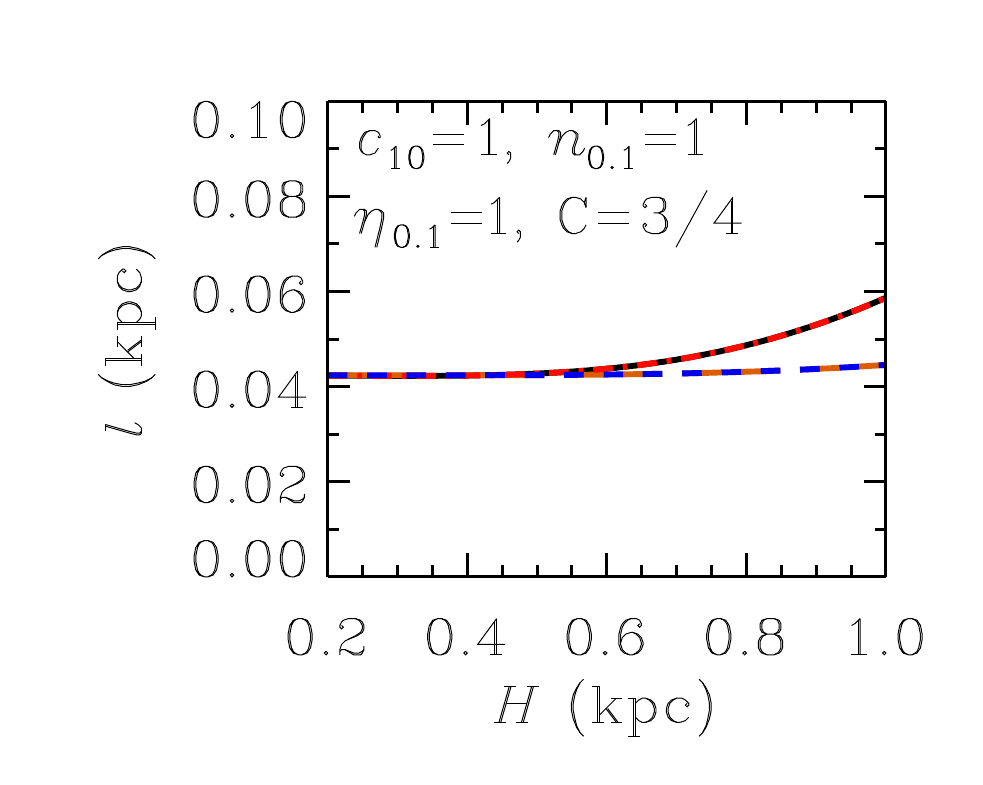}
  \includegraphics[width=28mm  ,clip=true,trim=  93 54 17 28]{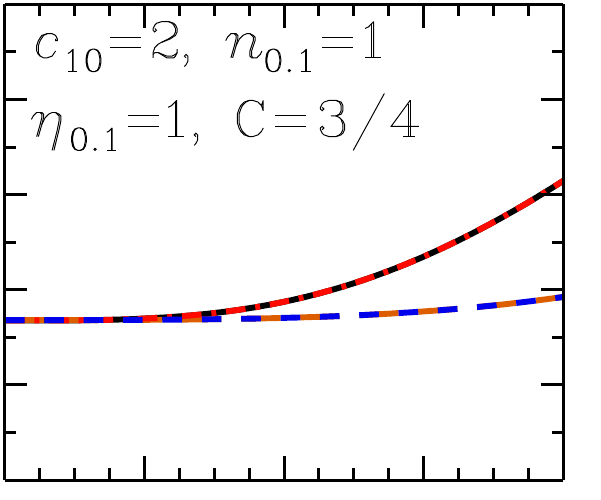}
  \includegraphics[width=28mm  ,clip=true,trim=  93 54 17 28]{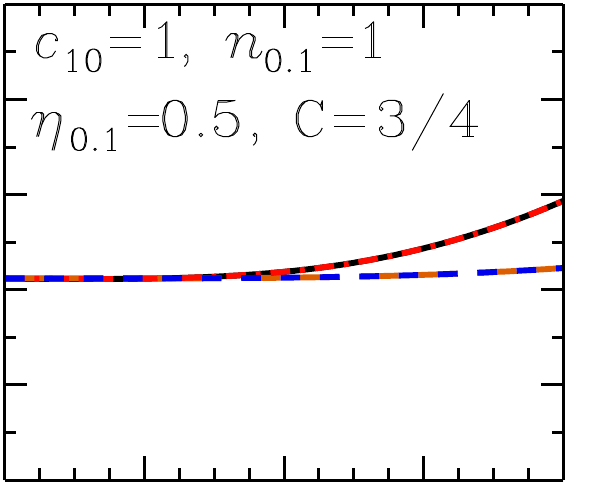}
  \includegraphics[width=28mm  ,clip=true,trim=  93 54 17 28]{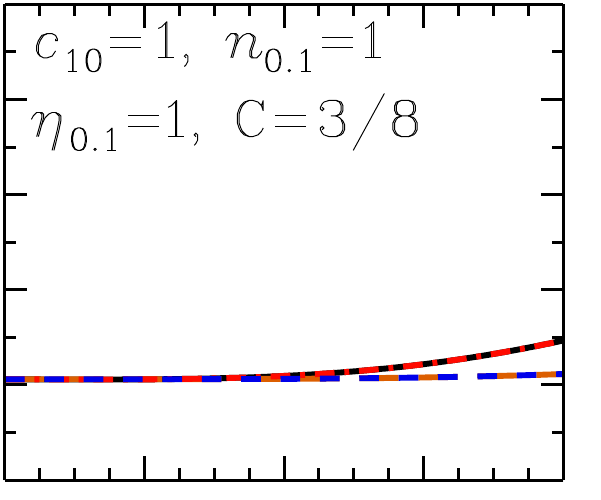}
  \includegraphics[width=28mm  ,clip=true,trim=  93 54 17 28]{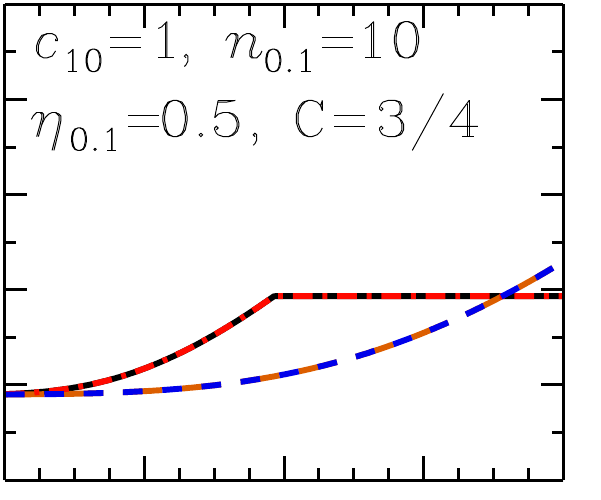}\\
  \includegraphics[width=39.2mm,clip=true,trim=  25 54 17 28]{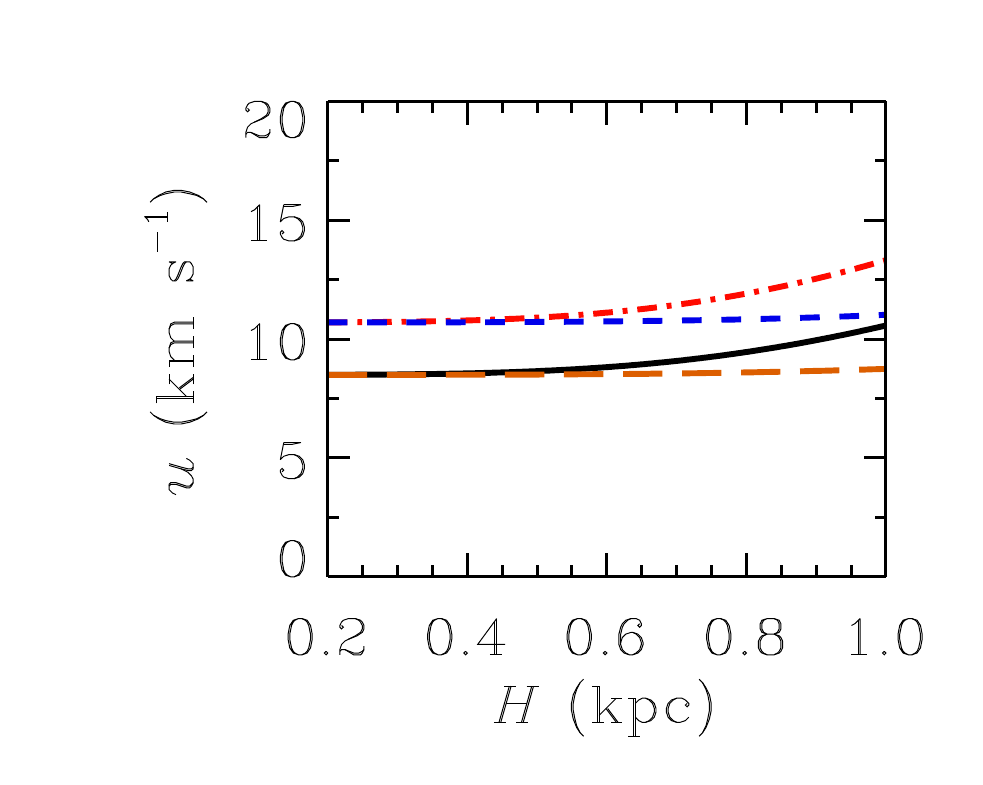}
  \includegraphics[width=28mm  ,clip=true,trim=  93 54 17 28]{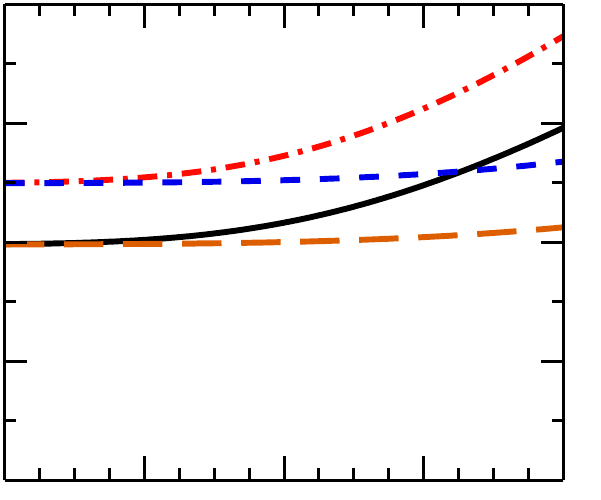}
  \includegraphics[width=28mm  ,clip=true,trim=  93 54 17 28]{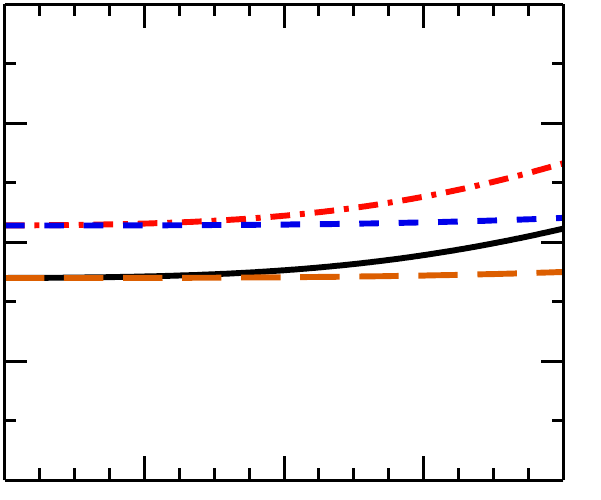}
  \includegraphics[width=28mm  ,clip=true,trim=  93 54 17 28]{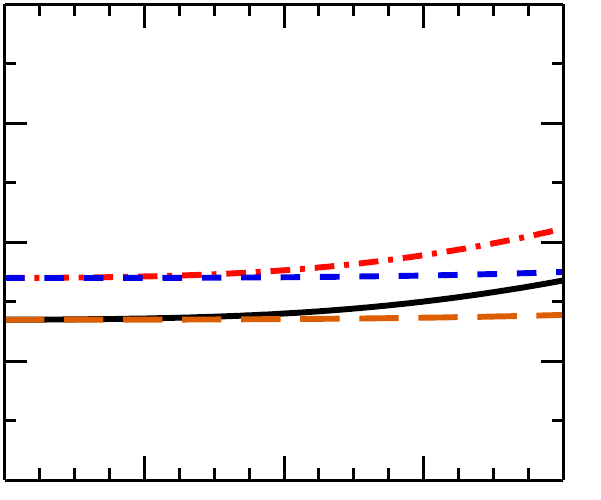}
  \includegraphics[width=28mm  ,clip=true,trim=  93 54 17 28]{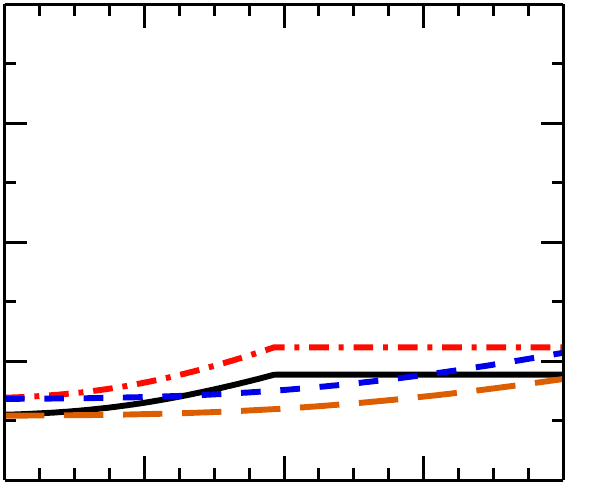}\\
  \includegraphics[width=39.2mm,clip=true,trim=  25 54 17 28]{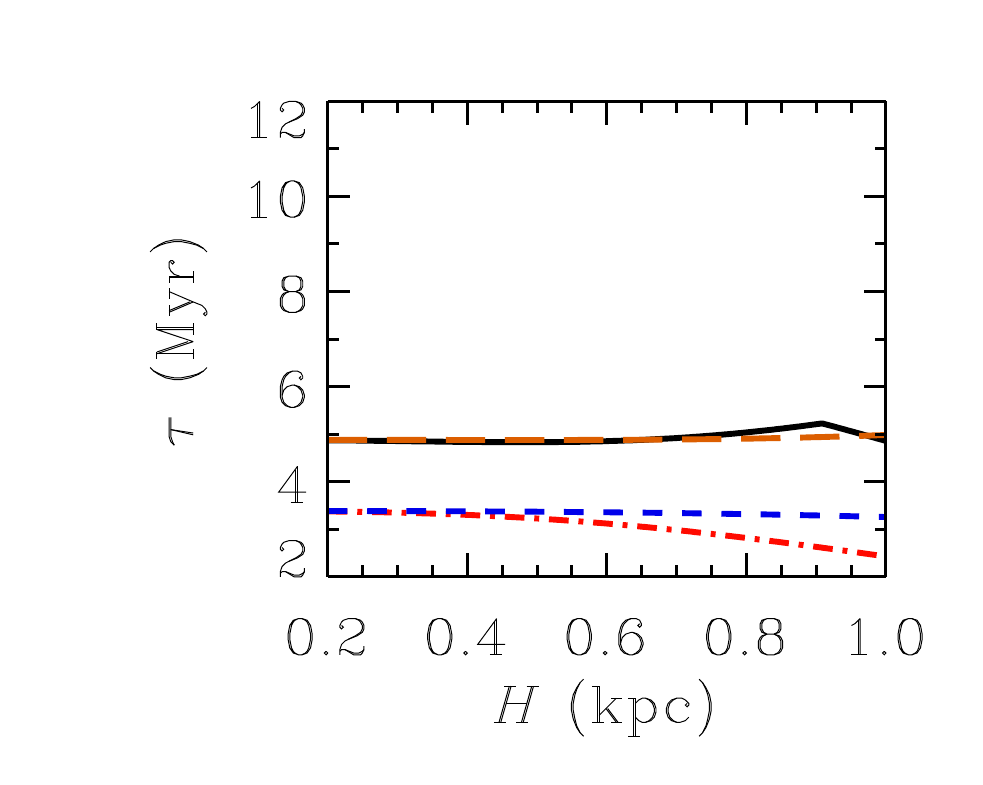}
  \includegraphics[width=28mm  ,clip=true,trim=  93 54 17 28]{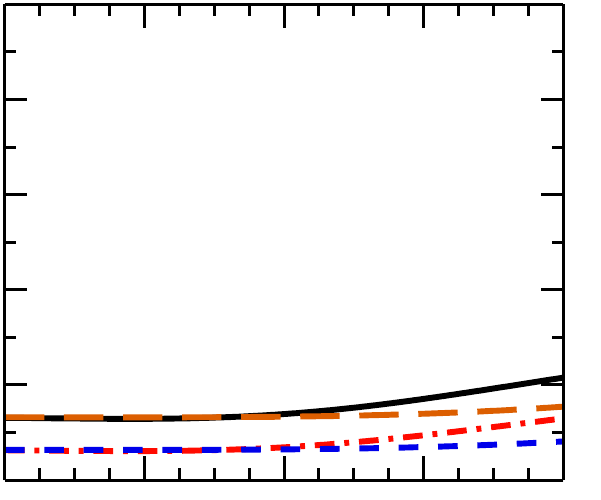}
  \includegraphics[width=28mm  ,clip=true,trim=  93 54 17 28]{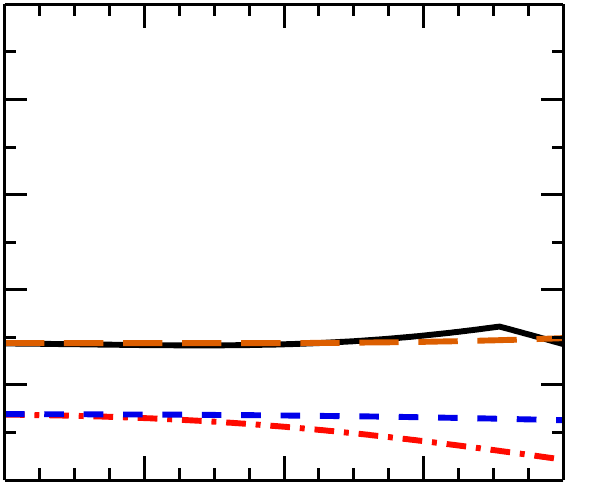}
  \includegraphics[width=28mm  ,clip=true,trim=  93 54 17 28]{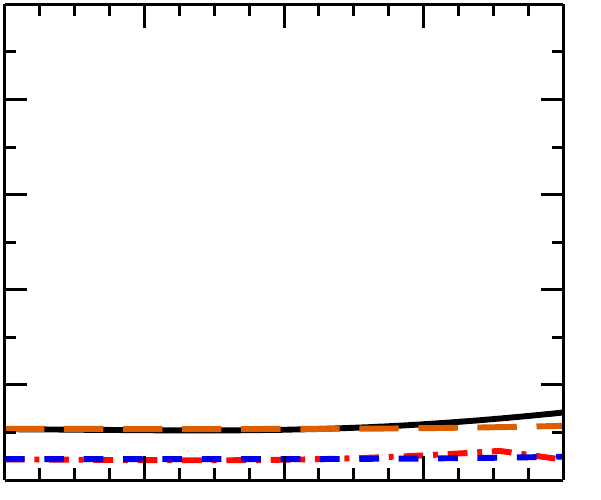}
  \includegraphics[width=28mm  ,clip=true,trim=  93 54 17 28]{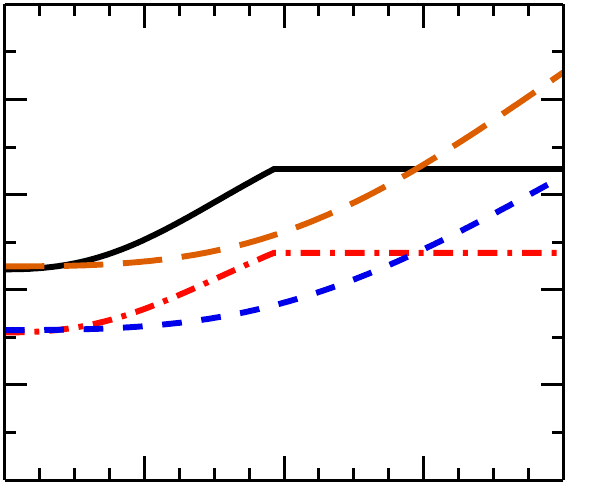}\\
  \includegraphics[width=39.2mm,clip=true,trim=  25 54 17 28]{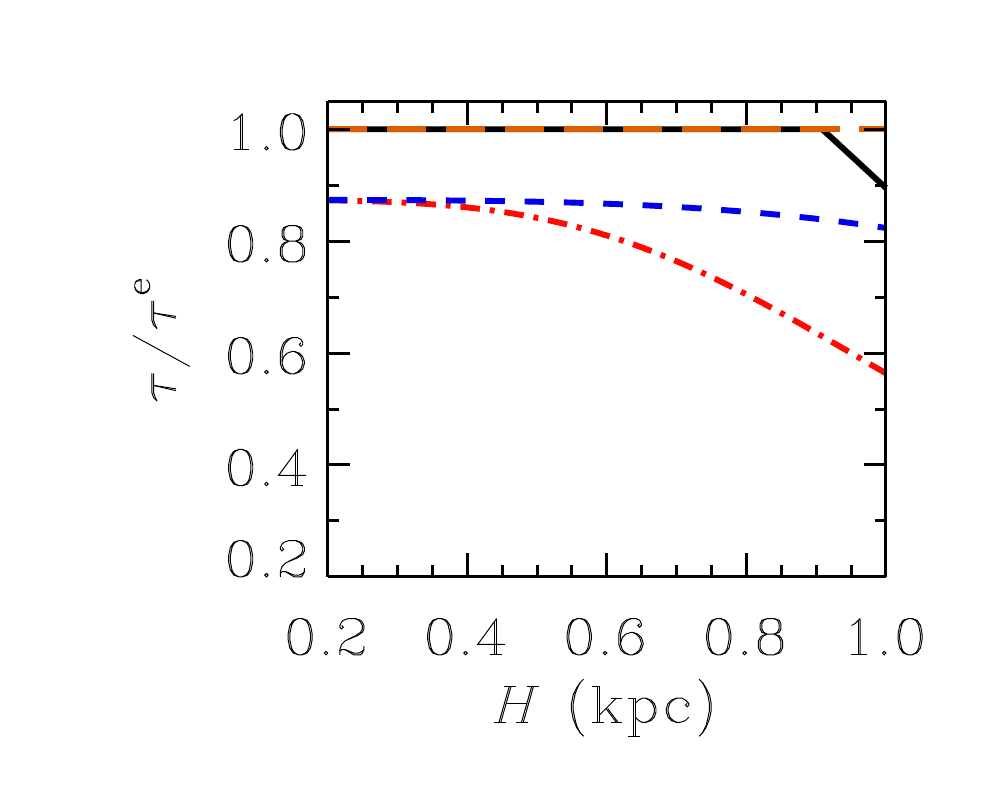}
  \includegraphics[width=28mm  ,clip=true,trim=  93 54 17 28]{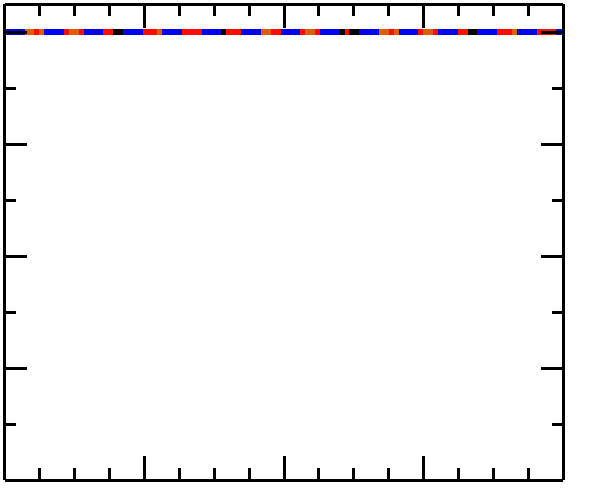}
  \includegraphics[width=28mm  ,clip=true,trim=  93 54 17 28]{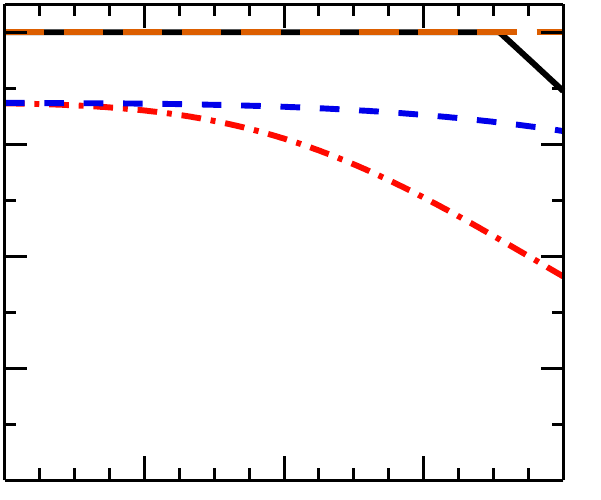}
  \includegraphics[width=28mm  ,clip=true,trim=  93 54 17 28]{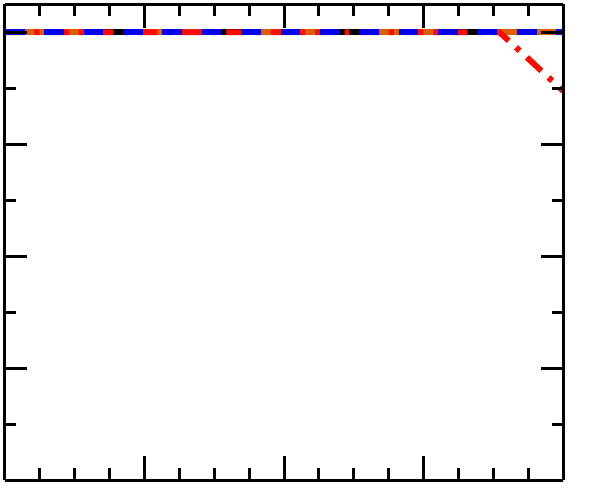}
  \includegraphics[width=28mm  ,clip=true,trim=  93 54 17 28]{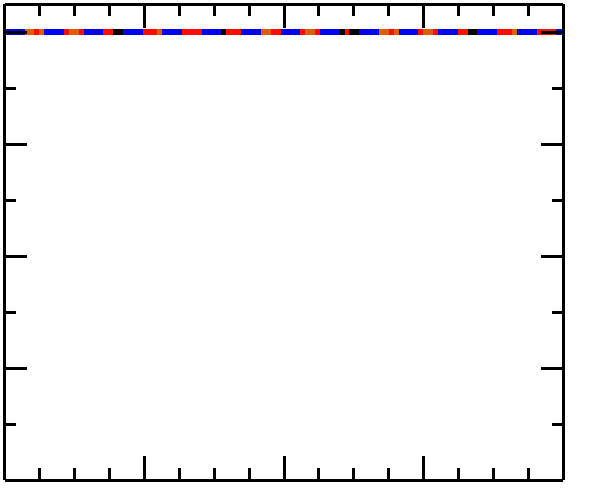}\\
  \includegraphics[width=39.2mm,clip=true,trim=  25 54 17 28]{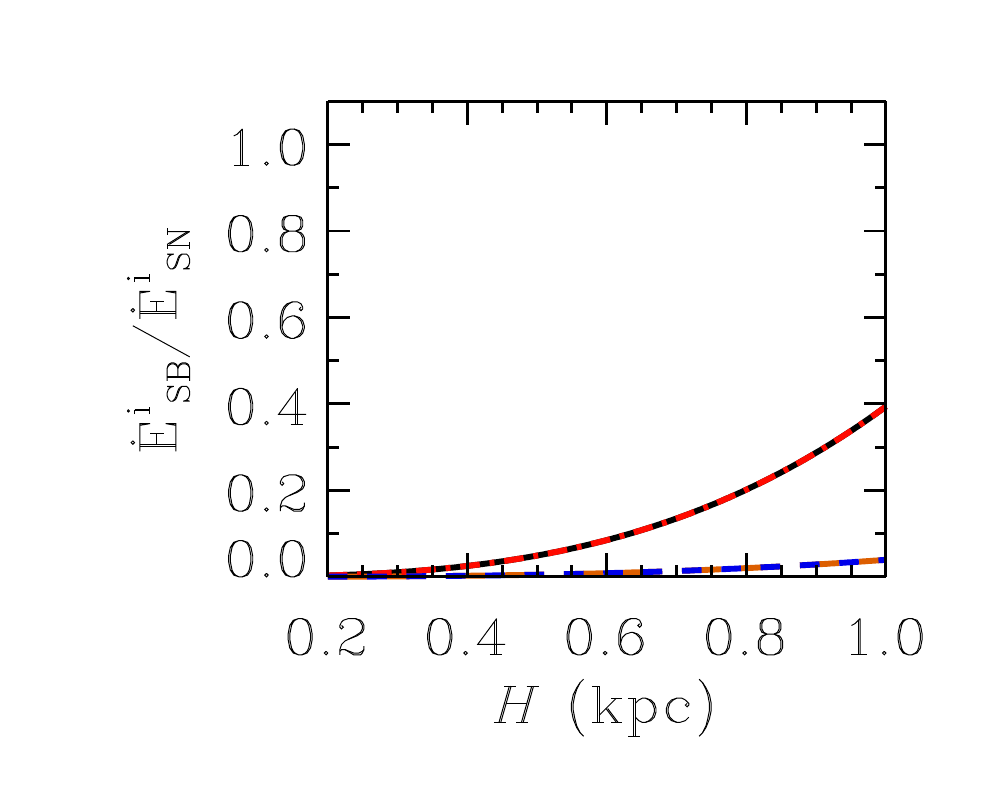}
  \includegraphics[width=28mm  ,clip=true,trim=  93 54 17 28]{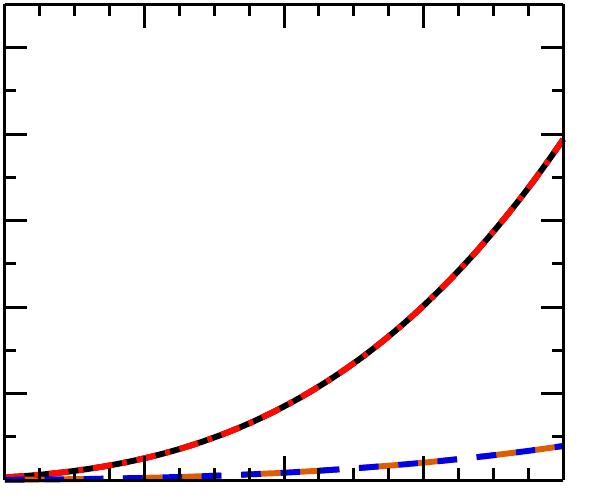}
  \includegraphics[width=28mm  ,clip=true,trim=  93 54 17 28]{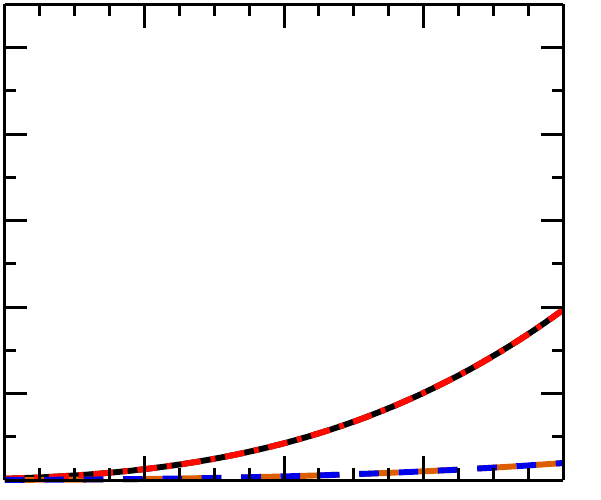}
  \includegraphics[width=28mm  ,clip=true,trim=  93 54 17 28]{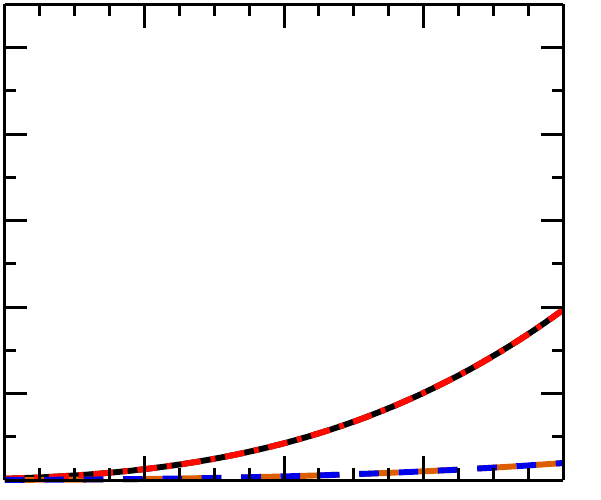}
  \includegraphics[width=28mm  ,clip=true,trim=  93 54 17 28]{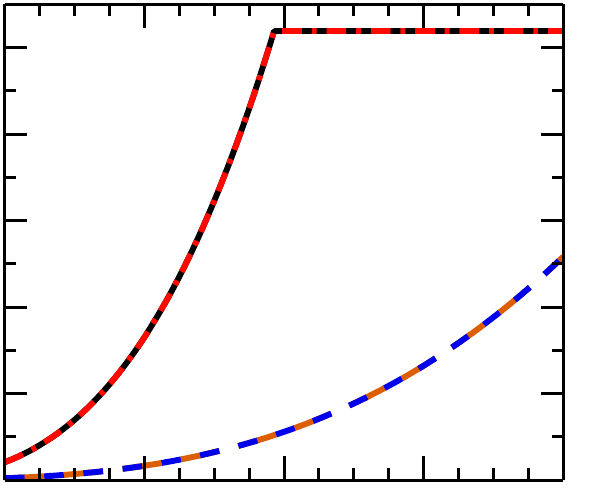}\\
  \includegraphics[width=39.2mm,clip=true,trim=  25 14 17 28]{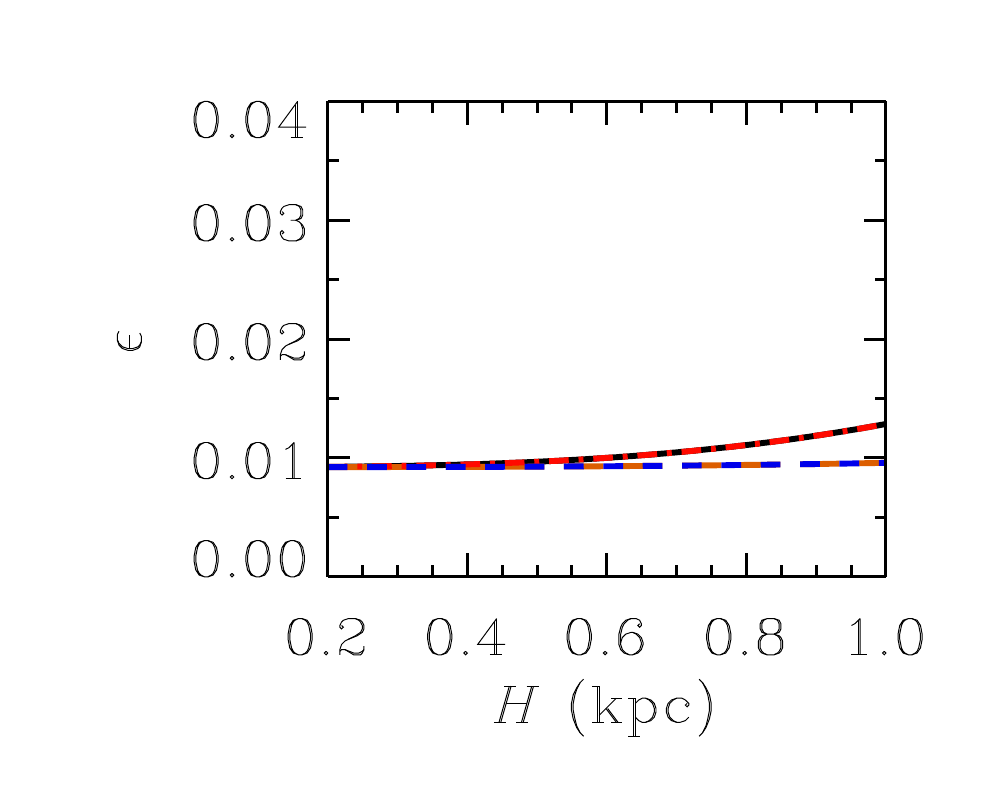}
  \includegraphics[width=28mm  ,clip=true,trim=  93 14 17 28]{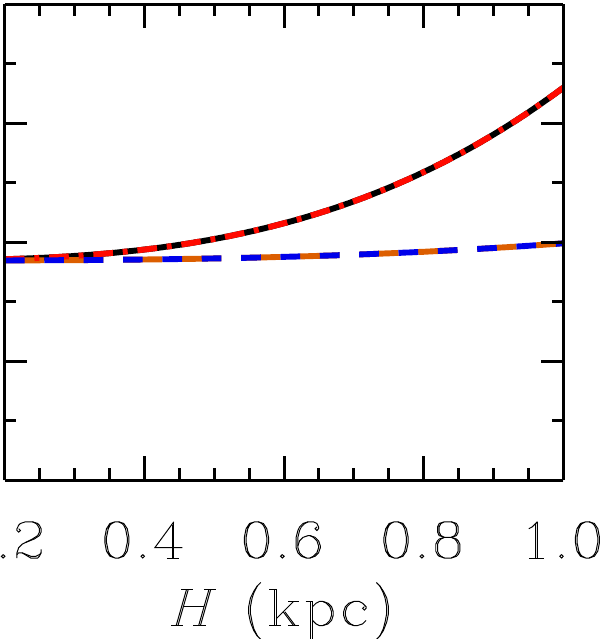}
  \includegraphics[width=28mm  ,clip=true,trim=  93 14 17 28]{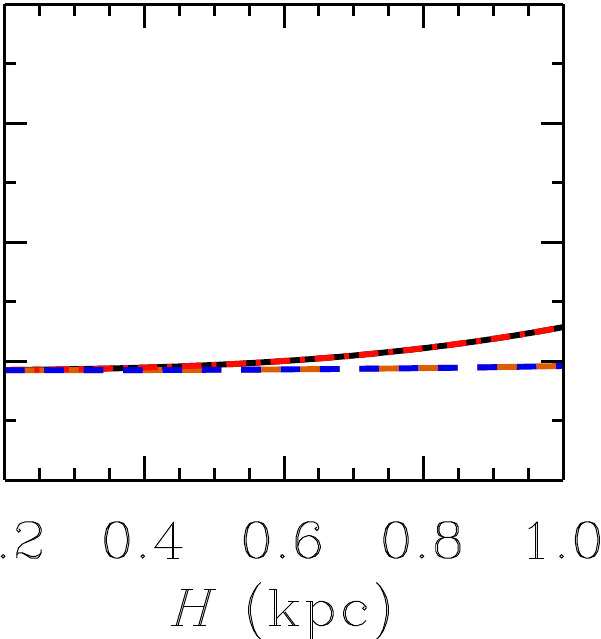}
  \includegraphics[width=28mm  ,clip=true,trim=  93 14 17 28]{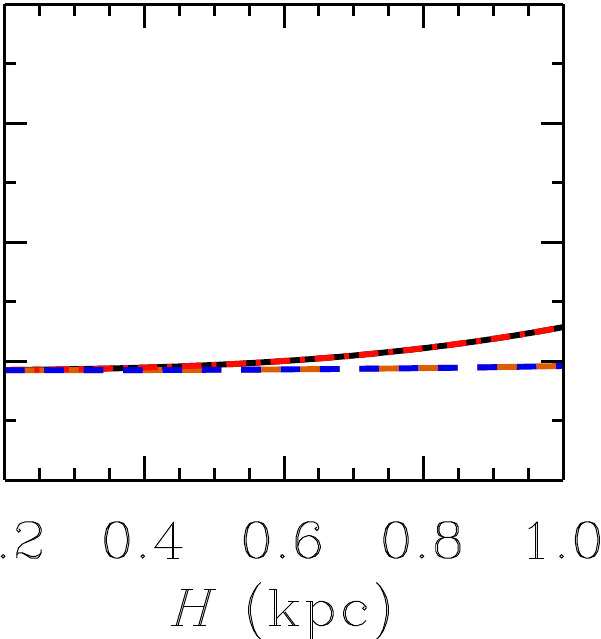}
  \includegraphics[width=28mm  ,clip=true,trim=  93 14 17 28]{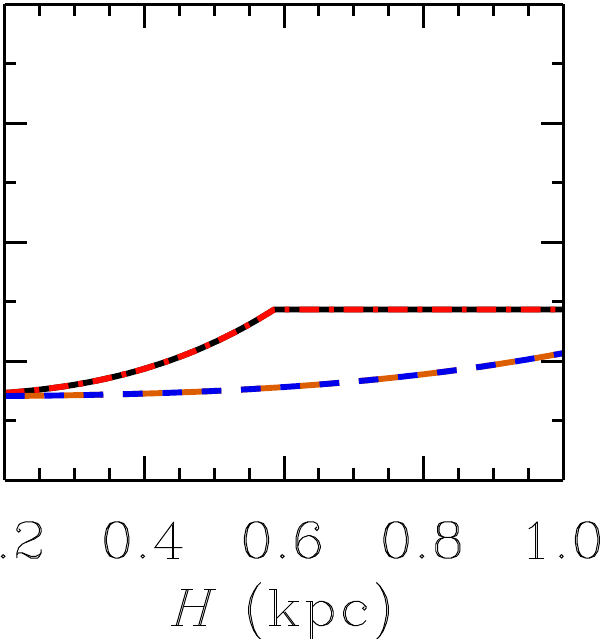}\\
  \caption{Similar to Figure~\ref{fig:pspace} but for the case $\xi=1/3$.
           \label{fig:pspace_xi1over3}
          }
\end{figure*}

\subsection{Results}
With this in mind, we present results of a model that is the same as that presented in Sections~\ref{sec:turb} and \ref{sec:pspace} 
except that $\xi=1/3$ instead of $1$.
Results of this alternative small-$\xi$ model are shown in Figure~\ref{fig:pspace_xi1over3},
which is otherwise similar to Figure~\ref{fig:pspace}.
The efficiency $\eps$ is lower in the $\xi=1/3$ model (bottom row) as compared to the $\xi=1$ model
because SBs blow out early in their expansions and thus transfer very little energy to the ISM.
In fact, the results are very similar to those obtained if turbulence driving by SBs is completely neglected,
but $f\SB=3/4$ is still assumed. 
This case has not been plotted, but it is just the $H\rightarrow0$ limit of the plots. 
When blowout occurs, the differences between the $\xi=1/3$ and $\xi=1$ cases are more significant for larger values of $H$,
where energy injected by SBs is a larger fraction of the total injected energy.
Only in the last column of Figure~\ref{fig:pspace_xi1over3}, where $n=1\cmcmcm$, 
can blowout be suppressed at large $H$ due to the high ambient density.

Adopting $\xi=1/3$ generally leads to smaller $l$ due to the reduction in large-scale driving,
and smaller $u$ due to the reduction in energy input and in $l$, compared to $\xi=1$. 
The value of $\tau$ is not greatly affected.
For $\xi=1/3$ these quantities depend only weakly on $H$ and $N\SB$, since these parameters only affect SBs.
The values obtained when $\xi=1/3$ and other parameters fiducial are $l\approx42\pc$, $u\approx9\kms$, $\tau=\tau\eddy\approx5\Myr$ and $\eps\approx0.009$ 
(first column solid black at $H=0.4\kpc$),
as compared with $l\approx74\pc$, $u\approx12\kms$, $\tau=\tau\renov\approx4\Myr$ and $\eps\approx0.016$ for $\xi=1$ (Figure~\ref{fig:pspace}).

\subsection{Two-layer Model}
\label{sec:two-layer}
A further possibility would be to generalize the model to consist of two distinct gas layers,
namely the cloud and Lockman layers, each with a different set of properties.
This treatment would allow for the possibility that average turbulence parameters 
vary with distance from the midplane $|z|$.
Naturally $l$ (and $\xi$) would increase with $|z|$ because SBs 
would bulge out into the more tenuous outer layer.
However, in a two-layer model one would need to consider how SN energy gets divided between the two layers.
One would also need to consider the possibility of individual SNe blowing out of the thin layer and into the thick layer.
Another possibility is to include explicitly the stratification of the ISM. 

\section{Limitations and Opportunities for Extending the Model}
\label{sec:discussion}
On account of keeping the model reasonably simple,
certain details have necessarily been omitted.
Expanding SN shocks may reflect off of interstellar clouds to produce secondary shocks \citep{Spitzer82}. 
The distribution function $P(\mu)$ of secondary shocks with Mach number $\ge\mu$
was calculated by \citet{Bykov+Toptygin87}, who assumed secondary shocks to be weak. 
The distribution function derived diverges unphysically for $\mu\rightarrow1$,
which leads to a vanishing renovation time (Equation~\ref{tau_SN_renov}).
Clearly $\tau\renov$ would decrease if secondary shocks were included,
but including them would require a new model that remains valid as $\mu\rightarrow1$,
which is beyond the scope of our simple treatment here.
On the other hand, we have also neglected mutual isolated SNe-SNe, isolated SNe-SB or SB-SB interactions,
as we have assumed that every spherical shock propagates independently of neighbouring shocks.
Such interactions probably make shock expansion less efficient and would thus lead to an increase in $\tau\renov$,
which could help to offset the omission of secondary shocks.

We have also chosen to omit the possible influence of galactic shear on the correlation time.
In their galactic dynamo model, \citet{Zhou+Blackman17} adopt $\tau^{-1}= (\tau\eddy)^{-1} +q\Omega$,
where $\Omega(r)$ is the large-scale angular velocity of gas at radius $r$, and $q(r)=-d\ln\Omega/d\ln r$. 
Here $q=1$ corresponds to a locally flat rotation curve.
In the vicinity of the Solar neighbourhood, $\Omega\sim20$--$30\kmskpc$, 
which translates to a shearing timescale of $\sim30$--$50\Myr$.
This is large compared to our estimates for $\tau$, so shear would not cause an important reduction in $\tau$.
However, this effect might be important for parts of the phenomenologically relevant parameter space.

A more detailed model would also allow for a distribution of $N\SB$,
rather than considering only the extreme case of isolated SNe and SBs each containing the same number of SNe, 
as we have done here \citep{Oey+Clarke97,Ferriere98,Nath+20}.
Whereas we have treated isolated SNe and SBs as different kinds of object
with different similarity solutions governing their evolutions, 
there would in reality be a transition from one type to the other as $N\SB$ is increased from $1$ to larger values.

Another possibility is to include effects of the magnetic field on the expansion of SBs \citep[e.g.][]{Evirgen+19},
which would provide nonlinear feedback in a dynamo model.

We have assumed that SB blowout occurs instantaneously when $R\SB=\xi H$, with $\xi=1$ for our fiducial model.
At this time, the SB merges with the ISM, driving turbulence.
A more sophisticated treatment would model the blowout phase 
from the time the SB begins to lose pressure support to the time it breaks up near the midplane.
Using the similarity solution for the SB temperature as a function of radius \citep{Maclow+Mccray88},
we compute the sound crossing time for the SB at blowout to be $\sim4\Myr$ for our fiducial parameter values,
as compared with $t\blowout\uSB=15\Myr$, so modeling this phase could lead to important differences.

Our model assumes a uniform ISM of a given disc semi-thickness $H$, so does not differentiate between the various ISM phases.
Thus, it effectively averages over the individual locations of isolated SNe, both in terms of distance from the midplane 
and occurrence inside or outside molecular clouds.
Furthermore, it does not distinguish between the average properties of the ambient medium encountered by SBs as opposed to that encountered by isolated SNe,
or between the average properties of the medium in which SNRs and SBs expand as opposed to that in which the turbulence is primarily driven.
Refinements to the model to address theses shortcomings would entail introducing more parameters,
and it is not clear whether this would be warranted given the various uncertainties.

More fundamentally, our model could also be extended to include turbulence driving by mechanisms other than SN feedback.

\section{Summary and Conclusions}
\label{sec:conclusions}
Various astrophysical phenomena, including galactic dynamos, are sensitive to the parameters of interstellar turbulence.
However, determining the values of these parameters, 
as well as their dependencies on other galactic or interstellar medium parameters, has remained an elusive goal.
We have applied standard similarity solutions for SNRs and SBs to a model of SN-driven interstellar turbulence, 
and obtained simple analytic expressions for the velocity correlation scale $l$, 
RMS turbulent speed $u$, velocity correlation time $\tau$, 
and SN to turbulent energy transfer efficiency $\eps$.

Our main motivation for this work was to extend dynamo models
such that they can be parameterized by quantities that are more accessible than $l$, $u$ and $\tau$,
namely, the underlying model parameters summarized in Table~\ref{tab:params}.
In this way, $l$, $u$ and $\tau$ would become intermediate quantities computed within the dynamo model, 
rather than input parameters.
As it is beyond the scope of the present work, we leave the implementation of this idea for the future. 
This could be done using already available dynamo models \citep[e.g.][]{Chamandy+15,Chamandy16,Chamandy+16,Rodrigues+19a},
but the underlying ISM parameters of Table~\ref{tab:params}, including their spatial and temporal dependencies,
would need to be modelled independently and/or constrained using observations.
The relations between turbulence parameters and underlying ISM parameters that we have derived are testable and may have many applications besides dynamos.
However, care must be taken to account for separate relations between the underlying parameters themselves.
No attempt is made to include these additional dependencies in our model.

In Figure~\ref{fig:pspace}, we present solutions over a large region of the underlying parameter space. 
We obtain the following approximate ranges for the quantities computed: 
$l\sim20$--$175\pc$, $u\sim3$--$23\kms$, $\tau\sim1$--$13\Myr$ and $\eps\sim0.01$--$0.04$.
For our fiducial set of underlying parameter values, applicable to the Solar neighbourhood, 
we obtain $l\approx74\pc$, $u\approx12\kms$, $\tau\approx4\Myr$ and $\eps\approx0.016$,
which can be compared with ``canonical'' order-of-magnitude estimates of $l\sim100\pc$, $u\sim10\kms$, $\tau\sim10\Myr$ and $\eps\sim0.04$,
and recent local ISM simulations of \citet{Hollins+17} which find $l\approx74\pc$, $u\approx13\kms$ and $\tau\approx5\Myr$.
The close agreement with the latter results is partly just a coincidence, 
but our results are in broad agreement with other simulations as well \citep[e.g.][]{Avillez+Breitschwerdt07,Gressel+08a,Gressel+Elstner20}.

Our model includes both isolated SNe and SBs, with a fraction $f\SB=3/4$ of SNe residing in SBs.
Turbulent energy is deposited in the ISM once the expansion velocity of an SNR or SB reaches the ambient sound speed.
We find that isolated SNe and SBs can both contribute significantly to the turbulent energy injection in the ISM.
This suggests that both of these sources should be accounted for in models of interstellar turbulence.%
\footnote{\citet{Yoo+Cho14} study MHD simulations with forcing on two scales
and find that even a relatively small amount of energy injection on the larger scale 
can have important effects on the properties of the turbulence.}
However, one is free to choose to include only one or the other of these components.
In particular, the model is simpler and contains many fewer parameters if clustering of SNe to form SBs is neglected ($f\SB=0$).

The evolution of SBs depends on the density scale height $H$ of the warm diffuse ambient gas 
(the Lockman layer in the Milky Way), 
and SB blowout happens in our model when the SB radius in the midplane is of order the ambient scale height \citep{Maclow+Mccray88}.
For $H\lesssim0.5\kpc$, 
SBs tend to blow out of the disc, and consequently, a smaller fraction of their energy ends up in interstellar turbulence.
Blowout is assumed to result in a rapid loss of pressure support near the midplane,
and a sudden reduction of the expansion speed to the ambient sound speed, at which point the SB merges with the ISM.
If blowout happens early, i.e. when the SB expansion speed $\dot{R}\SB\gg c\sound$,
then the energy deposited into the ISM is only a small fraction $\sim H^3/R\SB^3(t\sound\uSB)$
of what would have been injected had the SB been able to expand up to the time $t\sound\uSB$ when $\dot{R}\SB(t\sound\uSB)=c\sound$.
In this case, isolated SNe dominate the energy injection into turbulence in our model.

We also computed the fraction of SN energy that ends up in turbulence, 
and found this efficiency factor to typically lie in the range $\eps=0.01$--$0.04$.
This agrees closely with the range found by \citet{Bacchini+20} for the warm atomic gas.
The efficiency of conversion of SN energy to turbulent energy is found to be lower when SBs are included, 
compared to a model for which all SNe are isolated.
This is because SBs can blow out and also because SBs drive turbulence less efficiently than isolated SNe even when they do not blow out,
as seen in the bottom row of Figure~\ref{fig:pspace_pureSNSB}.
Since most SNe can be contained within SBs, the reduction in the overall efficiency can be significant.
When SN clustering is absent ($f\SB=0$) in our model, all SNe are isolated.
This results in values of $l$ and $\tau$ that are smaller than the $f\SB=3/4$ case, 
and values of $\eps$ that are larger, while the opposite is true if $f\SB=1$.

It could be argued that large driving scales comparable to $H$ cannot be present 
because the resulting turbulence would then disrupt the thin HI cloud layer, 
which has a scale height of $\sim130\pc$ in the Solar neighbourhood.
This motivated us to consider a variation on the model
in which an SB blows out earlier, when its radius in the midplane is equal to $\xi H$ with $\xi=1/3$ instead of $1$.
As $\xi$ is reduced, SBs end up being less important in transferring energy to turbulence in the disc.
For $\xi=1/3$, we obtain values of $l$ up to a few times lower than in the $\xi=1$ case,
and values of $u$ up to two times smaller, but $\tau$ remains about the same.\\

\vspace{12pt}

\authorcontributions{Conceptualization, methodology and writing: L.C.~and A.S. All authors have read and agreed to the published version of the manuscript.}

\funding{This research was funded by the Leverhulme Trust Grant RPG-2014-427 and the STFC Grant ST/N000900/1 (Project 2).}

\acknowledgments{We thank Kandaswamy Subramanian, Luiz~Felippe~S.~Rodrigues, Eric Blackman, Andrew Fletcher,
James Hollins, Can Evirgen, Amit Seta and Blakesley Burkhart for useful discussions. 
We are grateful to Jennifer Schober and Amit Seta for providing insightful comments on an earlier version of the manuscript.
We are also grateful to the two referees for providing constructive suggestions for improving the paper.
L.C.~acknowledges travel support from Newcastle University and support from NSF Grant AST-1813298.}

\conflictsofinterest{The authors declare no conflict of interest.}

%-------------------------------------------------------------------------------------------
\bibliography{refs}
\end{document}